\documentclass{nature}

\newcounter{firstbib}

\usepackage{graphicx}
\usepackage{deluxetable}
\usepackage{tablefootnote}
\usepackage{url}
\usepackage{amsmath,times}
\usepackage{lineno}
\usepackage{multirow}

\usepackage{caption}
\usepackage{subcaption}

\newcommand{\Msun}{~{\rm M}_\odot}

\newcommand{\gcm}{\rm ~g~cm^{-3}}

\newcommand{\kms}{\rm ~km~s^{-1}}
\newcommand{\ergs}{\rm ~erg~s^{-1}}

\newcommand\ion[2]{#1$\,${\small \MakeUppercase{\romannumeral#2}}}%

\newcommand\farcs{\mbox{$.\!\!^{\prime\prime}$}}%
\newcommand{\package}[1]{\texttt{#1}}
\newcommand{\swift}{\textit{Swift} }

\newcommand{\FeII}{Fe~{\sc ii}}
\newcommand{\MgII}{Mg~{\sc ii}}

\newcommand{\CaII}{\ion{Ca}{2}}
\newcommand{\AlIII}{\ion{Al}{3}}
\newcommand{\ZnII}{\ion{Zn}{2}}
\newcommand{\CrII}{\ion{Cr}{2}}
\newcommand{\MnII}{\ion{Mn}{2}}
\newcommand{\NIV}{\ion{N}{4}}
\newcommand{\FeIV}{\ion{Fe}{4}}
\newcommand{\HeII}{\ion{He}{2}}

\newcommand{\CIV}{\ion{C}{4}}
\newcommand{\CIII}{\ion{C}{3}}

\newcommand{\NIII}{\ion{N}{3}}

\newcommand{\NIVscat}{N\,{\sc iv}\,$\lambda 1718$}
\newcommand{\NIIIscat}{N\,{\sc iii}\,$\lambda 1747$}
\newcommand{\CIIIscat}{C\,{\sc iii}\,$\lambda 2297$}

\usepackage[dvipsnames]{xcolor}

\usepackage{amssymb}
\usepackage{amsmath}
\usepackage{tablefootnote}

\newcommand{\aap}{Astron. Astrophys.}

\newcommand{\apj}{Astrophys. J.}

\newcommand{\apjl}{Astrophys. J. Letters}
\newcommand{\apjs}{Astrophys. J. Supplements}

\newcommand{\pasp}{Publications of the Astronomical Society of the Pacific}

\title{Resolving the explosion of supernova 2023ixf in Messier~101 within its complex circumstellar environment}

\author{E.~A.~Zimmerman,$^{1,29*}$
I.~Irani,$^{1,29*}$
P.~Chen,$^{1}$
A.~Gal-Yam,$^{1}$
S.~Schulze,$^{2}$
D.~A.~Perley,$^{3}$
J.~Sollerman,$^{4}$
A.~V.~Filippenko,$^{5}$
T.~Shenar,$^{6}$
O.~Yaron,$^{1}$
S.~Shahaf,$^{1}$
R.~J.~Bruch,$^{7, 1}$
E.~O.~Ofek,$^{1}$
A.~De~Cia,$^{8, 9}$
T.~G.~Brink,$^{5}$
Y.~Yang,$^{5}$
S.~S.~Vasylyev,$^{5}$
S.~Ben~Ami,$^{1}$
M.~Aubert,$^{10}$
A.~Badash,$^{1}$
J.~S.~Bloom,$^{5}$
P.~J.~Brown,$^{11}$
K.~De,$^{12, 13}$
G.~Dimitriadis,$^{14}$
C.~Fransson,$^{4}$
C.~Fremling,$^{15, 16}$
K.~Hinds,$^{3}$
A.~Horesh,$^{17}$
J.~P.~Johansson,$^{2}$
M.~M.~Kasliwal,$^{16}$
S.~R.~Kulkarni,$^{16}$
D.~Kushnir,$^{1}$
C.~Martin,$^{18}$
M.~Matuzewski,$^{18}$
R.~C.~McGurk,$^{19}$
A.~A.~Miller,$^{20, 21}$
J.~Morag,$^{1}$
J.~D.~Neil,$^{16}$
P.~E.~Nugent,$^{22, 5}$
R.~S.~Post,$^{23}$
N.~Z.~Prusinski,$^{18}$
Y.~Qin,$^{16}$
A.~Raichoor,$^{22, 5}$
R.~Riddle,$^{15}$
M.~Rowe,$^{11}$
B.~Rusholme,$^{24}$
I.~Sfaradi,$^{17}$
K.~M.~Sjoberg,$^{25, 26}$
M.~Soumagnac,$^{27, 22}$
R.~D.~Stein,$^{16}$
N.~L.~Strotjohann,$^{1}$
J.~H.~Terwel,$^{14, 26}$
T.~Wasserman,$^{1}$
J.~Wise,$^{3}$
A.~Wold,$^{24}$
L.~Yan,$^{15}$
K.~Zhang,$^{5, 28}$
}

\begin{document}

\maketitle

\begin{affiliations}
  \item Department of Particle Physics and Astrophysics, Weizmann Institute of Science, 76100 Rehovot, Israel
  \item The Oskar Klein Centre, Department of Physics, Stockholm University, AlbaNova, SE-106 91 Stockholm, Sweden
  \item Astrophysics Research Institute, Liverpool John Moores University, IC2, Liverpool Science Park, 146 Browlow Hill, Liverpool L3 5RF, UK
  \item The Oskar Klein Centre, Department of Astronomy, Stockholm University, AlbaNova, SE-106 91 Stockholm, Sweden
  \item Department of Astronomy, University of California, Berkeley, CA 94720-3411, USA
  \item Departamento de Astrof'isica, Centro de Astrobiolog'ia (CSIC-INTA), Ctra.\ Torrej'on a Ajalvir km 4, 28850 Torrej'on de Ardoz, Spain
  \item The School of Physics and Astronomy, Tel Aviv University, Tel Aviv, 69978, Israel
  \item European Southern Observatory, Karl-Schwarzschild Str. 2, 85748 Garching bei München, Germany
  \item Department of Astronomy, University of Geneva, Chemin Pegasi 51, 1290 Versoix, Switzerland
  \item Université Clermont Auvergne, CNRS/IN2P3, LPC, F-63000 Clermont-Ferrand, France
  \item Department of Physics and Astronomy, Texas A\&M University, 4242 TAMU, College Station, TX 77843, USA
  \item MIT-Kavli Institute for Astrophysics and Space Research, 77 Massachusetts Ave., Cambridge, MA 02139, USA
  \item NASA Einstein Fellow
  \item School of Physics, Trinity College Dublin, The University of Dublin, Dublin 2, Ireland
  \item Caltech Optical Observatories, California Institute of Technology, Pasadena, CA 91125, USA
  \item Division of Physics, Mathematics and Astronomy, California Institute of Technology, Pasadena, CA 91125, USA
  \item Racah Institute of Physics, The Hebrew University of Jerusalem, Jerusalem 91904, Israel
  \item Cahill Center for Astronomy and Astrophysics, California Institute of Technology, 1200 E. California Blvd, MC 249-17, Pasadena, CA 91125, USA
  \item W. M. Keck Observatory, 65-1120 Mamalahoa Hwy, Kamuela, HI 96743, USA
  \item Department of Physics and Astronomy, Northwestern University, 2145 Sheridan Rd, Evanston, IL 60208, USA
  \item Center for Interdisciplinary Exploration and Research in Astrophysics (CIERA), Northwestern University, 1800 Sherman Ave, Evanston, IL 60201, USA
  \item Lawrence Berkeley National Laboratory, 1 Cyclotron Road, Berkeley, CA 94720, USA
  \item Post Observatory, Lexington, MA 02421, USA
  \item IPAC, California Institute of Technology, 1200 E. California Blvd, Pasadena, CA 91125, USA
  \item Department of Astronomy, Harvard University, Cambridge, MA 02138, USA
  \item Isaac Newton Group (ING), Apt. de correos 321, E-38700, Santa Cruz de La Palma, Canary Islands, Spain
  \item Department of Physics, Bar-Ilan University Ramat-Gan 52900, Israel
  \item Department of Astronomy \& Astrophysics, University of California, San Diego, La Jolla, CA, USA
  \item These authors contributed equally: E.~A.~Zimmerman, I.~Irani
\end{affiliations}


\clearpage

\begin{abstract}

Observing a supernova explosion shortly after it occurs can reveal important information about the physics of stellar explosions and the nature of the progenitor stars of supernovae (SNe)\cite{Waxman2017}. When a star with a well-defined edge explodes in vacuum, the first photons to escape from its surface appear as a brief shock-breakout flare. The duration of this flare can extend to at most a few hours\cite{Waxman2017,Bersten2018} even for nonspherical breakouts from supergiant stars\cite{Goldberg2022,Morag2023}, after which the explosion ejecta should expand and cool. Alternatively, for stars exploding within a distribution of sufficiently dense optically thick circumstellar material, the first photons escape from the material beyond the stellar edge, and the duration of the initial flare can extend to several days, during which the escaping emission indicates photospheric heating\cite{Ofek2010}. The difficulty in detecting SN explosions promptly after the event has so far limited data regarding supergiant stellar explosions mostly to serendipitous observations\cite{Bersten2018,Garnavich2016} that, owing to the lack of ultraviolet (UV) data, were unable to determine whether the early emission is heating or cooling, and hence the nature of the early explosion event. 
Here, we report observations of SN\,2023ixf in the nearby galaxy Messier 101, covering the early days of the event. Using UV spectroscopy from the {\it Hubble Space Telescope (HST)} as well as a comprehensive set of additional multiwavelength observations, we trace the photometric and spectroscopic evolution of the event and are able to temporally resolve the emergence and evolution of the SN emission. We derive a reliable bolometric light curve of the event and show that it indicates an initially rising temperature and luminosity as the supernova explosion shock breaks out from a dense layer of material surrounding the star with a radius $R_{0}\approx 1.9\times10^{14}\,$cm, significantly larger than typical supergiants. Our {\it HST} spectra reveal the dynamics of the shock and the expanding emitting material, as well as a rich spectrum of absorption lines from heavy elements mixed into the gas surrounding the exploding star. Our data uniquely provide a temporally-resolved description of a massive-star explosion, characterise the pre-explosion evolution of the doomed star, and demonstrate a method to directly measure the pre-explosion chemical composition of supernova progenitors.  

\end{abstract}


On 2023, May 19, SN\,2023ixf 
was discovered in the nearby (distance $6.9 \rm ~Mpc$\cite{Riess2022}) galaxy Messier 101 by K. Itagaki\cite{Itagaki2023} at 17:27:15 (UTC dates are used throughout this paper) and was reported to the Transient Name Server (TNS; https://www.wis-tns.org/object/2023ixf) at 21:42:21. The new object is located at right ascension  $\alpha=14^h03^m38^s.56$ and declination $\delta=54^{\circ}18^{\prime}41^{\prime\prime}.94$ (J2000), in the outskirts of its host (Methods~\S1).
Upon receiving this report, we rapidly obtained a classification spectrum\cite{Perley2023} at 22:23:45. The spectrum showed narrow emission lines often seen\cite{Bruch2022} in early-time spectra of Type II SNe, known as flash features\cite{Gal-Yam2014, Yaron2017}, including\cite{Perley2023} H, He~I, He~II, N~III, N~IV, C~III, and C~IV. 
We initiated a multiwavelength follow-up campaign for this event using the methodology of ref\cite{Gal-Yam2011}. Details about the collected observations are provided in Methods~$\S2-\S4$. In particular, we triggered UV spectroscopic observations using our {\it HST} target-of-opportunity (ToO) program that proved critical to reliably track the early UV evolution of this object.
Additional studies of this object have been promptly conducted and presented; a summary of these studies is provided in Methods $\S5$.   

\paragraph{The extended shock-breakout phase:}

\begin{figure}
\centering
\begin{tabular}{cc}
\hspace*{-3cm}\includegraphics[width=20cm]{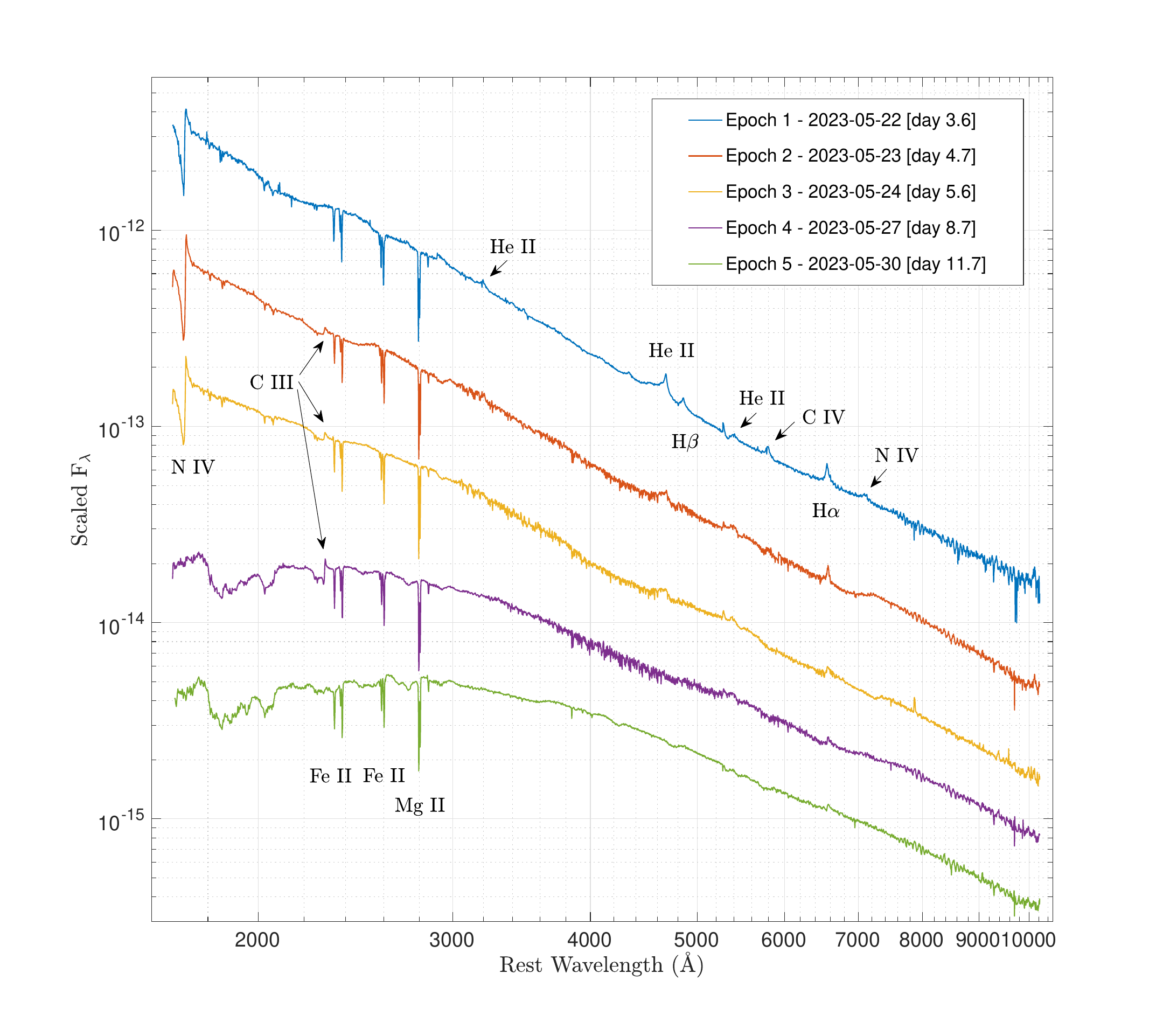}
\vspace*{-3cm}
\end{tabular}
\caption{{\it Hubble Space Telescope} UV spectra trace evolving shock-ionised flash lines and reveal metal absorption lines that probe the progenitor metallicity of SN\,2023ixf. All 5 epochs of coadded UV-optical spectra are shown in log space to enhance the visibility of narrow features across the entire spectrum.
Prominent flash features and UV absorption lines are marked, notably \CIII\ $\lambda 2297$ and \NIV\ $\lambda 1718$.
  \label{fig:HST}}
 \end{figure}

\begin{figure}
 \centering
\begin{tabular}{cc}
\hspace*{-2cm}\includegraphics[width=18cm]{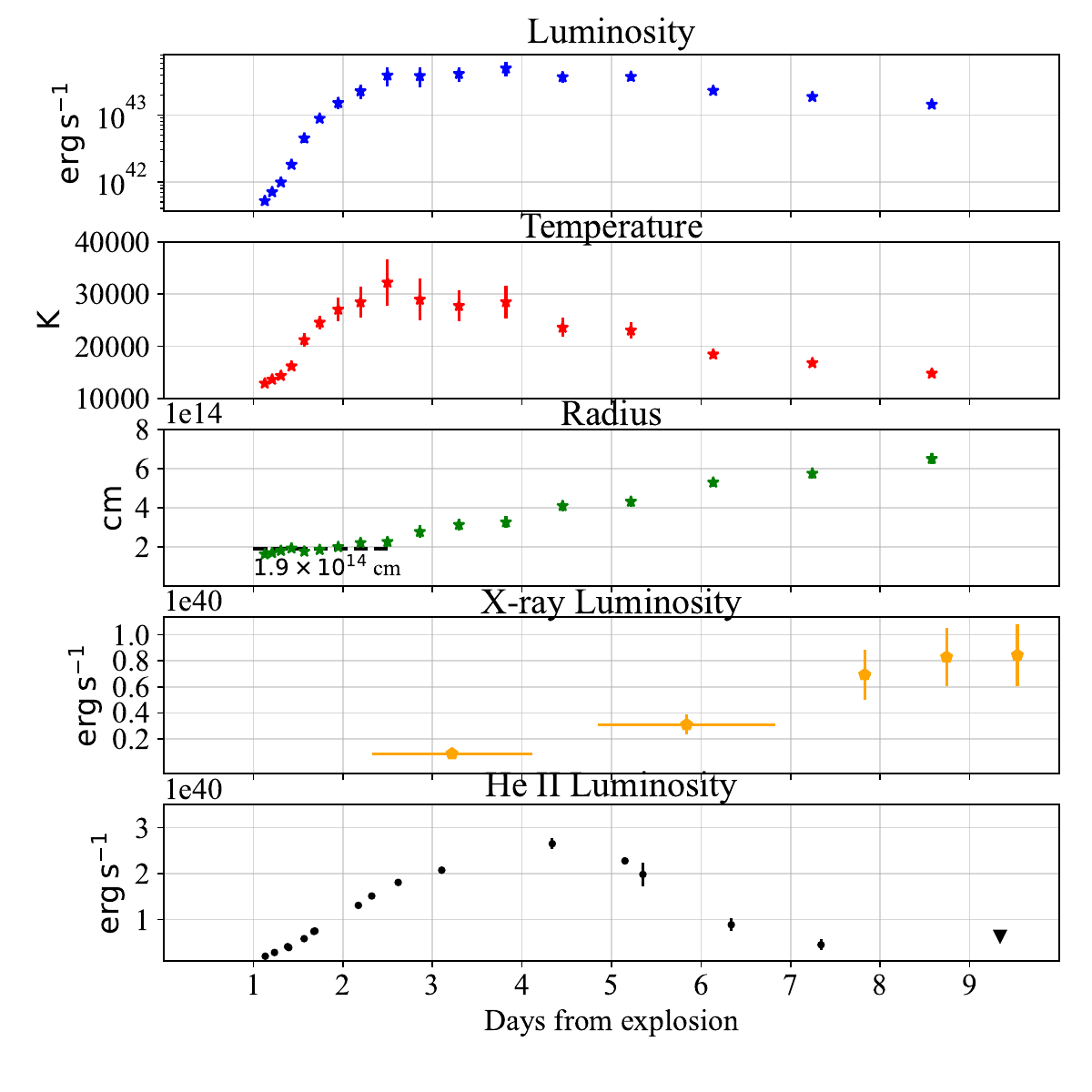}
\vspace*{-3cm}
\end{tabular}
\caption{ 
Bolometric evolution of SN\,2023ixf (top three panels, see titles) reveals the heating light curve expected from the extended breakout flare of an explosion embedded within thick CSM. A rising X-ray luminosity (fourth panel) combined with high-ionisation line flux decrease (traced by He~II here, last panel)  trace the hardening of the SN SED as the explosion shock progresses into thinner CSM. 
  \label{fig:line_vs_xra}}
 \end{figure}

As our {\it HST} spectra illustrate (Fig.~\ref{fig:HST}), the emission from the SN peaked during the first few days in the far-UV. The proximity of this event led to saturation of the UV photometers onboard the {\it Neil Gehrels Swift Observatory}, hindering standard photometric measurements. Anchoring our continuous optical-UV {\it HST} spectra to unsaturated visible-light photometry, we were able to extract reliable synthetic UV photometry and used it to independently confirm our custom analysis of saturated \swift\ observations (Methods $\S2$), and provide reliable UV coverage of the rise and fall of the SN light.   
Using the critical UV data, combined with visible-light and infrared (IR) observations (Methods $\S2$), we calculate a bolometric light curve (Methods $\S2$) that is shown in Fig. \ref{fig:line_vs_xra} along with the derived blackbody temperature and photospheric radius. 

Following the core collapse of a massive star, a shock wave propagates at velocity $v_{\rm s}$ from the centre of the star outward, heating the material it goes through. As the shock front reaches an area where the optical depth 
$\tau$ of the material above the shock is low enough ($\tau<c/v_{\rm s}$), the first photons escape the system, informing distant observers of the onset of an SN explosion, and the shock dissipates\cite{Waxman2017}. For stars with a well-defined outer radius $R_*$, the duration of this shock-breakout burst\cite{Waxman2017} is $R_*/c$, which for supergiant stars ($R_*>10^{13}$\,cm) is at most a few hours, after which the shocked material expands and cools. During most of this shock-cooling phase, the bolometric luminosity is approximately constant, the emitting material expands, and hence its temperature decreases\cite{Morag2023}. Alternatively, if the exploding star is engulfed by optically-thick circumstellar material (CSM), the shock breakout occurs at a much larger radius $r_0$ and the shock-breakout phase extends for a few days\cite{Ofek2010}.
During that period, photons from increasingly deeper layers escape, leading to a rise in luminosity. The emitting radius is $r_{\rm s} = r_0 + v_{\rm s} t$; for typical values of $v_{\rm s} \approx 10,000\kms$, initially $r_0 > v_{\rm s} t$, so $r_{\rm s}$ rises slowly. The temperature is therefore expected to rise. In this manner, early bolometric observations of SNe can differentiate between stars exploding in low-density environments, which expand and cool, and those exploding within a thick CSM, which are expected to initially get brighter and hotter, with an approximately constant emitting radius. 

Inspecting Fig.~\ref{fig:line_vs_xra}, we find that during the first three days of the SN evolution, we measure an almost constant photospheric radius, with a mean value of $(1.9\pm 0.2) \times10^{14}$\,cm. During this phase, the temperature rises from $13,000$\,K to $\sim32,000$\,K 
at peak, and the blackbody luminosity 
rises from $5.2\times10^{42}$ $\ergs$ to $3.9\times10^{43} \,\ergs$. For an extended shock breakout, the rise time is expected to be\cite{Ofek2010} $t_{\rm rise}=r_{\rm s}/v_{\rm s}$. Adopting the constant-radius value we measure as the shock-breakout radius $r_{\rm s} = (1.9\pm 0.2) \times10^{14}$\,cm, and approximating the shock velocity $v_{\rm s}$ by our measured photospheric velocity ($v_{\rm s} \approx v_{\rm phot}=8000\kms$; Methods $\S3$), we find $t_{\rm rise}\approx2.75\pm0.3$\,d, in excellent agreement with the data, suggesting that the radiation-mediated explosion shock is indeed breaking out from a dense CSM distribution\cite{Ofek2010}. 

The total bolometric luminosity at breakout can be estimated as the deposition of kinetic energy into the shocked CSM ($L_{\rm bo}\approx M v_{\rm s}^{2}/t_{\rm rise})$, where we assume a shock velocity of $8000\kms\ $ and the rise time $t_{\rm rise}$ as the relevant duration. 
We can therefore estimate the mass of the shocked CSM contained within the breakout radius $r_0$ to be $M\approx L_{\rm bo} t_{\rm rise}/v_{\rm s}^{2} \approx (7.4 \pm 2) \times 10^{-3} \Msun$, which suggests a characteristic density of $\rho \approx 5\times10^{-13} \gcm$. Pre-explosion studies of the progenitor (Methods $\S5$) indicate that its size was much smaller\cite{Van Dyk2023} than $r_0$ . Assuming an expansion velocity of $\approx 30\kms$ , derived from the blueshifted center of the narrow emission lines (Methods $\S3$), we find the mass-loss rate responsible for the confined CSM to be $\dot{M}\approx0.011\pm0.005\, \Msun\, \rm yr^{-1}$. The velocity and mass-loss rate also match independent estimates\cite{Jacobson-Galan2023}.
Following the end of the extended shock breakout from the dense inner CSM, the blackbody emission radius increases rapidly, and the temperature and luminosity drop. The continued rise of the light curves in visible light (Methods $\S2$; Fig.~\ref{fig:lightcurve}) is due to the emission peak moving redward from the far-UV and the growing radius.

\paragraph{Origin and evolution of the narrow emission lines:}

As the shock breakout is extended, the radiation-mediated shock gradually exits the thick CSM behind the breakout region, transitioning into a collisionless shock in the CSM plasma above\cite{Katz2012,Margalit2022}. This shock heats the CSM it passes through, while the continuously escaping photons cool the CSM electrons through inverse-Compton scattering. We show this cooling process to be highly efficient (Methods $\S6$), leading to a spectrum peaking in the extreme UV (EUV; $\gtrsim60$\,eV) at radii of up to $7\times10^{14} \,\rm cm$. This EUV radiation is sufficient to produce the highly ionised (He~II, N~IV, C~IV) narrow flash features appearing in the early-time spectra of SN\,2023ixf, and the
continuous source of ionising photons is actually required by the total flux of lines from the ionised CSM that we see, as the integrated luminosity of narrow-line components requires each atom to be ionised many times (Methods \S 3). For instance, to emit the total energy of $(1.66\pm 0.04) \times10^{45} \,\rm erg$ in H$\alpha$ between 1.38 and 6 days after explosion would require $0.92\pm0.02 \,\rm M_{\odot}$ of hydrogen if only one H$\alpha$ photon is emitted per atom, which is ruled out by our estimate of the total CSM mass ($M <0.01 \Msun$).

During the initial extended shock breakout, the expanding shock in the CSM increases the EUV flux, ionising the CSM further. This is observed through our sequence of day $\sim 1$--2 spectra. We see an increase in species ionisation, as \NIII\ $\lambda\lambda 4634,4641$, \CIII\ $\lambda\lambda 4647,4650 ,\lambda 5696$, and He~I $\lambda 5876,\lambda6678$ decrease in strength and disappear, while higher ionisation \NIV\ $\lambda\lambda 7109,7123$, \CIV\ $\lambda\lambda 5801,5812$, and \HeII\ $\lambda 4686$ lines continue to increase in flux (Fig. \ref{fig:line_vs_xra}). At this early stage, all lines manifest two components, a narrow core (full width at half-maximum intensity (FWHM) $\lesssim100 \kms$ at day 2.2), and a broad Lorentzian-shaped base that originates from electron scattering\cite{Huang2018}.

In the UV, our {\it HST} spectra (Fig.~\ref{fig:HST}) reveal two lines with P~Cygni profiles developing in the near-UV (NUV). \NIV $\lambda 1718$ with an absorption minimum at $\sim 1500 \kms$ is apparent already in our first {\it HST} spectrum on day 3.6, and \CIII\ $\lambda2297$ appears on day 4.7 and rapidly develops an absorption minimum that evolves to higher velocities of up to $\sim 1500\kms$ by day 8.7. 

Apart from H$\alpha$, the optical narrow lines disappear by day 5 (Methods $\S3$) as the shock propagates out of the efficiently Compton-cooled region, hardening the bremsstrahlung spectrum. The broad electron-scattering wings, however, last until day 6, since they originate from scattered photons, which can be delayed by up to the dynamical size of the shock transition layer $R_{\rm bo}/v_{\rm bo}\approx 2.5$ days. The NUV \NIV\ $\lambda 1718$ line disappears between days 5.5 and 8.7, consistent with the optical-line timescale, while the \CIII\ $\lambda 2297$ line that is remarkably detected until at least day 8.7. \CIII\ $\lambda 2297$ arises from a transition from an excited state populated by the EUV \CIII\ $\lambda 412$ transition, which itself results from the \CIV\ absorption of an electron into an excited \CIII\ state (a process known as dielectronic recombination)\cite{Hillier2023}. Thus, the persistence of this \CIII\ $\lambda 2297$ transition reflects a parent population of excited \CIV\ ions, lasting until at least day 8.7, likely still ionised by the hardening X-ray flux (Methods $\S8$). It would be very informative to model the exact mechanism through which these highly-ionised species persist; however, such a model is beyond the scope of this work.

\paragraph{Detection of a CSM density drop:}

\begin{figure}
 \centering
\begin{tabular}{cc}
\hspace*{-1cm}\includegraphics[width=18cm]{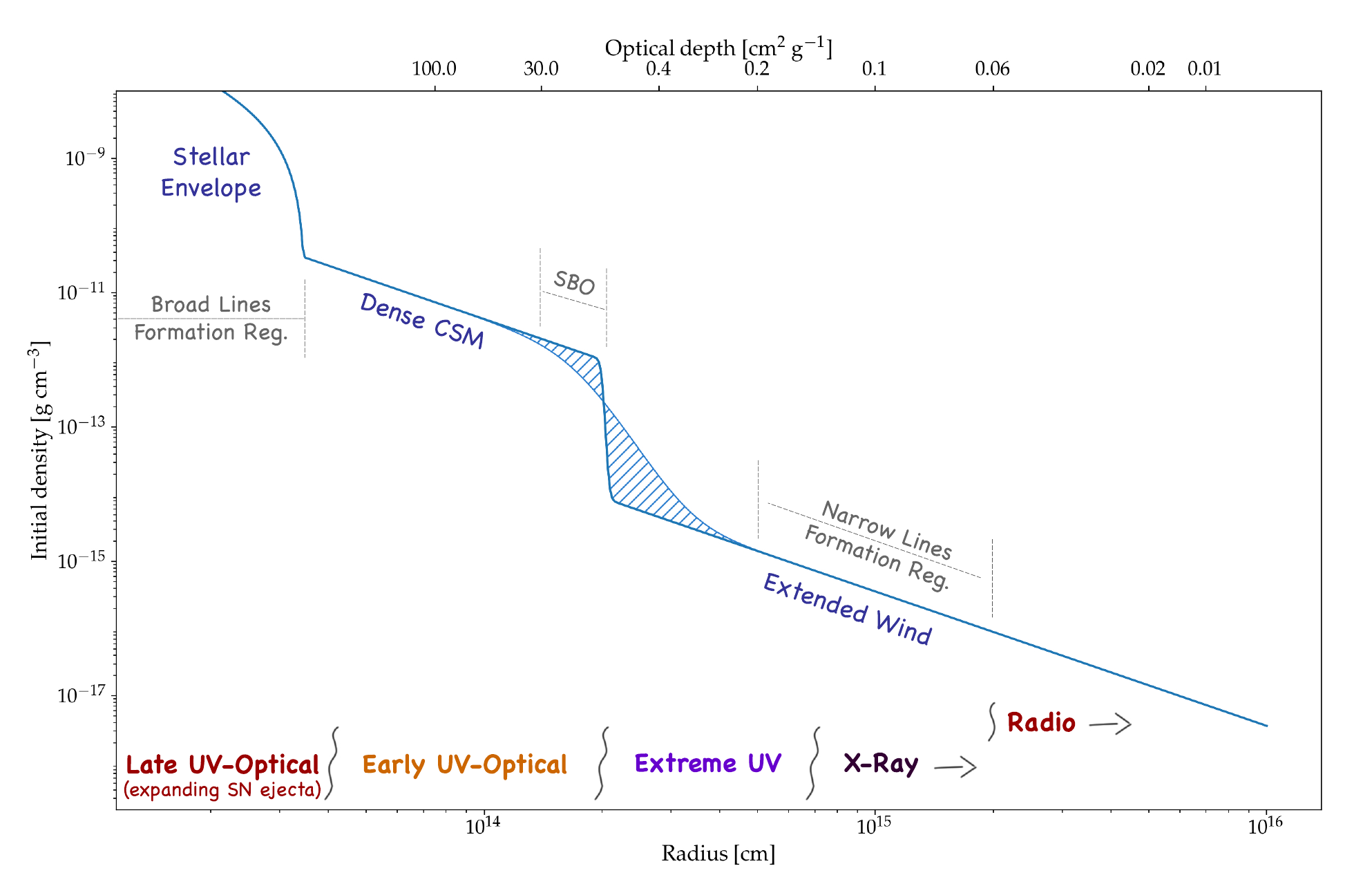}
\vspace*{-2.5cm}
\end{tabular}
\caption{A full mapping of the CSM structure. The stellar envelope inferred from pre-explosion images extends\cite{Van Dyk2023} to 
$\sim 5\times 10^{13} \,\rm cm$, above which thick ($\rho \approx 10^{-12} \gcm$) and confined ($r\lesssim 2\times10^{14}\, \rm cm$) CSM extends the shock breakout. The X-ray column-density measurements then suggest a drop in density around the shock-breakout region. A collisionless shock forms within the extended wind (after the drop in density near the shock-breakout radius), heating the CSM. At confined radii ($R\lesssim 7\times 10^{14}\, \rm cm$), this heating is suppressed by Compton scattering of the SN-emitted UV-optical photons, resulting in an EUV-dominated spectrum. However, at more distant radii, Compton cooling is no longer efficient, resulting in hard X-ray bremsstrahlung emission, followed by radio synchrotron emission from the heated electrons.
  \label{fig:CSM_structure}}
 \end{figure}

By day 4 after explosion, the collisionless shock continues to propagate into further layers of CSM that are not efficiently Compton-cooled. The hardening nonthermal spectrum from this lower-density CSM manifests in an escaping flux of X-rays that are detected starting day 4 (Fig.~\ref{fig:line_vs_xra}). 
Additional evidence for a sharp drop in density is provided by a decreasing neutral hydrogen column density deduced from the {\it NuSTAR} X-ray observations\cite{Grefenstette2023}, reducing X-ray absorption. These column densities translate into a CSM density of $4\times10^{-16}\gcm$ at a distance of $2\times10^{15}\, \rm cm$, about three orders of magnitude below our density measurement of $10^{-12}\gcm$ at $2\times10^{14}\, \rm cm$.
Fig.~\ref{fig:line_vs_xra} shows that the X-ray flux remains approximately constant as the collisionless shock continues to travel into the lower-density CSM.
As the energy source for this emission remains the deposition of kinetic energy into shocked CSM layers ($L\propto \dot{M}v_{\rm s}^{2}/t=2\pi r^{2} \rho v_{\rm s}^{3}$), the constant nature of the X-ray luminosity requires (for a constant velocity) a density profile falling as $\rho\propto r^{-2}$ (Methods $\S 4$). Assuming this density profile, the density extrapolated back to the shock-breakout radius would be $\rho \approx 10^{-14} \gcm$. This density is 1.5 orders of magnitude smaller than the density inferred above at the shock-breakout region. 
We present the full mapping of the CSM structure in Fig. \ref{fig:CSM_structure}, inferred from the different messengers described above.

\paragraph{Detection and implication of radiative acceleration:}

During their escape, photons interact with the CSM through radiative acceleration, dominated by the bound-free and bound-bound cross-sections (Methods \S 7). We observe this acceleration as an increase in the FWHM of the H$\alpha$ narrow component from $\lesssim 100\kms$ initially to $\sim 2000 \kms$ on day 11, and in the acceleration of \ion{He}{2} to $\sim 900 \kms$ on day 4.1, both of which are consistent with the report by ref\cite{Smith2023}.

The last narrow CSM spectral component surviving until day 15 is the H$\alpha$ P~Cygni profile that can be excited thermally and does not require either EUV or X-ray radiation to form. 
This feature persists as regular broad Type II SN photospheric features emerge in the SN spectrum, eventually disappearing as the outer CSM is swept up by the ejecta on day 15, at a distance of $\sim 10^{15}\, \rm cm$, assuming an ejecta velocity of $\sim 10,000 \kms$ inferred from the developing photospheric H$\alpha$ P~Cygni profile. 
This distance is consistent with calculations of the radiative acceleration of hydrogen in the CSM (Methods $\S7$). Assuming standard red supergiant (RSG) wind velocities (10--20$\kms$), this outer layer of CSM was expelled by the progenitor star a few decades before explosion.

Radiative acceleration explains the measurement of high-velocity ($\gtrsim 1000\kms$) flash P~Cygni profiles in H$\alpha$, the acceleration in the \ion{He}{2} and the NUV \NIV, and \CIII\ lines without requiring a recent stellar eruption to accelerate material to these velocities prior to the SN. This reconciles the SN observations with the nondetection of recent precursor eruptions from the progenitor\cite{Van Dyk2023}.
The strong acceleration observed here can only be achieved if the wind is optically thin, and provides additional support for the drop in density following breakout.

\paragraph{Chemical composition:}

The initial chemical composition of an exploding star has a strong impact on the properties of the resulting explosion. However, as all modern well-studied supernovae (SNe) occurred in external galaxies where properties of individual stars are very challenging to measure, direct observations of the chemical composition of SN progenitors do not exist. Instead, statistical studies have used the integrated composition of the entire host galaxy, or of nearby star-forming regions, as proxies for the properties of the actual exploding stars. Such studies are hindered by chemical mixing within galaxies\cite{Kuncarayakti2018}. 

The NUV spectra obtained of SN\,2023ixf reveal a plethora of narrow absorption lines from iron-group elements, commonly found in the interstellar material (ISM) of galaxies, including \ion{Fe}{2}, \ion{Mg}{2}, \ion{Mn}{2}, \ion{Zn}{2}, and \ion{Fe}{3} (see Supplementary Table \ref{tab:ism_lines}). These lines are seen in addition to commonly observed optical ISM lines, such as the Na~I~D and Ca~II H\&K doublets. While the UV lines are blended with the expected Milky Way (MW) absorption, their centre lies at the redshift of M101, and there is no significant MW Na~I~D absorption in our high-resolution optical spectra that spectrally resolve the velocity difference between M101 and the MW, suggesting that they originate mostly from gas in M101. 

The UV lines show no temporal development (change in equivalent width) during our observations. However, monitoring of the Na~I~D doublet with telluric-corrected high-resolution spectra (Methods $\S 8$) reveals that
the Na~I~D doublet is separated into two components -- a strong central line spanning $0$ to $-10 \kms$ and a weak blueshifted absorption component at $-30 \kms$ (Methods $\S 8$), a velocity shift similar to that of the SN narrow ($\sim 100 \kms$) flash features. This component shows a decrease in equivalent width starting on day $43$ after explosion,  suggesting that this component is influenced by the SN and hence originates from the immediate CSM of the progenitor star. 
The SN lies in a relatively isolated location, away from detected \ion{H}{2} or star-forming regions (Method $\S1$) that could account for strong ISM absorption, supporting a CSM origin for the detected lines.

We probe the column density and line-of-sight metallicity by applying the curve-of-growth technique\cite{Wilson1939} to the observed lines. The column-density measurements are presented in Supplementary Table~\ref{tab:ism_lines} (Methods $\S 8$). The ratio of different metals in the CSM likely reflects the composition of the progenitor prior to explosion. To calculate standard metallicity $Z$ (i.e., ratio of elements to hydrogen), a hydrogen column density is needed. Unfortunately, our UV spectra do not extend far enough into the UV to cover the Ly$\alpha$ line, which would have allowed us to derive this quantity. However, using an empirical correlation\cite{De Cia2016} between the Zn/Fe ratio and $Z$, We estimate that the natal metallicity of the progenitor of SN\,2023ixf, reflected in the CSM around it, was $Z=0.95$ solar.  

As a sanity check, we use integral-field-unit Keck Cosmic-Web Imager (KCWI)\cite{Morrissey2018} observations of the nearest star-forming region (80\,pc west of the SN location), and a vigorously star-forming region 310\,pc north of the SN. We measure the most prominent emission lines 
([\ion{O}{3}]\, $\lambda\lambda$4959, 5007, 
H$\alpha$, H$\beta$, and [\ion{N}{2}]\, $\lambda$6584) for both regions (Methods $\S 8$), and infer the metallicity of these regions using the O3N2 and R3 strong-line metallicity indicators. 
We find metallicities of $Z=1.00\pm0.06$ and $Z=0.68\pm0.01$ solar for the region to the west and north of the SN explosion site, respectively. While a moderate metallicity gradient is present across this small region of the host galaxy, the range of metallicity provides an indirect indicator of the metallicity for the progenitor of SN\,2023ixf. 
A solar to slightly sub-solar metallicity agrees with the inference from the UV absorption line analysis above.

Future {\it HST} observations, in particular including coverage of the far-UV, could further improve on the measurements we present here.
Routine early detection of events such as SN\,2023ixf by the forthcoming ULTRASAT space mission\cite{Shvartzvald2023} and follow-up UV spectroscopy using the proposed UVEX mission\cite{Kulkarni2021} could provide in the near future direct measurements of the chemical composition of SN progenitors prior to the explosion .



\clearpage
 
\clearpage

\clearpage


\clearpage
 
\clearpage
\begin{methods}

\subsection{1. The environment of SN\,2023ixf within its host:} \label{host_environment}
The SN\,2023ixf explosion site is located 9\,kpc from the centre of M101 (redshift $z=0.00084$)\cite{de Vaucouleurs1991} on the far side of an outer spiral arm (Extended Data Figure \ref{fig:Galaxy}). This region of the host galaxy is a site of ongoing star formation, with islands of enhanced star formation and diffusely distributed lower-level star-formation activity extending to the SN site (Extended Data Figure \ref{fig:Galaxy}c). The most nearby well-defined star-forming region lies only 80\,pc west of the SN explosion site. Another vigorously star-forming region lies about 310\,pc north of the SN explosion site. In $\S 8$, we show that these star-forming regions close to SN\,2023ixf are of sub-solar and solar metalicities. 

\subsection{2. Photometry:}

All phases in this manuscript refer to ${\rm JD} = 2,460,083.3$ as the SN explosion time, as inferred from early analysis of amateur-astronomer pre-explosion observations of M101\cite{Yaron2023}.

We obtained a full multiband light curve of SN\,2023ixf, covering the early rise, plateau, and radioactive decline. We present the full light curve in Extended Data Figure \ref{fig:lightcurve}. The data will be available from the WISeREP database and in a digital format as supplementary material. Standard reduction procedures are described in the supplementary online methods, as are our methods for reduction of saturated {\it Swift} photometry using the methods of ref\cite{Page2013}.

\paragraph{$^{56}$Ni mass and ejected mass estimate:}
By $t=90$\,days, the light curve has fully settled onto the radioactive $^{56}$Co tail (Extended Data Figure \ref{fig:lightcurve}). The $^{56}$Ni mass can be directly estimated from the bolometric luminosity (e.g., refs\cite{Bersten2009,Valenti2016}), which is equal to the energy deposition from $^{56}$Co decay, assuming full trapping of $\gamma$-ray photons as is typical for SNe II at this phase\cite{Sharon2020}. We calculate bolometric corrections at late times using spectra of SN\,2017ahn\cite{Tartaglia2021}, which shows similar persistent narrow features\cite{Yamanaka2023} and peak luminosity, with a slightly shorter plateau duration. We stitch together nebular spectra of this event obtained after the fall from the plateau in the optical ($t=72$\,days) and IR ($t=65$\,days) to match each other. We then calculate the bolometric correction to the $uBgriz$ photometry from the combined spectrum, extended assuming a blackbody extrapolation outside the observed band. This estimate is within 15\% of the bolometric luminosity calculated directly using a blackbody extrapolation to the SN\,2023ixf $uBgriz$ photometry. We find that the bolometric light curve of SN\,2023ixf is well described by the energy deposition from a $^{56}$Ni mass of $M_{^{56} \rm Ni} = 0.071\pm 0.005\, {\rm M}_{\odot}$. A 5\% systematic error is included to account for the scatter in the IR contribution ($>9000$\,\AA)  to the bolometric flux at $t=90$\,days\cite{Lyman2014}. Our estimate places SN\,2023ixf at the upper end of the Type II $^{56}$Ni mass distribution\cite{Valenti2016}. We show the full bolometric light curve, bolometric correction, and $^{56}$Ni deposition in Supplementary Figure \ref{fig:bolometric_full}.

Using the bolometric luminosity and the deposition from $^{56}$Ni, we calculate the time-weighted integrated bolometric luminosity $ET=\int Lt\,dt-\int Q_{^{56}{\rm Ni}}t\,dt$, where $Q_{^{56}{\rm Ni}}$ is the deposition from $^{56}$Ni. Note that $ET = 0.1\,M_{\rm ej}R_{*}v_{\rm ej}$ is an observable that is directly related to the ejected mass $M_{\rm ej}$, progenitor radius $R_{*}$, and typical velocity $v_{\rm ej}$; it is dependent only on the hydrodynamical profiles and is insensitive to the exact details of radiation transport and CSM mass\cite{Shussman2016,Nakar2016}. We find $ET = (3.3\pm0.2)\times 10^{55}\, \rm{erg\,s}$. We estimate\cite{Nakar2016} $v_{\rm ej}=v_{\lambda5169,50\,{\rm d}}=3400\pm200\kms$ and $v_{\rm env}= 8000\kms$, and we use the four estimates existing for the progenitor radius from pre-explosion data\cite{Jencson2023,Soraisam2023,Qin2023}. Here we derived the radius from the effective temperature and luminosity of the progenitor using $R = [L/(4 \pi \sigma T_{\rm eff}^{4})]^{1/2}$.
We obtain a combined constraint on the ejected mass of $7.5\pm2.5\Msun$ and a stellar envelope mass of $M_{\rm env} = 0.5\pm0.2\Msun$. The measurements of ref\cite{Kilpatrick2023} are omitted, as they are both an outlier and lead to an estimate of $M_{\rm ej}=13.6\pm1.9\Msun$, which is inconsistent with their implied initial mass of $11\Msun$. To reconcile the initial mass implied by other works of $\sim17\Msun$ with our ejected-mass estimate, the progenitor needed to either lose a large fraction of its mass before explosion, or leave a black-hole remnant. The former option requires that the mass-loss rate of $\sim10^{-4} \Msun\ \rm yr^{-1}$ inferred from the CSM extended wind would persist for $10^4$--$10^5\ \rm yr$, most of the red-supergiant phase of the progenitor star.


\subsection{3. Spectroscopy:}

We obtained a total of 112 spectra of SN\,2023ixf, taken with the instruments and configurations shown in Supplementary Table~\ref{tab:spectra}. The sequence of spectra in different phases is shown in Extended Data Figures \ref{fig:spec_early} and \ref{fig:spec_late}, and in Supplementary Figure \ref{fig:spec_phot}. All reduction notes are available as supplementary methods.

\paragraph{Hubble Space Telescope:}

Immediately following the classification of SN\,2023ixf as a Type II event showing flash-ionisation features\cite{Perley2023}, less than 90\,min from the SN discovery, we initiated our disruptive {\it HST} program GO-17205 (PI: E. Zimmerman) as part of the follow-up campaign. Rapid response and observation scheduling resulted in the first epoch of {\it HST} spectroscopy being obtained less than 50\,hr after the SN discovery, providing the first-ever NUV spectrum of a Type II SN during the flash-feature phase. Furthermore, as photometry from \swift\ saturated after $t=2$\,days, the {\it HST} spectra were the only reliable source of UV information by providing synthetic UV photometry. This synthetic photometry validated the readout-streak \swift\ photometry reduction and, thus, was the basis of the entire SN UV light curve.

Five visits of SN\,2023ixf were carried out as planned, covering a total of 22 {\it HST} orbits. All visits in the program were weighted toward the NUV, obtaining several orbits with the G230LB grism, but also included coverage of the optical part of the spectrum with single-orbit exposures using the G430L and G750L grisms. A full list of the program {\it HST} visits and exposures is shown in Supplementary Table \ref{tab:HST_visits}. Our sequence of {\it HST} spectra is presented in Fig. \ref{fig:HST}, showing a coadded spectrum for each of our five {\it HST} program visits. Reduction notes for the {\it HST} spectra are presented in the supplementary methods.

A strong \NIV\ $\lambda 1718$ P~Cygni profile persists throughout the first three visits (days 3.5--5.5 after explosion), reaching very high velocities of $>2500 \kms$ (Extended Data Figure \ref{fig:UV_pcygni}). This phase is roughly consistent with the peak flux in the strong flash features (Balmer lines and \HeII\ $\lambda 4686$), providing the first observation of flash features in the NUV. 
Starting from the second visit (+4.63\,days), a \CIII\ $\lambda 2297$ P~Cygni profile starts to develop, which persists on top of the broad photospheric features that emerge in visit 4 (+8.5\,days). During this visit, the optical spectrum exhibits a ``blue continuum,'' while the UV spectrum develops photospheric features, implying that the continuum originates by day 8.5 from the SN ejecta rather than from the CSM. The broad photospheric features are enhanced through visit 5 (+11.5\,days), during which no flash features appear in the UV spectra. 

Notable in their absence in the UV spectra are \HeII\  $n \rightarrow 3$ series lines, clearly seen in the early-time optical spectra ($4\rightarrow3\,\, \lambda 4686$). This is likely due to insufficient sensitivity to detect these weaker lines, as evident from the marginal detection of the $5\rightarrow 3\,\lambda 3203$ line in visits 1 and 2.
Another notably missing feature in the UV spectra is the predicted \FeIV\ absorption-line forest at 1600--1800\,\AA\, which appeared in previously published model CSM UV spectra\cite{Groh2014}. To assess whether this absence is caused by a low Fe abundance or the CSM temperature regime, we calculate non-LTE (local thermodynamic equilibrium) stellar models using the Potsdam Wolf-Rayet (PoWR) code, shown in $\S 8$, that indicate that these lines are likely missing owing to the CSM temperatures. 

Along with the strong \NIV~and \CIII~emission lines in the UV spectra, we find prominent line-of-sight absorption lines of \FeII~and \MgII, as well as weaker \AlIII, \ZnII, \MnII, and \CrII\ (shown in a zoom-in in Fig. \ref{fig:HST}).
These are known ISM lines\cite{Morton2003} and are accompanied by absorption lines of Ca and Na seen in high-resolution optical spectra from HARPS-N. These absorption features also appear in spectra of other Type II SNe (see Extended Data Figure \ref{fig:Photospheric_UV}), but they appear stronger in SN\,2023ixf.

In Extended Data Figure \ref{fig:Photospheric_UV}, we compare the photospheric features of SN\,2023ixf to other early-time UV spectra of Type II SNe. The UV photospheric features of SN\,2023ixf show a resemblance to those appearing in the early UV spectra of SN\,2022wsp\cite{Vasylyev2023b}, SN\,2021yja\cite{Vasylyev2022}, and SN\,2022acko\cite{Bostroem2023a}, suggesting some uniformity among Type II-P SNe. On the other hand, the photospheric features of SN\,2023ixf develop later than in SN\,2022acko, for which we have early-time NUV spectra, and lack a strong emission feature at $\sim 1910$\,\AA\ appearing in the early NUV spectra of SN\,1999em\cite{Baron2000} and SN\,2005ay\cite{Gal-Yam2008}. To assess the diversity of Type II SN NUV photospheric spectra, a larger sample of early-time NUV spectra must be collected. As some uniformity is observed in at least four recent SNe, these photospheric features must arise from the progenitor natal chemical composition and not from species synthesised during the SN explosion itself, whose parameters differ among explosions. It would thus be very informative to probe the progenitor chemical composition by modeling these early-time NUV spectra.

\paragraph{Narrow-line velocities:}

Using our sequence of high-resolution ($R=45,000$) NOT/FIES spectra, we measure the velocities of the fully-resolved strongest narrow flash features, namely H$\alpha$, H$\beta$, \HeII\ $\lambda 4686$, \CIV\ $\lambda\lambda 5801,5812$, and \NIV\ $\lambda\lambda7109,7123$. For the Balmer lines, we also measure the corresponding \HeII\ $n\rightarrow 4$ transition velocity. We present our velocity measurements in Supplementary Table \ref{tab:velocity}.

Depending on the epoch, we fit different phenomenological models to the narrow features in order to measure the velocity of different spectral components. To remove the local continuum, we fit a third-degree polynomial to the continuum (masking the line), and divide the spectrum by it.
In Extended Data Figure \ref{fig:Narrow_lines}, we present our fit for the $t=2.23$\,days spectrum, for which we fit a model consisting of a Lorentzian base (representing the electron-scattering wing) with Gaussian components representing the narrow lines themselves. For \HeII\ $\lambda 4686$ we also add a blueshifted Lorentzian base, improving our overall fit to this line. We interpret this blue excess as originating from the scattering caused by the (by this epoch) ionised \NIII\ $\lambda4640$, which we have shown can be delayed by up to $t_{\rm bo}$ in evolution. Overall, we find all narrow lines to be consistent with velocities of $\sim 100\kms$ at this epoch. All narrow lines are shifted blueward by $\sim 30 \kms$, except for H$\alpha$, which may be contaminated by M101 emission. 

In Supplementary Figure \ref{fig:broad_narrow_lines}, we present our fit for the $t=4.3$ and $t=5.11$\,days spectra. We fit Lorentzian profiles to the flash features, as no clear narrow components are visible (except for \HeII\ $\lambda 4686$ at $t=4.3$\,d, for which we add a narrow Gaussian component to the fit as well).
Starting from $t=8.13$\,days, as no other narrow features appear in the optical spectra, we only fit a model to H$\alpha$, showing a narrow P~Cygni profile. We represent this profile with a redshifted Lorentzian emission component adjacent to a blueshifted Gaussian absorption component. To avoid introducing a saddle mimicking the absorption component, we use a first-degree polynomial to remove the continuum from these later spectra.
These fits are presented in Extended Data Figure \ref{fig:UV_pcygni}. We define the P~Cygni velocity to be 3 times the standard deviation of the Gaussian absorption component, as this corresponds to the absorption profile blue edge well in cases where this edge is clear (see arrows in Extended Data Figure \ref{fig:UV_pcygni}) and can also be measured when the blue-edge measurement is hindered by uncertain continuum levels. A velocity of $v\approx2000 \kms$ is inferred from the mean of all four epochs.
All fits were made using the {\tt lmfit}\cite{Newville2014} package.

\paragraph{Narrow-line fluxes:}

In order to probe the strength of the narrow lines, we measure their flux using the sequence of medium-resolution spectra. To do so, we fit a first-degree polynomial to the continuum in the emission-line vicinity and remove it from the spectrum. We define the borders of the nearby continuum by drawing its edge from a uniform distribution. The emission flux is then obtained using trapezoidal integration. We repeat this process 1000 times, drawing different edges to the continuum, and take the mean value as the measurement and the standard deviation as the measurement uncertainty.

The total emitted flux from \HeII\ and H$\alpha$ cannot be explained by a single ionisation event, as it corresponds to the emission of one photon per atom by $\sim 1\, \rm M_{\odot}$ of material. We present \HeII\ flux measurements in Fig. \ref{fig:line_vs_xra} and the other line fluxes in Supplementary Table \ref{tab:fluxes}.

\paragraph{Measuring distant CSM and ISM properties:}

Utilising high-resolution ($R=115,000$) spectra obtained with the northern High Accuracy Radial velocity Planet Searcher (HARPS-N), we observe the fully resolved \CaII~K ($\lambda 3934$) \& H ($\lambda 3968$), as well as \ion{Na}{1}~D ($\lambda\lambda 5890,5896$), ISM absorption lines.
We find that both Ca features show a significant MW component, which would not be resolved in low-resolution spectra. However, there are no significant Na~D MW components. Therefore, we conclude that some of the UV ISM lines observed with {\it HST}/STIS could be blended with a MW component. However, note that all UV ISM features are centred at the M101 redshift.
\CaII\ is no longer detected after the $t=21$\,days spectrum, as the SN light curve reddens, decreasing the signal at blue wavelengths. 

We continue to monitor the \ion{Na}{1}~D component until $t=85$\,days. Both absorptions in the doublet show four significant components centred at blueshifts of $v\approx0\kms,\,4\kms,\,8\kms,$ and a  shallower $30 \kms$ with respect to the M101 frame (see Extended Data Figure \ref{fig:NaID}). At each epoch, we measure the EW of each component. To do so, we remove the local continuum around the D1 and D2 lines separately by dividing the spectrum with a first-degree polynomial fit calculated after removing outliers.
We then use a bootstrapping algorithm to calculate the component EW, by applying different Gaussian noise to the spectra with the linear fit standard deviation to each run. We present the result of this calculation in Supplementary Table \ref{tab:NaD}.

Tracking the \ion{Na}{1}~D doublets, we find that starting from $t=43$\,days, both the $-30 \kms$ and $-8\kms$ components weaken in EW. A change in Na EW has been shown to suggest the presence of CSM around Type Ia SNe\cite{Patat2007}. While in the case of SNe~Ia, Na variability shows recombination of ionised Na to neutral Na (i.e., an increase in the \ion{Na}{1}~D absorption), in the case of SN\,2023ixf we see the ionisation of neutral Na. Nonetheless, ionisation of Na indicates that the changing Na components are in the vicinity of the SN explosion since the number of ionising photons decreases with distance squared. Further ionisation of extended CSM is but another indication of the hard spectral energy distribution (SED) of SN\,2023ixf, even at later phases.
As the SN is located away from its nearby star-forming region (Methods $\S 1$), and the progenitor star experienced elevated mass loss for up to $\sim 10^4$--$10^5$\,yr (Methods $\S 2$), it is likely that most of the strong ISM absorption lines originate from CSM around the progenitor star, suggesting that the metal abundance we measure from these lines traces the progenitor chemical composition (as shown in Methods $\S 8$).

To measure the EW of the UV absorption lines at each epoch, we remove the local continuum by fitting a third-degree polynomial to the area adjacent to the lines and fitting a Gaussian to each line. By fitting Gaussian profiles, we resolve blended components, allowing us to measure each line in a blended doublet individually. We used the {\tt lmfit} package\cite{Newville2014} to conduct the fits. Lastly, we integrate over the best-fit model, producing an EW for each line. An example fit is shown in Supplementary Figure \ref{fig:absorption_lines}. We use the EW of these lines to probe the SN metal abundances as discussed in $\S8$.

Unlike the optical \ion{Na}{1}~D, no significant change in the intensity of the UV absorption lines can be measured. However, many of the weak lines fall below the requisite signal-to-noise ratio (SNR) as the SN UV brightness fades, and all lines are unresolved at later epochs so that any variable components would be diluted by the strong saturated absorption.

\subsection{4. X-ray observations}

The \swift\ satellite obtained multiple observations of SN\,2023ixf at X-ray energies with its X-ray telescope (XRT)\cite{Burrows2005} in photon-counting mode. We present the early X-ray rise in Figure \ref{fig:line_vs_xra}, while the full X-ray light curve is shown in Extended Data Figure \ref{fig:XRTlc}. The individual measurements are summarised in Supplementary Table \ref{tab:xrt}. After an initial rise in the X-rays, SN\,2023ixf settles onto an X-ray plateau. As the luminosity of a cooling forward shock is given by $L = 2\pi r^2 \rho v_{\rm sh}^{3}$, and the shock velocity $v_{\rm sh}$ is only slowly decreasing (i.e., staying roughly constant at early times), a constant X-ray light curve suggests a steady mass-loss and a density profile of $\rho\propto r^{-2}$, where $r$ is the distance from the progenitor star. Some variability is seen in the X-ray plateau, suggestive of deviations from the $\rho\propto r^{-2}$ profile, which can arise due to variable progenitor mass loss (see Methods $\S5$ progenitor properties). We note that at the time of writing SN\,2023ixf is still X-ray bright. 
Detailed reduction notes of the XRT data are shown as Supplementary methods.

\subsection{5. Summary of additional studies of SN\,2023ixf at the time of writing:}

Located in the nearby, well-known galaxy M101, SN\,2023ixf attracted significant attention from the community. In this section, we summarise the main results published thus far.

\paragraph{Studies of the narrow lines:}
Several papers have discussed the early-time optical flash features and their temporal progression. Refs\cite{Jacobson-Galan2023,Bostroem2023b} present early flash spectrocsopy of SN\,2023ixf, and discuss the increase in the flash-feature ionisation (\ion{He}{1}, \NIII, \CIII $\rightarrow$ \HeII, \NIV, \CIV), unique to this event. 
Both studies also compare the early spectra to non-LTE models, reaching a mass-loss-rate estimate of the same order of magnitude as that inferred by us for the confined CSM up to the shock-breakout radius of $\dot{M}\approx2\times10^{-2} \Msun \rm ~yr^{-1}$. Ref\cite{Jacobson-Galan2023} also finds a density of $\rho\approx10^{-12} \gcm$ at a radius of $r\approx10^{14}\, {\rm cm}$, consistent with our results. Both studies suggest that the CSM is confined to $(0.5-1)\times10^{15}\,{\rm cm}$. We find that low-density CSM extends further to explain the persistent X-ray luminosity and narrow spectroscopic features until at least day 8.7 (\CIII\ $\lambda2297$) and day 15 (H$\alpha$). Ref\cite{Jacobson-Galan2023} presents their best-fit model (r1w6b) from Ref\cite{Dessart2017} extending to the UV. We illustrate a comparison of the NUV spectra we obtained to this model in Supplementary Figure \ref{fig:NUV_CMFGEN}.
Both studies also compare SN\,2023ixf to other SNe displaying flash features and note the similarity of SN\,2023ixf to SN\,1998S\cite{Leonard2000,Fassia2001}, SN\,2017ahn\cite{Tartaglia2021}, SN\,2020pni\cite{Terreran2022}, and SN\,2020tlf\cite{Jacobson-Galán2022}, all of which show long-lasting flash features compared to most other SNe\cite{Bruch2022}. We also note that SN\,2017ahn, SN\,2020pni, and SN\,2020tlf all exhibit a rise in their bolometric luminosity, suggesting that they experience an extended shock breakout in a wind, similar to SN\,2023ixf.

Ref\cite{Smith2023} studies the narrow lines in detail using high-resolution spectroscopy, noting that the narrow lines exhibit broadening between days 2 and 4. Our high-resolution velocity measurements from $t=2.3$\,days are consistent with this study, and the broadening of narrow lines is consistent with our results on radiative acceleration of the CSM. We add the velocity measurements by this study into Extended Data Figure \ref{fig:radiative_acceleration}. Ref\cite{Smith2023} also discusses the possibility of the CSM being asymmetric, citing the difference in blueshift between highly-ionised species and lack of narrow ($\sim100 \kms$) P~Cygni profiles. In our independent analysis, we also observe a blueshift of the narrow lines in the $t=2.3$ day NOT/FIES spectrum (see Extended Data Figure \ref{fig:Narrow_lines} and Supplementary Table \ref{tab:velocity}). We note, however, that UV spectroscopy reveals deep, narrow P~Cygni profiles, unlike the optical H$\alpha$ profile.

\paragraph{Spectropolarimetry:}

Ref\cite{Vasylyev2023a} conducted early-time spectropolarimetry of SN\,2023ixf and observes a high continuum polarization of $p_{\rm cont}\approx 1 \%$ until day $2.5$, dropping to $0.5\%$ on day $3.5$ after explosion. This timeline is consistent with our extended shock-breakout time $R_{\rm bo}/v_{\rm s} \approx 2.5$\,days. As discussed in ref\cite{Vasylyev2023a}, as the shock moves within optically thick CSM, the photosphere is expected to originate from a radius at which the optical thickness  $\tau\approx2/3$. This is consistent with the roughly constant radius obtained through our blackbody measurements, suggesting that at this radius ($\sim 1.9\times10^{14}\,{\rm cm}$), the CSM is aspherical. We note, however, that this asymmetry is only measured in layers above the breakout radius. The sudden change in polarization between days $2.5$ and $3.5$ is also consistent with a change in the CSM structure, which coincides with the drop in density we derive as the SN shock moves from the confined CSM into the extended wind (with the measurement of first X-ray emission). Ref\cite{Vasylyev2023a} also observes a depolarization in the centre of the narrow \HeII\ $\lambda 4686$ and H$\alpha$ emission, suggesting that they form above the early polarized photosphere. This is consistent with our results, showing that many of the narrow features form in the extended wind.

\paragraph{X-ray and other messengers:}

As demonstrated through our CSM model, X-ray and (mostly lack of) radio detections of SN\,2023ixf have been critical messengers in the mapping of CSM structure.

Ref\cite{Grefenstette2023} provided two epochs of X-ray observations with the {\it Nuclear Spectroscopic Telescope Array (NuSTAR)}. While both epochs are covered by the \swift/XRT as well, {\it NuStar} provides several important insights unavailable through XRT. Both epochs fit a bremsstrahlung spectrum, with a peak at $>25$ and $34^{+22}_{-12}\,{\rm keV}$, respectively, showing that the X-ray spectrum is hard. The inferred {\it NuSTAR} unabsorbed bolometric X-ray luminosity in both epochs of $L_{X}\approx 2.5\times 10^{40} \ergs$ is consistent with our XRT measurements, taking into account a bolometric correction.
Lastly, the decrease in soft X-ray absorption between the first {\it NuStar} epcoh ($\sim 4$\,days) and the second epoch ($\sim 11$\,days) traces the extended-wind profile and is consistent with the drop in density we expect after the shock-breakout region. We adopt the mass-loss rate and densities reported by this study as the extended wind density.

Ref\cite{Berger2023} reports limits on millimeter observations of SN\,2023ixf, consistent with the high mass-loss rate we infer for the confined CSM, but inconsistent with the lower mass-loss rates we infer for the extended wind. Ref\cite{Grefenstette2023}, however, shows that lower mass-loss rates consistent with our extended-wind model can also be allowed by the radio data.

\paragraph{Photometric evolution and models:}

Ref\cite{Hosseinzadeh2023} shows a fit to a recent shock-cooling model\cite{Morag2023} for photometry starting at $t=1$\,day. As we measure the SN heating for the first 2.5\,days, we find this result to be inconsistent with our data. While shock cooling can describe the unsaturated early-time \swift/UVOT photometry, it is in conflict with the full SN light curve, including our {\it HST} data and {\it Swift} streak-photometry, which shows a prolonged rise in all UV bands. 
Ref\cite{Hosseinzadeh2023} also discusses a change in the light-curve behaviour between the first day before discovery by Itagaki and later photometry. This is consistent with our pre-discovery P48 $g$-band photometry. Further discussion of the change in slope between very early times and the +1\,day light curve will be presented in a future paper\cite{Yaron2023b} analysing more amateur-astronomer photometry.

Ref\cite{Hiramatsu2023} performs blackbody SED fits to the $UBVRI$ light curve (i.e., excluding the saturated \swift\ data), reaching the conclusion that SN\,2023ixf experiences heating during the first $\sim 5$\,days after explosion. Ref\cite{Hiramatsu2023} therefore concludes that the SN goes through an extended shock breakout in a wind, consistent with our results. However, owing to the inaccessibility of the saturated UV light curve, the bolometric luminosity is only sampled in the Rayleigh-Jeans regime, causing an overestimation of the luminosity rise time (by $\sim2.5$\,days) and an underestimation of the SN heating (by $\sim 16,000\, {\rm K}$). Ref\cite{Hiramatsu2023} therefore concludes that the confined CSM extends much farther than our results [$(3-7)\times10^{14}\, {\rm cm}$ compared to $\sim 2\times10^{14}\, {\rm cm}$].

Ref\cite{Singh Teja2023} conduct a multiband study of SN\,2023ixf as well, presenting the \swift\ grism NUV observations and three epochs of photospheric (starting from $+6.9$\,days) FUV spectra using the UltraViolet Imaging Telescope (UVIT) onboard the {\it AstroSat} satellite. Ref\cite{Singh Teja2023} also perform a light-curve study without the saturated early-time UV data, inferring a more extended dense CSM (results consistent with those of ref\cite{Hiramatsu2023}).

\paragraph{Progenitor properties:}

Several studies have pointed out a candidate progenitor star using pre-explosion images\cite{Van Dyk2023,Jencson2023,Kilpatrick2023,Soraisam2023,Qin2023}, 
indicating that the progenitor candidate was an RSG experiencing periodic ($\sim 1000$\,days) mass-loss episodes.
Refs\cite{Jencson2023,Soraisam2023,Qin2023} also constrain the mass-loss parameters from the progenitor IR variability and SED fits. This mass loss should reflect the extended wind, as the confined CSM must have been ejected closer to the time of explosion owing to its much larger density. The inferred CSM densities from these studies are indeed consistent with the values we measure for the extended wind.
However, it is interesting to note that no eruptive IR source has been identified for the confined CSM (below the breakout radius), as IR observations were obtained a mere 10\,days before the explosion by ref\cite{Jencson2023}. Ref\cite{Van Dyk2023} also conducted a study of nearby \ion{H}{2} regions and found the metallicity to be solar to slightly subsolar, consistent with our findings in Methods $\S 8$.

An interesting possibility from the variable nature of the progenitor mass loss described in the above studies is the expectation for variability in the X-ray light curve.
This would arise due to the X-ray luminosity being a tracer of the density, as the source of the emission is momentum transfer to the CSM ($L_{\rm X-ray}\propto \rho v^{3}_{\rm s}$).
Such periodicity would be hard to trace in detail as the light-travel time from the shock edge increases by $R/c$, smearing any such effect. We note that some variability might exist in the X-ray light curve, which shows dimming of the X-ray flux between 10 and 20\,days after explosion, followed by a rebrightening (see Extended Data Fig. \ref{fig:XRTlc}).

\subsection{6. Evidence for shock breakout in a wind:}

\paragraph{Blackbody fits:}
Starting at the first time when multiband photometry is available at JD = 2,460,084.42, we linearly interpolate our photometry onto a grid with uniform steps in log time. For each grid point we fit the resulting SED with a blackbody (BB) function to recover the blackbody bolometric luminosity, radius, and temperature of SN\,2023ixf. We also compare the optical-UV {\it HST} spectra to a grid of BB temperatures and luminosities and calculate the best fit and the corresponding uncertainties, assuming a likelihood $\log{\mathcal{L}} \approx \chi^{2}$. Both methods are in good agreement. We report our results in Table \ref{tab:bb_tab} and in Supplementary Figure \ref{fig:BB_SEDs}. A full bolometric light curve is shown in Supplementary Figure \ref{fig:bolometric_full}. During the first three days of its evolution, the photospheric radius is approximately constant, with a mean value of $(1.9\pm 0.2)\times10^{14}$\,cm. During this phase, the temperature rises from $13,000$\,K to $\sim 32,000$\,K at peak, and the blackbody luminosity rises from $5.19\times10^{42}$\,$\ergs$ to $3.9\times10^{43}$\,$\ergs$. As the radius begins to increase, the temperature and luminosity drop, and the continued rise of the light curves is powered by cooling, as the SED peak moved from the unobserved FUV into initially the UV and later the optical bands.

\paragraph{Compton cooling of the shock-heated wind at early times:}
As the radiation-mediated shock breaks out of the dense CSM, the accelerated ejecta slam into the low-density and optically thin wind, creating a collisionless shock\cite{Katz2012,Waxman2017} that is mediated by collective plasma instabilities. The average proton temperature at the shocked region\cite{Katz2012, Margalit2022} is $T_{p} \approx 0.2$\,MeV. The electrons are heated by the shock passage and Coulomb interactions with the protons to temperatures of $\gtrsim60$\,keV. These energetic electrons inverse-Compton scatter on the flux of UV photons from the underlying breakout region, suppressing the X-ray flux at early times. Owing to the large density difference between the breakout region and the extended encompassing wind, there are not enough hot electrons to significantly impact the UV-optical spectral peak. Rather, the plasma energy is carried away by EUV photons created through inverse-Compton scattering. The spectrum will continuously harden as the radiation flux decreases with time and distance, making this cooling channel inefficient.

The cooling rate of a plasma with energy $E_{p}$ of hot electrons at $\tau<1$ and at a temperature $T$ by photons at much lower energies is given by 
\begin{align}
        dE_{p}=-\frac{\sigma_{T}}{4\pi r^{2}}\frac{4 T L_{\gamma} \,dt}{m_{e}c^{2}}\, ,
\end{align}
where $\sigma_{T}$ is the Thompson cross section and $m_e c^2$ is the electron rest-mass energy. Assuming $E_p=3T$, we find

\begin{align}
        \frac{d\log E_{p}}{dt}=-\frac{\sigma_{T}}{3\pi r^{2}}\frac{L_{\gamma}}{m_{e}c^{2}}\, .
\end{align}

Using the observed bolometric light curve, we calculate the cooling rate of the plasma by the breakout burst photons at various locations and times since the explosion. We show these results in Supplementary Figure \ref{fig:cooling_rate}; they  indicate that during the first day of the explosion, Compton cooling is not yet significant enough to prevent X-ray emission, but no observations exist at this time. During days $t=2$--6, X-ray emission is suppressed at the collisionless shock front, but the electrons remain hot enough to allow EUV emission. After day 6, the density drops and X-ray emission is no longer suppressed by inverse-Compton scattering. Our results are broadly consistent with the observed \HeII\ emission (ionised efficiently by $\lesssim 100$\,eV photons) before day 6 and the emergence of X-rays on days 4--5. The continued presence of \CIII\ $\lambda2297$ after the disappearance of \HeII\ is consistent with the hardening of the illuminating spectrum rather than an effect of recombination time, as discussed in the main text.

\subsection{7. Radiative acceleration:}
Radiative acceleration of CSM has been calculated\cite{Fransson1982} and shown to have occurred in previous SNe, such as SN\,2010jl\cite{Fransson2014}. Here, we calculate the velocity gained by radiative acceleration at a given radius using the time-dependent blackbody spectral luminosity that we measured (Methods $\S2$) for SN\,2023ixf.

The acceleration a spherical shell of material at $\tau<1$ experiences due to a flux $f_{\nu}$ passing through it is given by
\begin{align}
        \frac{d^2r}{dt^2}=\frac{1}{4\pi r^{2}\mu c}\int\limits _{0}^{\infty}L_{\nu}\sigma_{\nu}\,d\nu\, ,
\end{align}
\noindent
where $L_{\nu}$ is the spectral luminosity of the internal source, $\mu = 0.7m_p$ is the average particle mass assuming solar mixture, and $\sigma_{\nu}$ is the total cross section at a given frequency. To estimate the total cross section, we use the open-source opacity table described in ref\cite{Morag2023} and based on atomic line lists\cite{kurucz_atomic_1995} by assuming CSM with a density of $10^{-15}\, \rm {g\, cm^{-3}}$ and a temperature of $5\, \rm eV$, which would generate a thermal electron velocity distribution consistent with the FWHM of the electron-scattering wings of the narrow emission lines observed during the first few days. The effective cross section due to bound-bound and bound-free processes dominates the effective cross-section for acceleration (but not for diffusion) and is much larger than the Thompson cross section. While the CSM is not in LTE with the incoming radiation from the underlying SN, the cross section is most affected by bound-free interactions in the plasma. A full steady-state solution of the fractions of different plasma species is outside the scope of this work.

In Extended Data Figure \ref{fig:radiative_acceleration}, we show the velocities measured from various narrow lines as they evolve until photospheric features emerge. Our calculation demonstrates that with a fraction of neutral elements achieved at $5\, \rm eV$, the absorption cross section is sufficiently high to fully explain the observed line velocities with radiative acceleration from the SN UV--optical emission at a single location. 
The first high-resolution spectrum capable of resolving lines with $v<100 \kms$ was taken at $t=2.3$\,days. At this time, radiative acceleration was already high enough to dominate the observed velocity, indicating that the CSM velocity is not measured from these features. As a consequence of our calculation, we also determine the position and radiative acceleration experienced by shells immediately above the breakout region. Given sufficient time, matter below the shock-breakout region (instead of the ejecta) would accelerate enough to shock the matter. In our case, significant acceleration does not occur before day 3. By this time, the ejecta themselves will sweep up all matter, which can be accelerated to $>5000 \kms$, indicating a collisionless shock had, in fact, formed above the breakout region.

\subsection{8. Constraining the progenitor Metallicity:}
\paragraph{Nearby H~II regions:}
Using the integral-field-unit Keck Cosmic Web Imager KCWI\cite{Morrissey2018}, we measure prominent emission lines from the two nearby star-forming regions 80\,pc west and 310\,pc north of the SN explosion site (see $\S 1$). Supplemantary Table \ref{tab:hii_lines} summarises the line fluxes of the most prominent emission lines ([\ion{O}{3}]\,$\lambda\lambda$4959, 5007, H$\alpha$, H$\beta$, and [\ion{N}{2}]\,$\lambda$\,6584) of the two regions. Using the O3N2 and R3 strong-line metallicity indicators and the calibrations from ref\cite{Curti2017}, we infer metallicities of 
$1.00\pm0.06$ and $0.68\pm0.01$ solar for the regions to the west and north of the SN explosion site (respectively), indicating that there is 
a moderate
metallicity gradient across this small region of the host galaxy.

In ref\cite{Croxall2016}, multi-object spectroscopy of M101 was performed with the Large Binocular Telescope. They obtained a spectrum of the \ion{H}{2} region at $\alpha = 14^{\rm hr}03^{\rm m}38.562^{\rm s}$, $\delta = +54^\circ 18' 41.94''$ (J2000) that is $\sim 10''$ ($\sim340$\,pc) north of SN\,2023ixf (spectrum ID NGC5457+225.6-124.1). Their deep observation also recovered emission from the weak auroral line [\ion{O}{3}]\,$\lambda$\,4363, which allows a measurement of the metallicity using the $T_e$ method. These authors determine a metallicity of 0.64 solar, corroborating our value.


Most SNe will be detected at larger distances than SN 2023ixf, and the H II regions to the north and west would be indistinguishable from the ground. Adopting the weighted mean of the two regions from our KCWI observation ($0.69\pm0.01$) as a proxy of the metallicity at the SN site. 

\paragraph{ISM lines:}

To infer the column densities from the ISM EWs, we use the curve-of-growth (CoG) technique\cite{Unsold1930, Wilson1939}, though see ref\cite{Prochaska2006} for the limitations and perils of this method. We present the resulting column densities in Supplementary Table~\ref{tab:ism_lines} and the fit to the CoG in Supplementary Figure \ref{fig:COG}. From the series of \ion{Fe}{2} absorption lines, we measure an instrumental plus intrinsic line-broadening parameter $b$ of $70 \kms$. Assuming that all other species have the same $b$ value, we can measure the column densities $N$ of all species. To determine the metallicity of the SN line of sight, the neutral hydrogen column density of the line of sight has to be known. However, the \textit{HST}/STIS spectra do not cover Ly$\alpha$, commonly used for measuring $N$(\ion{H}{1}). Hence, in order to estimate the metallicity, we use an empirical relation [Zn/Fe] =0.73\,[Zn/H] + 1.26 from ref\cite{De Cia2016}. Given the observed [Zn/Fe] $=1.25$, we estimate the metallicity of the neutral ISM to be $Z=0.95$ solar, using the solar abundances for H and Zn from ref\cite{Asplund2009a}. While this relation has a scatter of $0.3$ dex, it rules out extremely low progenitor metallicity ($\lesssim 0.5$ solar), while large extrasolar metallicities are ruled out by the environment measurements, which are consistent with the latter estimate.

In addition, we calculate the mass dust-to-metals ratio DTM $\approx 0.43$, based on the observed [Zn/Fe] and using the formalism of ref\cite{Konstantopoulou2023}, also used by ref\cite{Heintz2023}. This DTM value is very close to the Milky Way value (0.45). Finally, we analyse the metal pattern using the method of ref\cite{De Cia2021}, but adapted for the case where no H measurements are available\cite{Ramburuth-Hurt2023}. We find no evidence for alpha-element enhancement or Mn underabundance in the gas, their metal columns aligning well with what is expected from only dust depletion. This suggests that the neutral ISM in this galaxy has not been enriched by a recent burst of star formation. Either the star formation in this galaxy has been rather continuous (with contributions from SNe~Ia as well, in addition to core-collapse SNe), or the recent star formation did not yet have the time or extent to enrich the neutral ISM in the host galaxy. 

\paragraph{Comparison to PoWR models:}
A full modelling of our spectral datasets requires detailed non-LTE time-dependent radiative-transfer calculations in expanding atmospheres\cite{Dessart2010}, which is beyond the scope of this work. Here, we use stationary non-LTE PoWR models\cite{Hamann2003, Sander2015} to obtain a better understanding of the conditions that lead to the formation of diagnostic spectral lines in the UV, and to constrain the iron content in the progenitor's wind.  Briefly, the {\tt PoWR} code solves the one-dimensional non-LTE radiative transfer and rate equations in an expanding atmosphere while ensuring energy conservation. Models are defined via the chemical abundances, surface effective temperature $T_*$, bolometric luminosity $L$, mass-loss rate $\dot{M}$, terminal wind speed $v_\infty$, and clumping factor $D$ (which describes the ratio of the clumped material to an equivalent smooth wind). The surface effective temperature $T_*$ is defined at a mean Rosseland optical depth of $\tau_{\rm Ross} = 20$; the photospheric effective temperature at $\tau_{\rm Ross} = 2/3$ is denoted as $T_{2/3}$. The stellar radius $R_*$  follows from the Stefan-Boltzmann relation.  The wind is assumed to follow a standard $\beta$ law\cite{CAK1975} with $\beta = 1$. While the models do not provide a full representation of the physical conditions in the CSM (e.g., ionising bremsstrahlung radiation), we can use the effective temperature as a proxy for the ionisation balance in the emitting layers.

A more convenient parameter for describing the wind is given by the transformed radius\cite{Schmutz1989},

\begin{equation}
 R_\text{t} = R_* \left[ \frac{v_\infty}{2500\,{\rm km}\,{\rm s}^{-1}\,}  \middle/  
 \frac{\dot{M} \sqrt{D}}{10^{-4}\,{\rm M}_\odot\,{\rm yr}^{-1}}  \right]^{2/3}\, .
\label{eq:Rt}
\end{equation}
For given $T_*$, $R_{\rm t}$, and chemical abundances, the strength of emission lines remains similar, regardless of the luminosity.

We do not attempt to accurately reproduce the spectral line profiles, which would likely require more sophisticated velocity fields than the $\beta$ law adopted here. Instead, we focus on the impact of varying $T_*$ on the iron diagnostics. Supplementary Figure\,\ref{fig:PoWRTeff} shows a comparison between the UV data and three PoWR models. The PoWR models were computed with abundances following ref\cite{Dessart2017} and a solar iron composition,  $\log R_{\rm t} = 1.5\,[{\rm R}_\odot]$, $\log L = 5.3\,[{\rm L}_\odot]$, $v_\infty = 1500\,\kms$, and $D=4$, as well as  varying values of $T_* = 35$, 40, and 45\,kK.  As Supplementary Fig.\,\ref{fig:PoWRTeff} illustrates, the cooler model produces a wealth of Fe\,{\sc iv} lines in the range 1650--1900\,\AA. However, this model also greatly overpredicts the strength of the \CIIIscat\ line, and predicts the presence of lines such as \NIIIscat, which is not observed. A better match is obtained for models with $T_* \approx 45\,$kK (amounting to $T_{2/3} \approx 43\,$kK). 
However, models with $T_* \gtrsim 45\,$kK no longer show significant Fe\,{\sc iv} lines, which inhibits their usage as abundance diagnostics. At these temperatures, the Fe\,{\sc v} lines, which dominate the range 1400--1500\,\AA, would have provided useful abundance diagnostics, but unfortunately this range cannot currently be observed with disruptive ToO {\it HST} observations. We conclude that the subsolar iron abundance implied by the nearby \ion{H}{2} regions is consistent with the data, but we cannot provide more accurate constraints using the available spectra.

\end{methods}





 \clearpage
 
\begin{addendum}

\item[Supplementary Information is available for this paper]

\item[Correspondence] Correspondence and requests for materials
should be addressed to Erez~A.~Zimmerman~(email: erezimm@gmail.com).

\item
This preprint has not undergone peer review or any post-submission improvements or corrections. The Version of Record of this article is published in Nature, and is available online at https://doi.org/10.1038/s41586-024-07116-6 .

We thank Prof. Boaz Katz and Prof. Eli Waxman for their advice on the physical interpretation of SN\,2023ixf.
This work is based on observations obtained with {\it HST} as part of proposal GO-17205 (PI E. Zimmerman).
Based in part on observations obtained with the Samuel Oschin 48-inch Telescope and the 60-inch Telescope at the Palomar Observatory as part of the Zwicky Transient Facility (ZTF) project. ZTF is supported by the U.S. National Science Foundation (NSF) under grant AST-2034437 and a collaboration including Caltech, IPAC, the Weizmann Institute of Science, the Oskar Klein Center at Stockholm University, the University of Maryland, Deutsches Elektronen-Synchrotron and Humboldt University, the TANGO Consortium of Taiwan, the University of Wisconsin at Milwaukee, Trinity College Dublin, Lawrence Livermore National Laboratories, IN2P3, University of Warwick, Ruhr University Bochum and Northwestern University. Operations are conducted by COO, IPAC, and UW.
The SED Machine is based upon work supported by the NSF under grant 1106171.
Based in part on observations made with the Nordic Optical Telescope (NOT), owned in collaboration by the University of Turku, and Aarhus University, and operated jointly by Aarhus University, the University of Turku, and the University of Oslo (representing Denmark, Finland, and Norway, respectively), the University of Iceland, and Stockholm University, at the Observatorio del Roque de los Muchachos, La Palma, Spain, of the Instituto de Astrofisica de Canarias. These data were obtained with ALFOSC, which is provided by the Instituto de Astrofisica de Andalucia (IAA) under a joint agreement with the University of Copenhagen and NOT.
Some of the data presented herein were obtained at the W. M. Keck Observatory, which is operated as a scientific partnership among the California Institute of Technology, the University of California, and the National Aeronautics and Space Administration (NASA); the Observatory was made possible by the generous financial support of the W. M. Keck Foundation.
This work includes observations obtained at the Liverpool Telescope, which is operated on the island of La Palma by Liverpool John Moores University in the Spanish Observatorio del Roque de los Muchachos of the Instituto de Astrofisica de Canarias with financial support from the UK Science and Technology Facilities Council.
A major upgrade of the Kast spectrograph on the Shane 3\,m telescope at Lick Observatory, led by Brad Holden, was made possible through generous gifts from the Heising-Simons Foundation, William and Marina Kast, and the University of California Observatories. Research at Lick Observatory is partially supported by a generous gift from Google.
This work benefited from the OPTICON telescope access program (\url{https://www.astro-opticon.org/index.html}), funded from the European Union's Horizon 2020 research and innovation programme under grant agreement 101004719.
Based in part on observations made with the Italian Telescopio Nazionale Galileo (TNG) operated on the island of La Palma by the Fundación Galileo Galilei of the INAF (Istituto Nazionale di Astrofisica) at the Spanish Observatorio del Roque de los Muchachos of the Instituto de Astrofisica de Canarias.
This research used data obtained with the Dark Energy Spectroscopic Instrument (DESI). DESI construction and operations is managed by the Lawrence Berkeley National Laboratory. This material is based upon work supported by the U.S. Department of Energy, Office of Science, Office of High-Energy Physics, under Contract No. DE–AC02–05CH11231, and by the National Energy Research Scientific Computing Center, a DOE Office of Science User Facility under the same contract. Additional support for DESI was provided by the U.S. National Science Foundation (NSF), Division of Astronomical Sciences under Contract No. AST-0950945 to the NSF’s National Optical-Infrared Astronomy Research Laboratory; the Science and Technology Facilities Council of the United Kingdom; the Gordon and Betty Moore Foundation; the Heising-Simons Foundation; the French Alternative Energies and Atomic Energy Commission (CEA); the National Council of Science and Technology of Mexico (CONACYT); the Ministry of Science and Innovation of Spain (MICINN), and by the DESI Member Institutions: www.desi.lbl.gov/collaborating-institutions. The DESI collaboration is honored to be permitted to conduct scientific research on Iolkam Du’ag (Kitt Peak), a mountain with particular significance to the Tohono O’odham Nation. Any opinions, findings, and conclusions or recommendations expressed in this material are those of the author(s) and do not necessarily reflect the views of the U.S. National Science Foundation, the U.S. Department of Energy, or any of the listed funding agencies.
We made use of IRAF, which is distributed by the NSF NOIR Lab.
The Gordon and Betty Moore Foundation, through both the Data-Driven Investigator Program and a dedicated grant, provided critical funding for SkyPortal.
A.V.F.'s supernova group at U.C. Berkeley has been supported by Steven Nelson, Gary and Cynthia Bengier, Clark and Sharon Winslow, Sanford Robertson, Briggs and Kathleen Wood, the Christopher R. Redlich Fund, and numerous other donors.
T.S. acknowledges support from the Comunidad de Madrid (2022-T1/TIC-24117).

\item[Author Contributions] 

E.~A.~Z. is the PI of the {\it HST} proposal, triggered the {\it HST} observations, reduced the {\it HST} spectra, conducted follow-up observations, conducted spectroscopic and physical analysis, and wrote the manuscript.\\
I.~I. helped develop the manuscript, contributed to follow-up design and execution, conducted photometric analysis and physical analysis, reduced the saturated \swift\ data, and contributed to the physical interpretation of the data.\\
P.~C. contributed to follow-up design and execution, reduced photometry, and reduced the \swift\ spectra.\\
A.~G.-Y. leads the Weizmann research group, provided mentorship, and edited the manuscript.\\
S.~S. conducted spectroscopic analysis on the ISM lines and reduced the XRT data.\\
D.~A.~P. classified the supernova and provided LT data.\\
J.~S. helped develop the manuscript, provided the NOT data, and is a ZTF builder.\\
A.~V.~F. leads the U.C. Berkeley research group, is PI of the Lick program, and thoroughly edited the manuscript.\\
T.~S. modeled the SN spectra with PoWR and advised about spectral features.\\
O.~Y. helped develop the manuscript and created several figures.\\
S.~S. provided microtelluric models for the HARPS spectra and advised on statistical tests.\\
R.~B. contributed to follow-up design and execution, and helped develop the manuscript.\\
E.~O.~O. advised on the physical interpretation and helped develop the manuscript.\\
A.~D.~C. conducted an analysis of the ISM absorption lines.\\
T.~G.~B. obtained and reduced all of the Lick data.\\
Y.~Y. and S.~S.~V. helped develop the manuscript.\\
S.~B. contributed to follow-up design and execution.\\
M.~A. is part of the ZTF calibration team.\\
A.~B. advised on the measurement of absorption-line EWs.\\
J.~S.~B. and K.~Zhang obtained data for this study.\\
P.~J.~B. and M.~R. consulted on {\it Swift} data reduction and helped develop the manuscript.\\
M.~M.~K. and K.~D. provided infrared data from Gattini and IRTF.\\
G.~D., J.~H.~T., and K.~M.~S. obtained the INT spectrum.\\
C.~F. requested and shared the KCWI data.\\
A.~H. and I.~S. advised on radio data.\\
K.~H. and J.~W. reduced LT data.\\
J.~P.~J. obtained the NOT/FIES spectra.\\
S.~R.~K. is the ZTF PI and contributed Keck data.\\
D.~K., J.~M., and T.~W. advised on the physical interpretation and models.
J.~D.~N. Obtained and reduced the KCWI data.\\
C.~M., M.~M., N.~Z.~P., and R.~C.~M. are part of the KCRM commissioning team and were involved in obtaining the KCWI data. \\
P.~E.~N. provided the DESI data and contributed to follow-up design and execution. \\
L.~Y. and A.~A.~M. helped develop the manuscript and are coauthors of {\it HST} proposal GO-17205.\\
R.~S.~P. provided photometry from the Post Observatory, Mayhill, NM.\\
Y.~Q. and R.~D.~S. provided and reduced Keck spectroscopic data.\\
A.~R. reduced the DESI spectra. \\
R.~R. and B.~R. are ZTF builders.\\
M.~S. is a coauthor of {\it HST} proposal GO-17205.\\
A.~W. is part of the ZTF data-system team.\\

\item[Competing Interests] 
The authors declare that they have no competing financial interests.

\item[Data Availability Statement]
Photometry and spectra used in this study will be made available on WISeREP\cite{yg12,WISeREP}. A log of the available spectra can be found in Supplementary Table \ref{tab:spectra}. All scripts used to conduct the analyses presented in this paper are available from the corresponding author upon request. Opticon observations were obtained under Program ID OPT/2023A/001, PI Erez Zimmerman.

\item[Code Availability Statement]
Relevant software sources have been provided in the text, web locations provided as references, and are publicly available.

\end{addendum}

\setcounter{enumiv}{\value{firstbib}} 

\begin{extendeddata}
 
\begin{EDfigure}
    \centering
    \includegraphics[width=\textwidth]{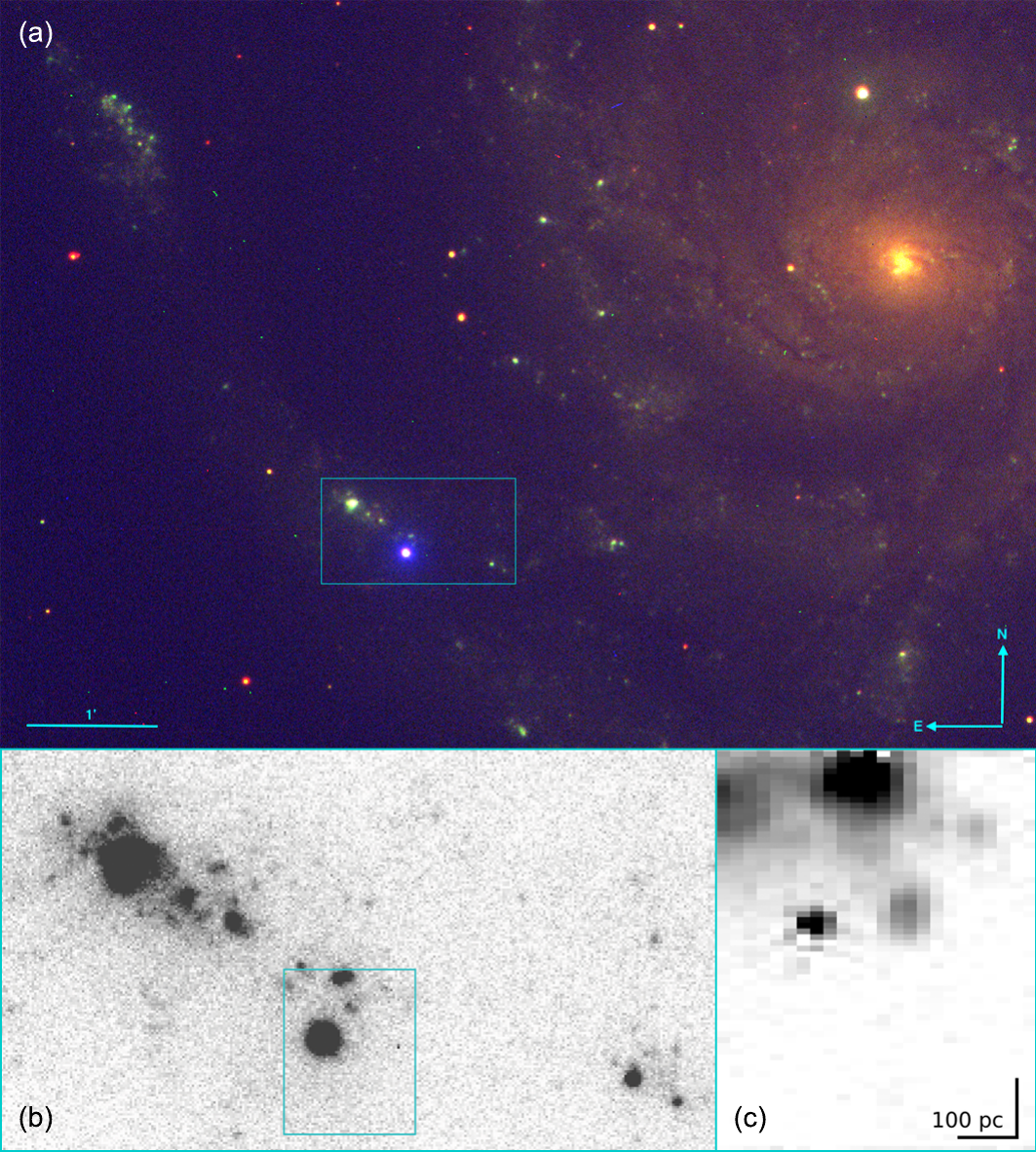}
    \caption{(a) A composite $ugi$ image taken by the Liverpool Telescope (LT) showing the location of SN\,2023ixf within the Messier 101 galaxy. The SN is clearly seen to be very blue (the most blue object within the inset rectangle). It is located within a spiral arm at the outskirts of the galaxy.
    (b) The LT $g$-band image showing the area marked in the composite image. Nearby star-forming regions are clearly seen.
    (c) H$\alpha$ image of the direct vicinity of SN\,2023ixf, as marked in the LT $g$-band image, constructed from observations with the integral field unit KCWI. The SN explosion site
    is embedded in a region of ongoing star formation, including diffusely distributed lower-level star-formation activity and islands of enhanced star formation.
    \label{fig:Galaxy}}
\end{EDfigure}

\clearpage

\begin{EDfigure}
\centering
\begin{tabular}{cc}
\includegraphics[width=\textwidth]{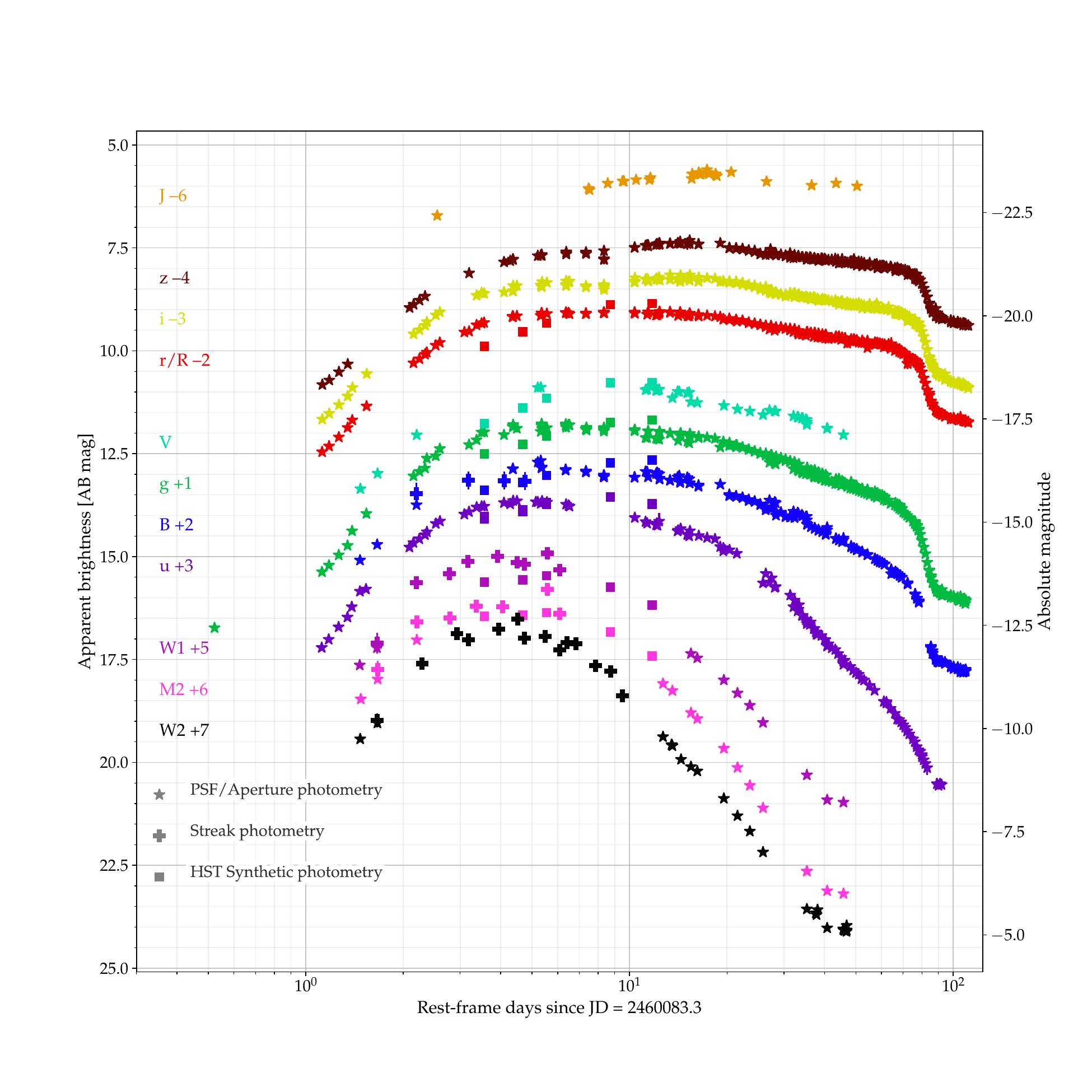}
\vspace*{-3cm}
\end{tabular}
\caption{UV--optical light curves of SN\,2023ixf using photometry reported in this work. PSF or aperture photometry are shown with star symbols, UVOT streak photometry is indicated using a plus symbol, and synthetic photometry from {\it HST} is indicated using squares. 
  \label{fig:lightcurve}}
 \end{EDfigure}
 
\clearpage

\begin{EDfigure}
 \centering
\begin{tabular}{cc}
\includegraphics[width=\textwidth]{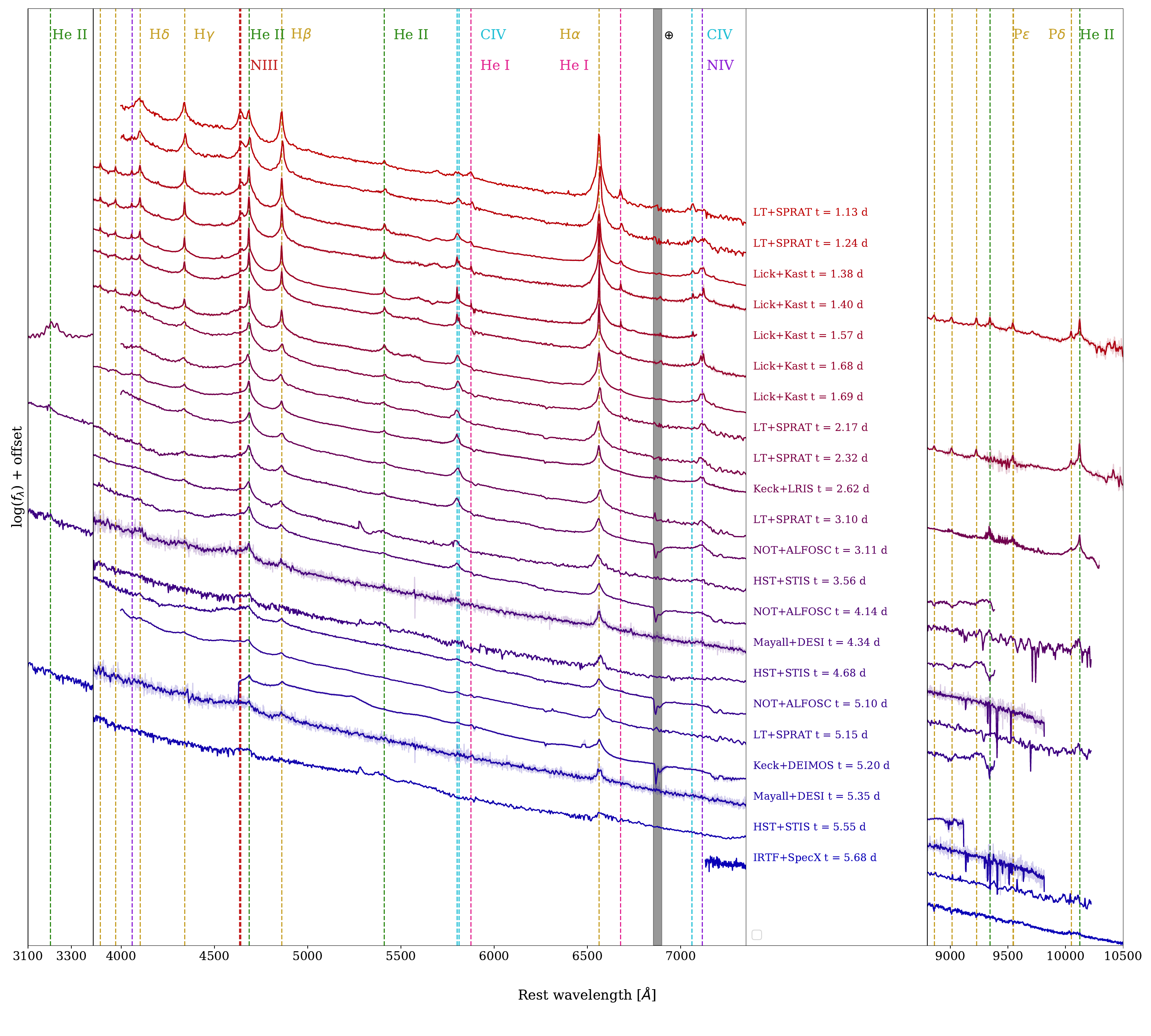}
\vspace*{-3cm}
\end{tabular}
\caption{The early-time spectral sequence of SN\,2023ixf. A plethora of narrow flash-ionisation lines is seen in the earliest spectra, including the H Balmer series, H Paschen series, several series of \HeII\ ($n\rightarrow3,4,5,6$), \ion{He}{1}, \CIV, \NIV, \NIII, and \CIII. Lower-ionisation species (\CIII, \NIII, \ion{He}{1}) weaken until they disappear in the optical spectra by day 2.
  \label{fig:spec_early}}
 \end{EDfigure}
 
\begin{EDfigure}
 \centering
\begin{tabular}{cc}
\hspace*{-1cm}\includegraphics[width=19cm]{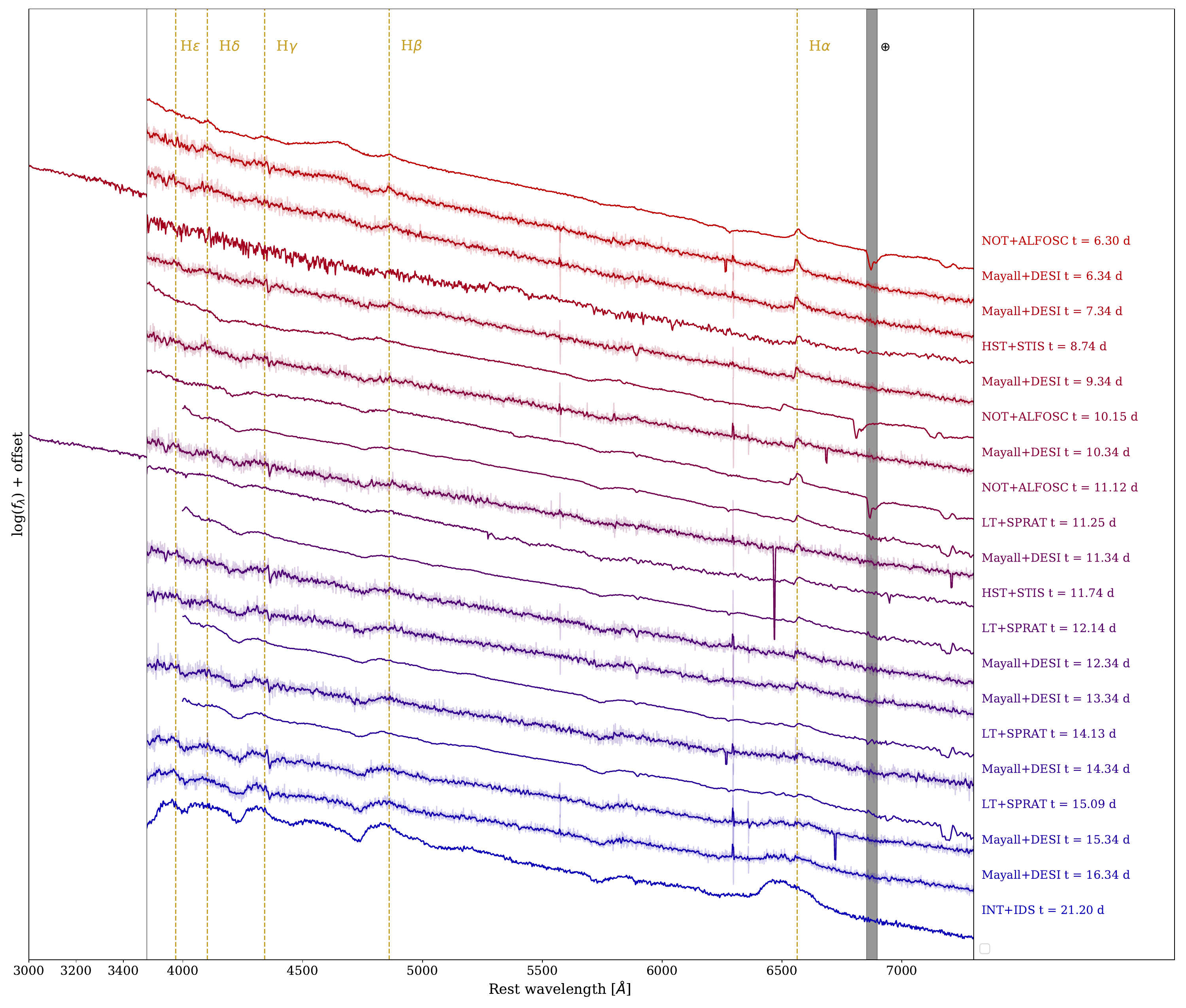}
\vspace*{-3cm}
\end{tabular}
\caption{The spectral sequence of SN\,2023ixf after $t=6$ days, and until broad features appear. All narrow lines except narrow $H\alpha$ P~Cygni disappear by day 6 in the optical. The narrow P~Cygni no longer appears by day $\sim 16$.
  \label{fig:spec_late}}
 \end{EDfigure}

 \clearpage

\begin{EDfigure}
 \centering
\vspace*{-2cm}\includegraphics[width=\textwidth]{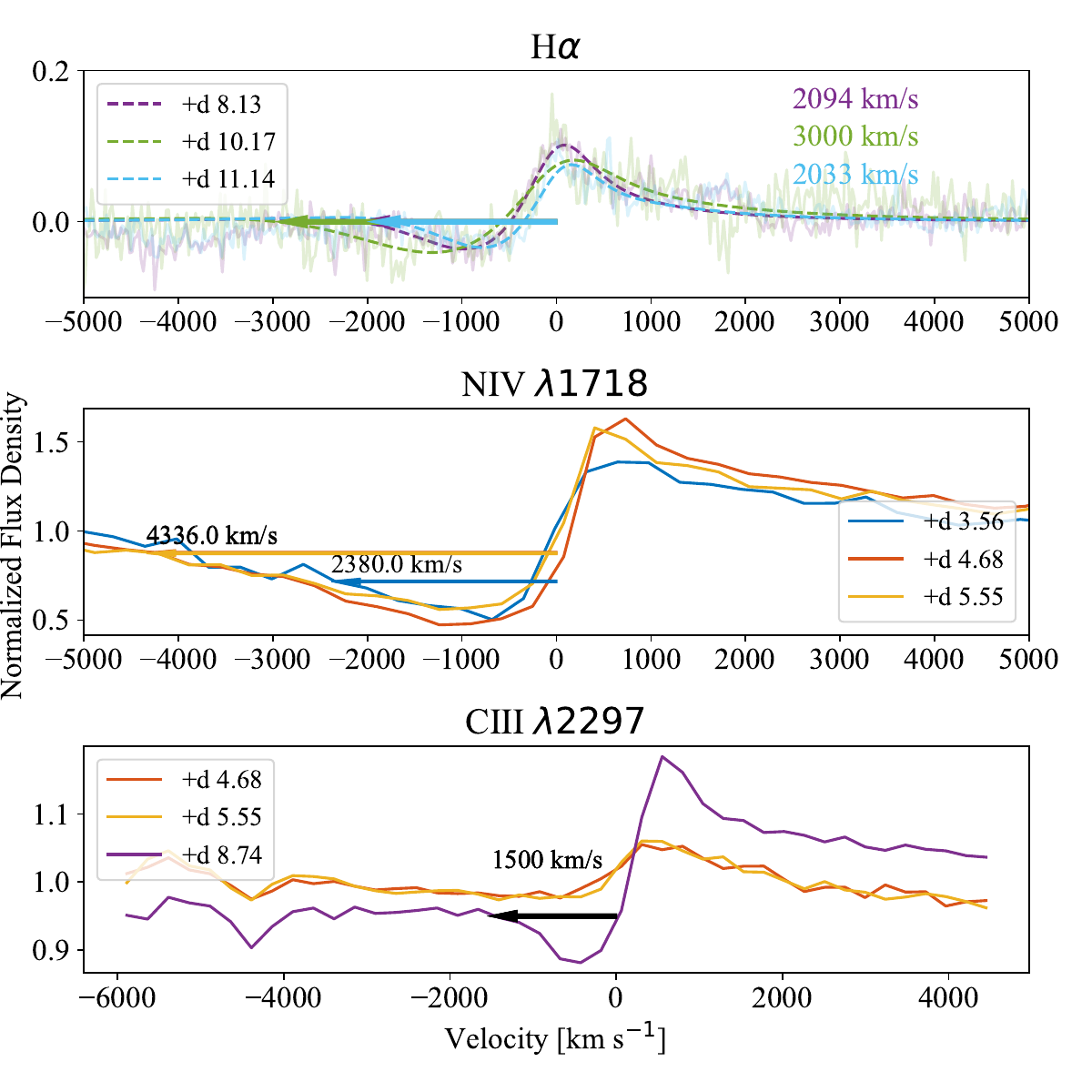}
\vspace*{-3cm}\caption{Narrow P~Cygni profiles from the NOT/FIES spectra (optical, H$\alpha$) and {\it HST/STIS} (UV, \NIV, \CIII). We model (dashed lines) the H$\alpha$ profile to measure the blue-edge velocity, which is shown in corresponding colours. The original spectrum (binned, transparent) is shown as well.
We do not model the \NIV\ feature as it lies at the edge of the STIS CCD, and one cannot determine the exact blue edge of the P~Cygni profile. We therefore adopt a phenomenological value of 3 times the minimum velocity of the profile as the blue-edge velocity. We also do not model the \CIII\ feature as it is blended with a photospheric absorption in the $t=8.74$\,day spectrum. However, the blue edge is clearly seen to be $\sim 1500 \kms$ (marked).
\label{fig:UV_pcygni}}
\end{EDfigure}

\clearpage

 \begin{EDfigure}
 \centering
\vspace*{-3cm}\hspace*{-2cm}\includegraphics[width=21cm]{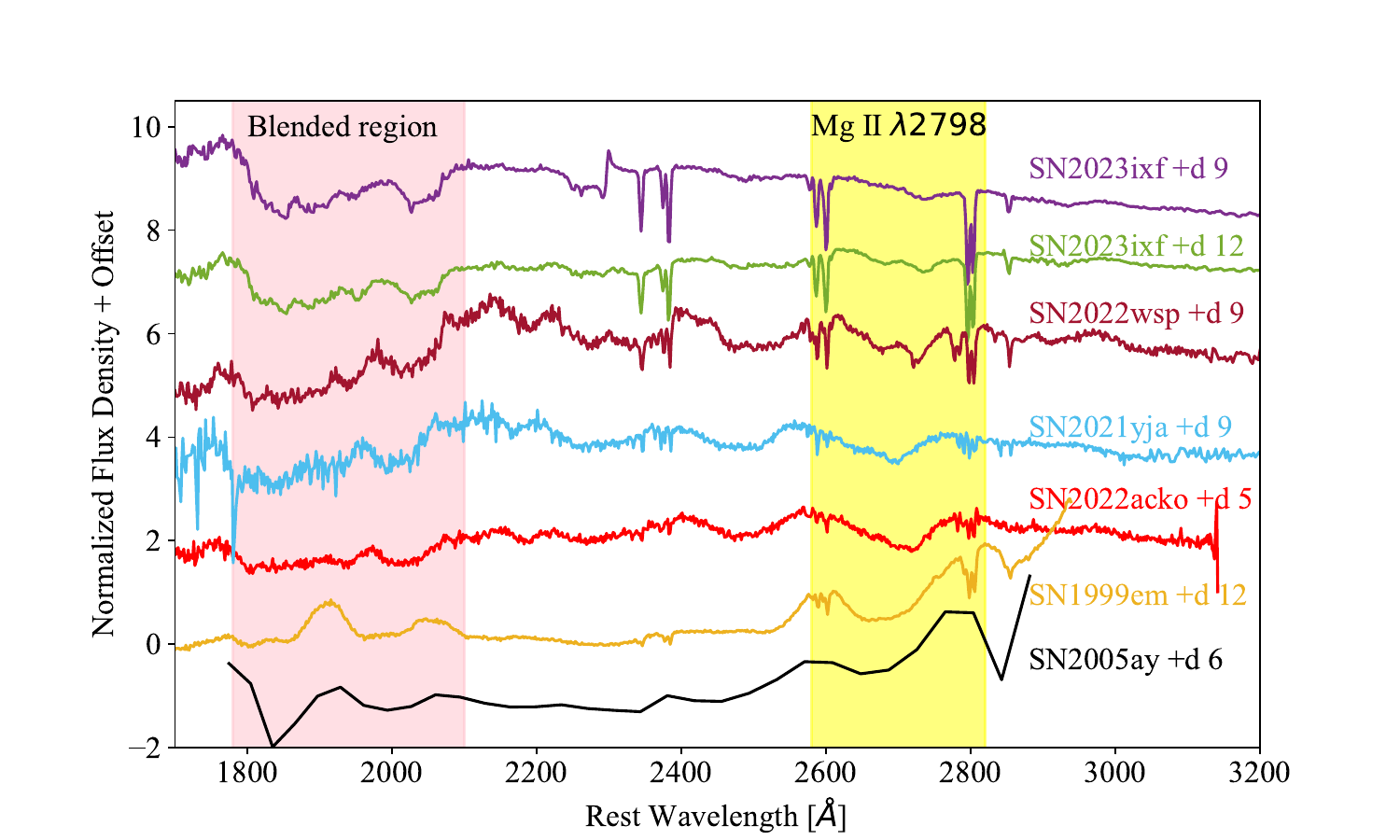}
\vspace*{-3cm}\caption{Comparison of photospheric UV spectra of SN\,2023ixf with the early-time UV spectra of SN\,2022wsp\cite{Vasylyev2023b}, SN\,2021yja\cite{Vasylyev2022}, SN\,2022acko\cite{Bostroem2023a}, SN\,1999em \cite{Baron2000}, and SN\,2005ay \cite{Gal-Yam2008}. All spectra were taken by {\it HST} except for the SN\,2005ay spectrum obtained with {\it GALEX}. The prominent \ion{Mg}{2} $\lambda 2798$ line is marked in yellow. Its double-peaked shape is similar to that of SN\,2022wsp, indicating an \ion{Fe}{2}~transition\cite{Vasylyev2023b}. The broad absorption across 1800--2100\,\AA\ is marked as well and arises from a mix of metals. The features in SN\,2023ixf are similar to those of SN\,2022wsp, SN\,2021yja, and SN\,2022acko, yet shifted due to different velocity regimes. The uniformity of these features suggests that they originate from the exploding star's natal chemical composition. However, a strong emission feature at $\sim 1910$\,\AA\ appears in the spectra of SN\,1999em and SN\,2005ay, suggesting some diversity in SN~II UV photospheric features exists as well.\label{fig:Photospheric_UV}}
\end{EDfigure}

\clearpage

\begin{EDfigure}
 \centering
\includegraphics[width=\textwidth]{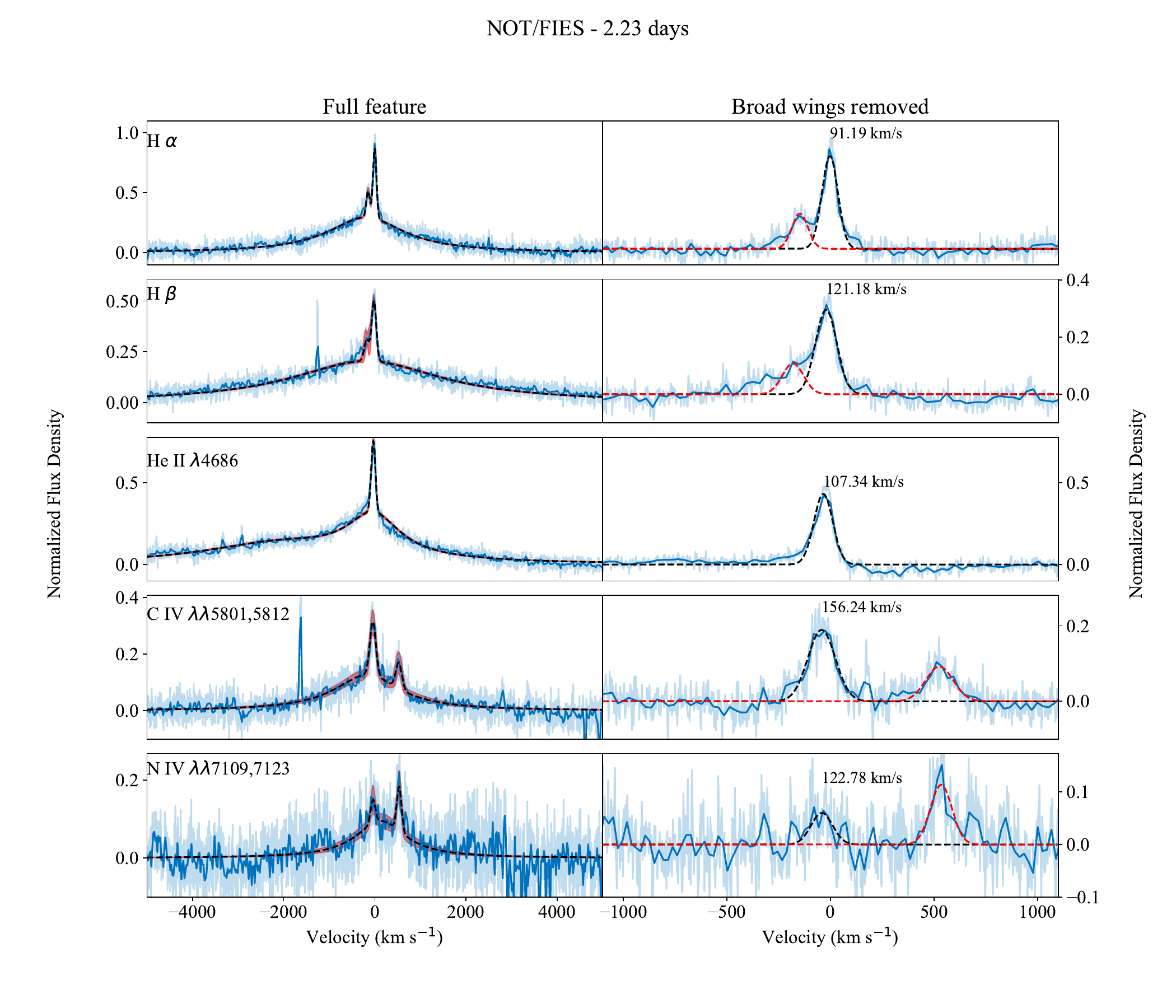}
\caption{Narrow flash features from the $t=2.3\,d$ NOT/FIES spectrum. We fit a model consisting of a Lorentzian base along narrow Gaussian components for the respective narrow lines. For \HeII\ we fit a second Lorentzian component to account for the blue edge.\label{fig:Narrow_lines}} 
\end{EDfigure}

\clearpage

\begin{EDfigure}
 \centering
\vspace{-2cm}\includegraphics[width=15cm]{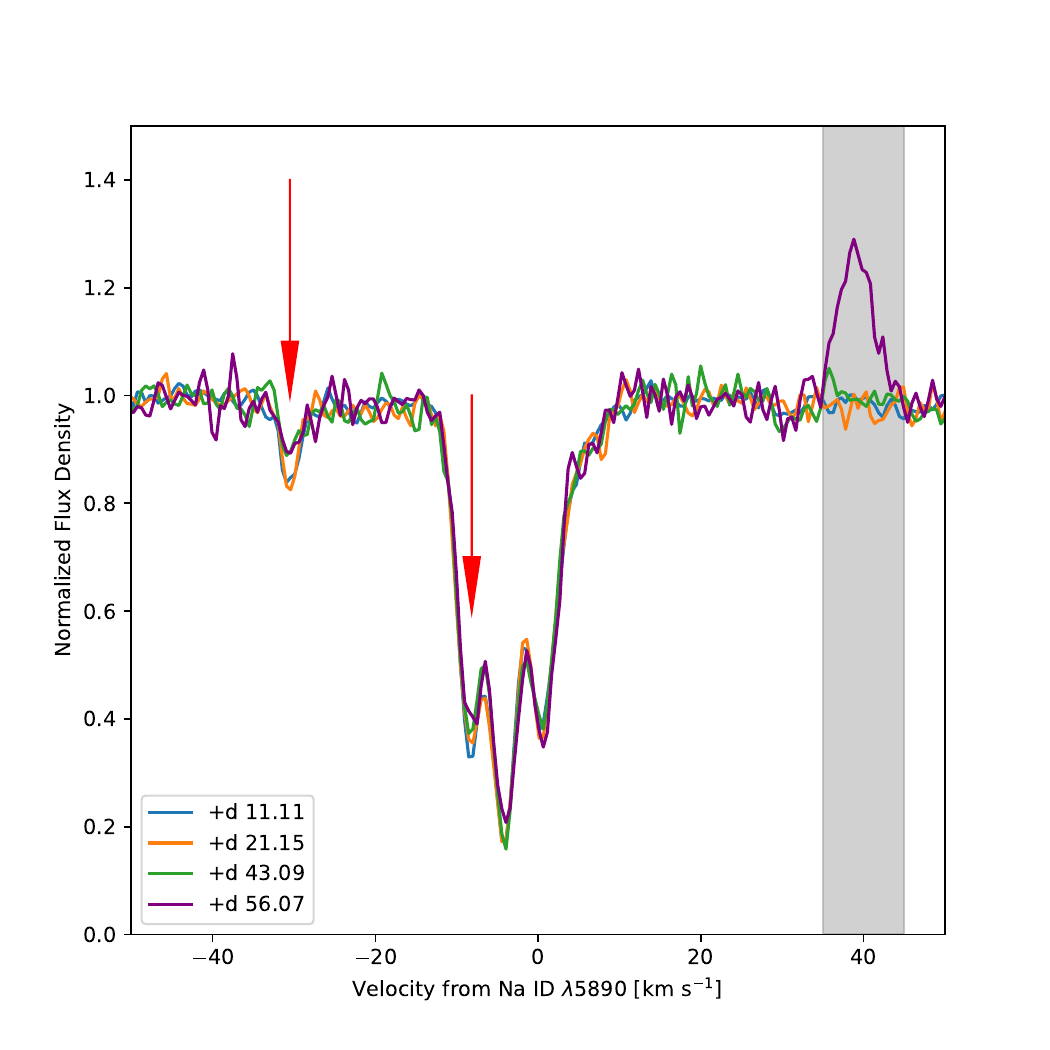}
\vspace{-2cm}\caption{The Na~I~D1 $\lambda 5890$ complex from the first four HARPS-N spectra. Three distinct clouds of Na~I~D are seen in the strong line-of-sight ISM line. Another weaker component is visible at $-30 \kms$. Components that show a significant decrease in EW are marked with a red arrow. The decrease in EW is caused by ionisation of neutral Na, which requires the changing component to come from the SN CSM. The component in the $t=56.07$\,day spectrum at $\sim 40 \kms$ arises from a calibration lamp due to the SN becoming fainter (marked in grey).\label{fig:NaID}}  
\end{EDfigure}

\clearpage

\begin{EDfigure}
 \centering
\hspace*{-2cm}\includegraphics[width=21cm]{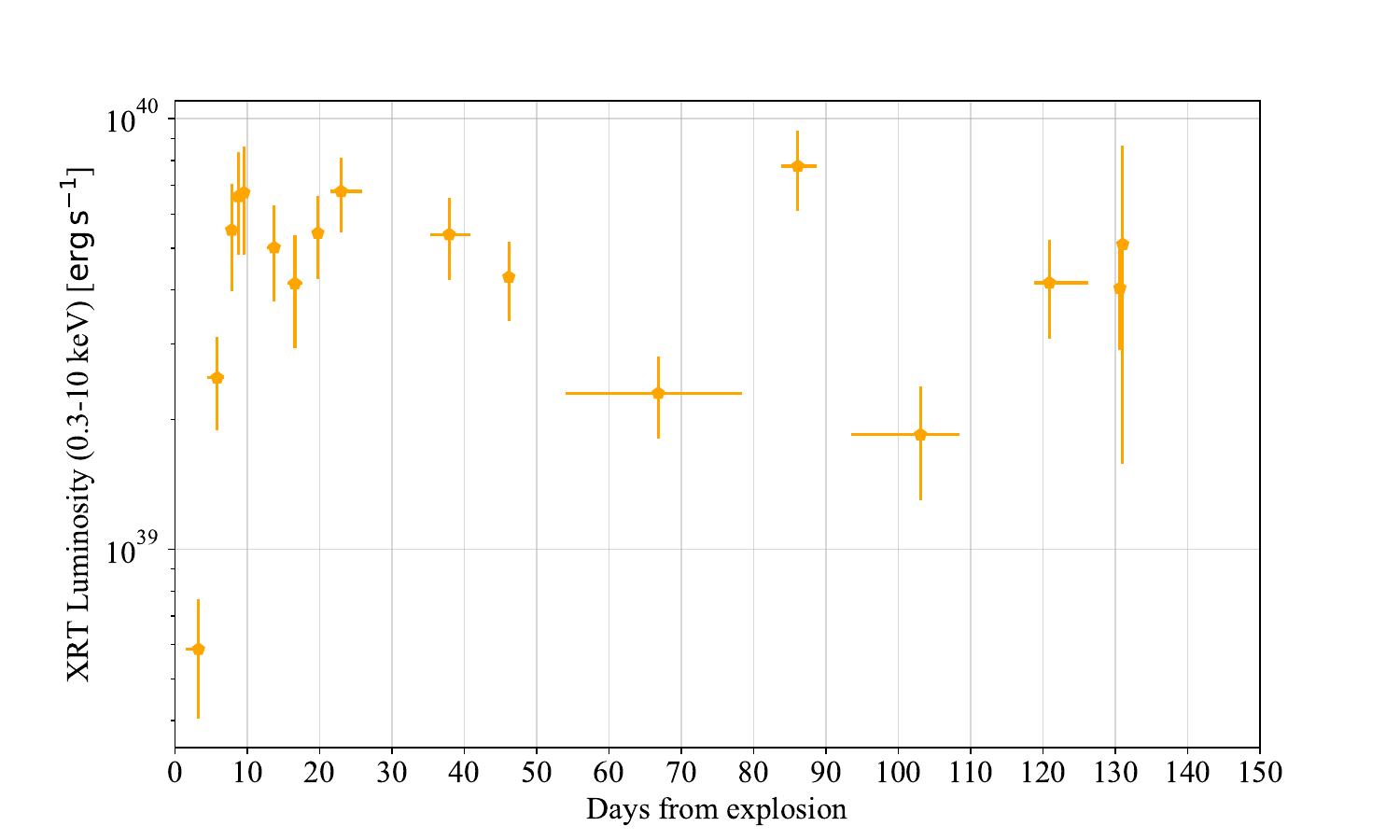}
\hspace*{-0.1cm}\caption{Absorbed X-ray luminosity from the XRT 0.--10\,keV band. Each point represents a binned measurement, with an error bar representing the measurement period. The relatively constant X-ray luminosity suggests an average CSM density profile of $\rho\propto r^{-2}$. The variability around the mean luminosity between $t\approx10\,$--$\,20$ days could correspond to variability in the CSM density, as suggested by progenitor papers (Methods $\S 5$).
\label{fig:XRTlc}} 
\end{EDfigure}

 \clearpage
 
\begin{EDfigure}
 \centering
\begin{tabular}{cc}
\hspace*{-1cm}\includegraphics[width=18cm]{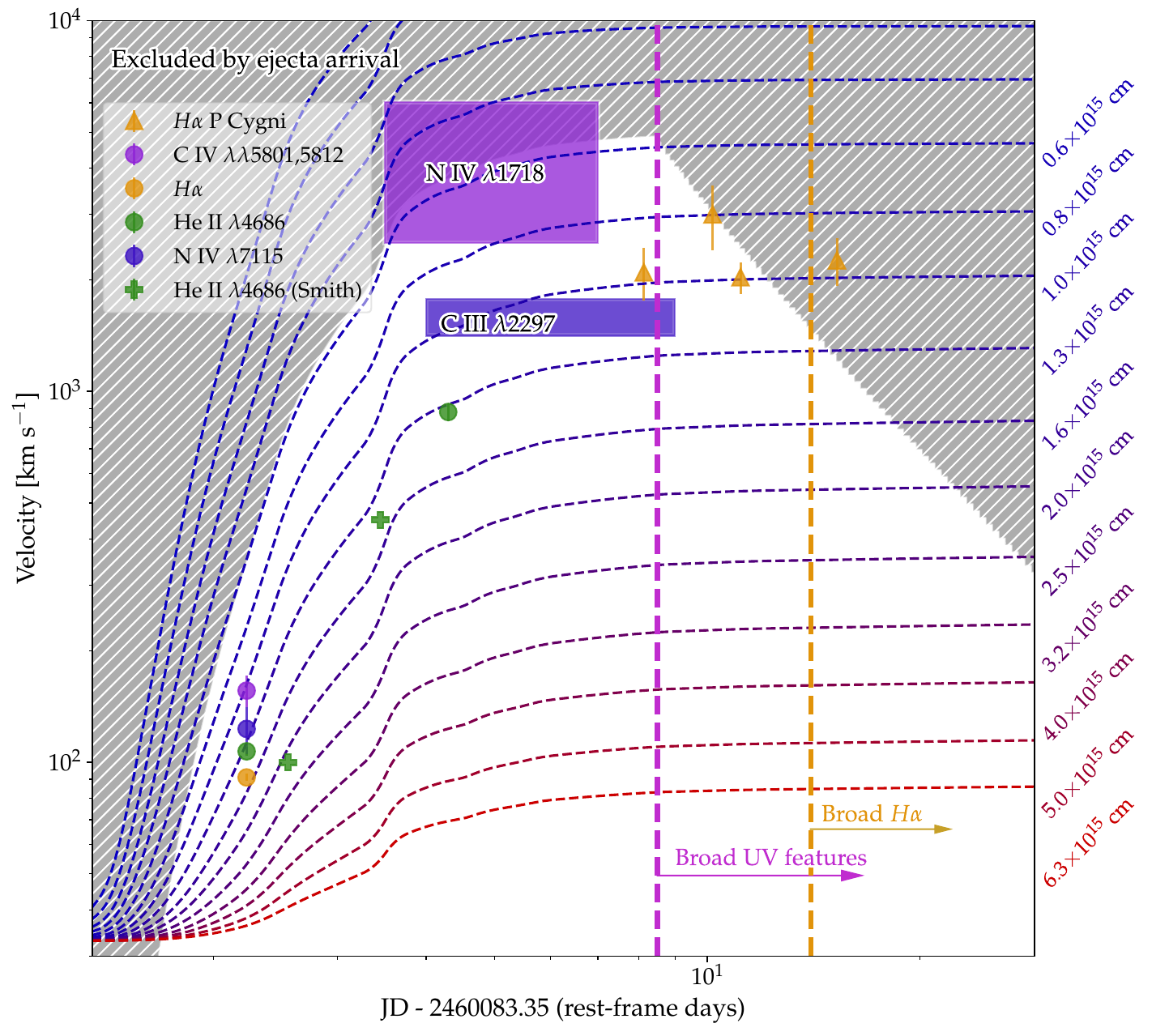}
\vspace*{-2cm}
\end{tabular}
\caption{Feature velocities and appearance times for H$\alpha$, \HeII\ $\lambda 4686$, \NIV\ $\lambda 1718$, and \CIII\ $\lambda 2297$. The dashed lines show the expected radiative acceleration of stationary matter at $\tau <1$, calculated from the cumulative radiated energy at different radii. We indicate the region excluded by ejecta starting at $R_{\rm bo}/2 = 0.8\times 10^{14}\, \rm cm$ with a velocity of $10^{9}\, \rm{cm\, s^{-1}}$, estimated from the blue edge of the photospheric spectrum. Approximate measurements from ref\cite{Smith2023} are added as well.
  \label{fig:radiative_acceleration}}
 \end{EDfigure}

\end{extendeddata}

\paragraph{Supplementary Methods}

\section{Photometric reduction:}

\paragraph{P48:}
A pre-discovery $g$-band image of SN\,2023ixf was obtained by the Zwicky Transient Facility (ZTF)\cite{Bellm2019,Graham2019} with the Palomar 48-inch Schmidt telescope (P48) on 2023-05-19 07:45:07. The image was reduced through an automatic pipeline\cite{Masci2019}. Follow-up photometry was not obtained with P48 owing to saturation.

\paragraph{Liverpool Telescope:}
We carried out an extensive campaign of imaging observations of SN\,2023ixf, beginning from 2023-05-19 at 22:06:49.78 (within a few hours of the initial report by Itagaki et al.) and continuing on all operational nights of the telescope thereafter.  Observations were acquired in all SDSS filters ($u$, $g$, $r$, $i$, $z$) with varying exposure times.  For the first few nights when the SN was rapidly rising we acquired 3--5 sets of observations each night with a cadence of about 1\,hr; after peak brightness our cadence dropped to 2 observations per night (cadence 4\,hr) and then 1 observation per night.  There are two significant gaps in our coverage: the gap shortly after peak brightness is due to inclement weather (high humidity); the subsequent gap was primarily due to a power-supply issue. 

In all epochs and filters to date, the SN is sufficiently bright that host-galaxy contamination is not a significant source of error, and we measure the magnitudes of the source using standard aperture photometry with a radius matched to the seeing conditions.  However, neither the SDSS nor PS1 photometric catalogues are reliable in this field owing to the complex M101 background.  We constructed our own photometric catalog of secondary standards using data from the night of discovery and observations of the standard stars PG1047, BD+18, and SA\,114 (on this night, several images of the field were taken at varying airmasses, the night was photometric, and exposures were long as the SN was still well below peak).  This secondary-standard catalogue was then used to calibrate all other observations of the field.

A small number of observations close to peak light are (mildly) saturated.  Frames for which the peak counts exceed 46,000 DN were identified and excluded.

The statistical error on virtually all measurements is negligible. The photometric uncertainty is approximated using the scatter in the magnitude measurements of the calibration stars relative to our secondary calibration.  This does not include systematic uncertainties in the calibration of the field, which are estimated to be about 0.05\,mag (but will affect all measurements in each filter in the same way).

\paragraph{P60/SEDM:}

We observed SN\,2023ixf with the robotic 60-inch telescope at Palomar Observatory (P60)\cite{Cenko2006} using the Spectral Energy Distribution Machine (SEDM)\cite{Blagorodnova2018}. We obtained 31 epochs of $g$, $r$, and $i$-band images and 30 epochs of $u$-band images between 2023 May 20 and 2023 June 30. Short exposure times of around 10\,s were adopted to avoid saturation when SN\,2023ixf was bright. The first-epoch images taken on 2023 May 20 were chosen as reference images in each band. We manually selected a few isolated stars in the reference image with negligible host flux contamination to build the effective point-spread function (PSF) for each image. We performed PSF photometry for all the detected sources in each image and relative photometry calibration against the reference image in each band. The relative photometry was performed with the sources used for PSF construction. Source detection, PSF construction, and PSF photometry were done with IRAF (IRAF is distributed by the National Optical Astronomy Observatory, which is operated by the Association of Universities for Research in Astronomy (AURA) under a cooperative agreement with the U.S. National Science Foundation NSF.) using the {\tt daophot} package\cite{Tody1986}. We then derived the photometric zero points of the reference image and applied the zero points to the relative photometry to get the final light curves. For photometry calibration, we used the AAVSO Photometric All-sky
Survey (APASS\cite{Henden2015}) catalogue for the $g$, $r$, and $i$ bands, and the secondary magnitudes derived for LT photometry above for the $u$ band. We checked and confirmed that the differences between APASS and LT secondary standards are smaller than 0.05\,mag in the $g$, $r$, and $i$ bands.

\paragraph{RC32 telescope at Post Observatory:}

We started the follow-up campaign of SN\,2023ixf on 2023 May 23 with an RC32 telescope operated by Post Observatory at Mayhill, New Mexico. The telescope has an aperture size of 80\,cm and is equipped with an FLI CCD. Observations were obtained with Astrodon Photometric filters, specifically Johnson\cite{Johnson} $B$ and Sloan\cite{Sloan} $g,\,r,\,i,\,z$. We followed the PSF photometry method as described in ref\cite{Chen2022}. The final $B$-band magnitudes in the Vega system were obtained by calibrating them to the APASS catalog, and the $g,\,r,\,i$, and $z$-band magnitudes in the AB system \cite{Oke_Gunn1983} by calibrating them to the ATLAS All-Sky Stellar Reference Catalog \cite{Tonry2018}.

\paragraph{Las Cumbres Observatory 0.4\,m telescope}
We used public images of SN\,2023ixf taken between 2023 May 20 and 2023 June 4 with a Las Cumbres Observatory 0.4\,m telescope\cite{Brown2013}. The reduced images with BANZAI \cite{McCully2018} were downloaded from the Las Cumbres Observatory data archive. We performed photometry in the same way as we did for RC32 images and obtained $B,\,V$ (Vega system) and $g,\,r,\,i$ (AB system) magnitudes. For $V$-band calibration, we used the APASS catalog.

\paragraph{\swift/UVOT photometry:}
Immediately following classification, UV photometry was acquired using the UV-Optical Telescope (UVOT) onboard the \textit{Neil Gehrels Swift Observatory} \cite{Gehrels2004,Roming2005}. UVOT observed the field 46 times during the first 3 weeks of the event and documented the rise and decline of the UV light curve. During most of the UV-bright phase of the evolution ($\lesssim 13$ AB mag), SN\,2023ixf saturated the UVOT images. However, the data were recoverable using readout-streak photometry. We reduced these data using the methods of ref\cite{Page2013}. We masked the raw images for bright and faint sources, corrected them for large-scale and small-scale sensitivity variations, and corrected them for sensitivity loss using the latest calibration files from September 2020. We then calculated the coincidence-loss-corrected mean count rates across the pixel columns and converted these to magnitudes using the zero point reported by ref\cite{Page2013}, depending on the appropriate readout mode of the detector. A 0.1\,mag systematic calibration error is assumed in all observations, and data are binned within intervals of 0.5\,day to reduce scatter.

In epochs during which observations did not saturate (both before and after peak brightness), we reduced images using the \swift \package{HEAsoft} toolset\cite{Heasoft}. Source counts were extracted using \package{uvotmaghist} from the individual exposures images using a circular aperture with a $5''$ radius. The background was estimated from several larger regions surrounding the host galaxy. These counts were then converted to magnitudes using the photometric zero points of ref\cite{Breeveld2011} with the latest calibration files from September 2020. We did not attempt to subtract the host-galaxy flux at the location of the SN under the assumption that the SN dominates the flux at the SN location, which is verified by the $\sim 4$--7\,mag decline observed by UVOT after peak brightness. By integrating the UVOT transmission filters, we estimate that the red leak from optical wavelengths becomes important ($\sim 2.5\%$) at UV minus optical colours of $\sim 4$\,mag.

\paragraph{Gattini:} The position of SN\,2023ixf was covered as part of survey operations of the Palomar Gattini-IR $J$-band survey\cite{Moore2019, De2020a}. We retrieved $J$-band photometry of SN\,2023ixf by performing forced PSF photometry at the SN position on the difference images generated in the data-reduction pipeline\cite{De2020b}. We show the resulting light curves in Extended Data Figure \ref{fig:lightcurve}.

\clearpage

\subsection{2. Spectroscopy reduction:\\}

We have made use of the {\tt skyportal}\cite{Coughlin2023} instance {\tt fritz} for observation planning and collaboration within ZTF for this study.

\paragraph{P60/SEDM:}

We obtained a series of spectra with the P60 using the SEDM (see previous section for photometry).
All spectra were automatically reduced \cite{Rigault2019}.

\paragraph{Liverpool Telescope:}
Following the classification spectrum, we obtained a spectral series of SN\,2023ixf with the SPRAT robotic spectrograph \cite{Piascik2014} mounted on the Liverpool Telescope (LT). A spectrum was taken on all operational nights of the telescope since the SN discovery. To improve the flux calibration and SNR of the spectra, we manually ran the reduction pipeline on all spectra instead of relying on the automatic pipeline. To improve the flux calibration, we reapplied standard-star correction with ESO reference standard-star spectra. As the trace was unusually bright, we could not have corrected the spectra for cosmic rays using a standard procedure such as the LACosmic algorithm. We therefore correct for cosmic rays by manually removing any obvious cosmic-ray outliers in the extracted spectra.

\paragraph{Lick Observatory:}

We obtained a series of early-time spectra, starting from $t=1.3$\,days, using the Kast double spectrograph\cite{Miller1994} mounted on the Shane 3\,m telescope at Lick Observatory. These observations were made using both arms of the instrument and in several configurations, providing a resolving power of $R=970$--1940 for the blue side and $R=690$--4180 for the red side (see Table \ref{tab:spectra} for exact details).
To minimise slit losses caused by atmospheric dispersion\cite{Filippenko1982}, the slit was oriented at or near the parallactic angle. The data were reduced following standard techniques for CCD processing and spectrum extraction\cite{Silverman2012} utilising IRAF routines and custom {\tt Python} and {\tt IDL} codes\cite{TheKastShiv}. Low-order polynomial fits to comparison-lamp spectra were used to calibrate the wavelength scale, and small adjustments derived from night-sky emission lines in the target frames were applied. The spectra were flux calibrated using observations of appropriate spectrophotometric standard stars observed on the same night, at similar airmasses, and with an identical instrument configuration.

\paragraph{Swift/UVOT grism:}
We requested {\it Swift}/UVOT grism spectra of SN\,2023ixf shortly after the discovery. The first epoch of UVOT  observation started on MJD = 60084.47 (1.67\,days after the estimated explosion time). We acquired the UVOT grism data taken between 2022 May 20 and 2022 July 5 from the Swift Archive Download Portal\cite{Swiftportal} and extracted the spectra with {\tt uvotpy}\cite{Kuin2014} which utilises calibration presented in ref\cite{Kuin2015}. The observation information is listed in Supplementary Table \ref{tab:grism_obsinfo} and the spectra are presented in Supplementary Figure \ref{fig:uvot_grism}. We summed all the exposures belonging to the same epoch (with the same OBSID) except for the first epoch (OBSID = 00032481003), for which we summed the second, third, and fourth exposures and kept the individual spectra from the first and the fifth exposure. As a result, we report 20 grism spectra of SN\,2023ixf in this paper. 

\paragraph{Dark Energy Spectroscopic Instrument (DESI):}
Starting from $t=4.3$\,days, the DESI collaboration\cite{desi2016a} observed and published spectra of SN\,2023ixf using the Dark Energy Spectroscopic Instrument (DESI) at the 4\,m Mayall telescope at Kitt Peak National Laboratory\cite{desi2023a}. Daily spectra were obtained when allowed by weather until $t=25.34$\,days, after which two more epochs were taken, resulting in a total of 21 spectra, which have been processed with the standard DESI pipeline\cite{Guy2023}. We use these spectra in our general spectroscopic analysis.
As these spectra are well flux-calibrated, we compared our photometric flux calibration to them as a sanity check for our process.

\paragraph{Hubble Space Telescope:}

We reduced all spectra using the standard {\tt STIStools} pipeline and its {\tt Python} interface. The reduction process included alignment and combination of the dithered spectra, cosmic-ray rejection, exposure combination, and fringe-flat correction for the G750L exposures. We manually removed any strong cosmic rays still showing after the pipeline process. While single-orbit combined spectra produce an hourly temporal resolution spectrum, we found no significant spectral development between orbits in any visit. Therefore, we chose to combine all exposures in single visits, yielding improved cosmic-ray and bad-pixel rejection.

During the {\it HST} visits, some guiding and saturation issues emerged. In visit 1, the guide-star acquisition was delayed, causing a failure of two exposures and an off-centre slit acquisition of all later exposures during the visit. This created an inaccurate absolute flux calibration during this visit, which we mitigated using a photometric flux correction of the optical grisms, applying the same shift to the NUV spectra during the visit. Visit 3 suffered guiding issues as well, resulting in the loss of five G230L exposures and one G750L exposure.

During visits 2--4 some parts of the spectrum were degraded owing to saturation, causing unreliable flux calibration because of the nonlinearity of the CCD response function. As these pixels are counted by the {\it STIStools} pipeline, they only represent a lower limit of the exact flux at the pixel. We note, however, that the severity of the issue mostly affects the G430L spectrum, covered by ground-based spectra, while the saturated NUV parts of the spectrum show counts much closer to the instrumental limit (of 33,000 counts). We also note that saturation does not affect the bluest wavelengths of the NUV spectra critical for blackbody fits (away from the Rayleigh-Jeans tail). While we cannot fully mitigate this issue, we find the overall quality of the NUV spectra to be good, as no clear nonlinearity is seen between the saturated red part of the NUV spectra and the unsaturated blue wavelengths. To correct the overall flux calibration, we applied a photometric correction to the optical parts of the spectra, adding the same shift to the NUV part of the spectrum.
Lastly, to avoid saturation in the fifth {\it HST} visit, we increased the CCD gain.

\paragraph{HARPS-N:}
We obtained a total of five spectra with the northern High Accuracy Radial velocity Planet Searcher (HARPS)\cite{Cosentino2012} mounted on the 3.58\,m Galileo National Telescope (TNG) in the Roque de los Muchachos Observatory, La Palma. These spectra were obtained through the OPTICON telescope access program 23A001 (PI: E. Zimmerman). The echelle spectrograph provides high-resolution ($R=115,000$) spectra. All spectra were reduced by the observatory using the standard pipeline. We used non-blaze-corrected spectra in our analysis to avoid artifacts from SN features that are broader than typical stellar features expected by the deblazing algorithm.

To account for atmospheric microtelluric absorption, we calculate a telluric model using the {\tt TelFit} code\cite{TelFit} with the atmospheric conditions at the observatory at the time of observations. For the $t=21$ and $t=56$\,day spectra, we manually increased the strength of the calculated telluric model by a factor of 2 and 2.5, respectively, as the untreated calculated model did not result in a complete removal of the identified telluric absorption. Since this correction is linear and applies to all calculated microtelluric absorption lines, we find the correction to be robust. 

\paragraph{Nordic Optical Telescope:}
We obtained a series of low-resolution spectra with the Alhambra Faint Object Spectrograph and Camera (ALFOSC)\cite{ALFOSC} on the 2.56\,m Nordic Optical Telescope (NOT) at the Observatorio del Roque de los Muchachos on La Palma (Spain). The spectra were obtained with grism \#4, providing a wavelength range of 3300--8500\,\AA, and either a 1\farcs0 or 1\farcs3 wide slit, depending on the weather conditions. The resolving power is $R \approx 340$ for the 1\farcs0-wide slit. The ALFOSC data were reduced using {\it PypeIt}\cite{Prochaska2020b,Prochaska2020a}, version 1.8.1. The reduction includes cosmic-ray rejection, bias corrections, flat fielding, and wavelength calibration using spectra of He-Ne comparison lamps obtained immediately after the target. The relative-flux calibration was done with spectrophotometric standard stars observed during the same night.

With the NOT, we also obtained high-resolution spectroscopy using the FIbre-fed Echelle Spectrograph FIES\cite{Telting2014a} in its medium-resolution mode, using fiber \#3 (diameter 1\farcs0) covering the wavelength interval 3700--8300\,\AA\ at $R \approx 46,000$.  The data were reduced using the FIEStool software\cite{Stempels2017} provided by the observatory. The reduction includes cosmic-ray rejection, bias corrections, flat fielding, wavelength calibration with a Th-Ar lamp, and merging the individual orders. The relative-flux calibration was done with spectrophotometric standard stars observed during the same night.

\paragraph{IRTF:} 
On 2023-05-25, we obtained a near-infrared (NIR) spectrum of SN\,2023ixf using SpeX on the NASA Infrared Telescope Facility\cite{Rayner2003} as part of program 2023A070 (PI K. De). The observations were acquired in the SXD mode, consisting of a series of dithered exposures in the ABBA pattern, amounting to a total exposure time of 600\,s. The spectra were reduced and extracted using the \texttt{spextool} package\cite{Cushing2004}. The extracted spectra were flux calibrated and corrected for telluric absorption with observations of the A0\,V star HIP\,65280 using the \texttt{xtellcor} package\cite{Vacca2003}.

\paragraph{KCWI:}

On 2023-06-05, we observed SN\,2023ixf with the Keck Cosmic Web Imager\cite{Morrissey2018} (KCWI) on a night of
commissioning for the newly installed red arm (Keck Cosmic Reionization Mapper, KCRM).  The
Medium slicer with a field of view of $16'' \times 20''$ was in place.  The blue arm was
configured with the BL grating yielding an average resolution of $R \approx 1800$ with a bandpass
of 3580--5578\,\AA.  The red arm was configured with the RL grating yielding an average
resolution of $R \approx 1000$ and a bandpass of 6465--9777\,\AA.  We acquired four images
with the blue arm and exposure times of 10\,s each, and three exposures with the red arm and
exposure times of 60\,s each. A nearby flux standard was observed right after the SN. The red-arm SN exposures were median stacked to mitigate cosmic
rays.

The data were reduced using a development version of the Keck KCWI pipeline\cite{KCWIpipeline} to perform instrument signature removal and the construction of geometrically-corrected and flux-calibrated data cubes.  The wavelength solution for the BL grating had an average residual of $0.089 \pm 0.012$\,\AA, and the RL grating solution yielded an average residual of $0.100 \pm 0.014$\,\AA.  The flux-calibration residuals had a root-mean-square (RMS) scatter of $\sim 2 \%$ in both channels.

\paragraph{INT/IDS:}
We obtained one spectrum with the Intermediate Dispersion Spectrograph (IDS) on the 2.5\,m Isaac Newton Telescope. The spectrum was reduced using standard \textsc{iraf/pyraf} and IDL/Python routines for bias/overscan subtractions and flat fielding, while the wavelength solution was derived with comparison lamps and verified against bright night-sky emission lines. The final flux calibration and telluric-line removal were performed using spectrophotometric standard-star spectra obtained on the same night.

\paragraph{Keck/LRIS:}

We acquired a set of three spectra using the Low Resolution Imaging Spectrometer (LRIS)\cite{Oke1995} on the W. M. Keck I telescope as part of program C256 (PI: S. Kulkarni), during the same night. The first spectrum was taken on 2023-05-21.29, approximately 2.46\,days after the explosion, and the last spectrum was taken at 2.58\,days post-explosion. All spectra were taken using a $1''$ slit oriented at the parallactic angle to minimise slit losses due to atmospheric dispersion. We chose a short exposure time of 20\,s to prevent saturation issues. The observational configuration provided a median resolving power of $R \approx 2000$ on the blue side (3000--5600\,\AA) and $R\approx 6700$ on the red side (5600--10,000\,\AA).

We reduced the data using {\tt LPipe}\cite{Perley2019}. First, the science exposures are flat-fielded using dome flats taken immediately after the observing run, cleaned for cosmic rays, and sky-subtracted. Owing to the lack of spectral evolution during the night, we stack three separate exposures into one. Then, a one-dimensional spectrum was extracted from the stacked science exposure, wavelength calibrated using comparison-lamp frames taken before the observing run, and flux calibrated using standard stars taken during the same night. Finally, the flux-calibrated spectra from both the blue and red sides were joined.

\paragraph{Keck/Deimos:}

On 2023-05-23, we obtained three spectra of SN\,2023ixf using the DEep Imaging Multi-Object Spectrograph (DEIMOS)\cite{Faber2003} on the 10\,m Keck II Telescope (PI: R. Stein). We performed observations with multiple long-slit configurations, to maximise wavelength coverage. We obtained one medium-resolution 30\,s spectrum with the 600ZD grating, and two higher-resolution 30\,s spectra with the 1200G grating (one with the GG455 filter, and another with the redder OG550 filter). 
The data were reduced using {\it PypeIt} \cite{Prochaska2020a, Prochaska2020b}, with the dedicated DEIMOS instrument default settings. Flux calibration was performed with the built-in {\it PypeIt} DEIMOS sensitivity functions.

\paragraph{Extinction estimate:}
To probe the host-galaxy extinction, we adopt the relations in ref\cite{Poznanski2012} between the equivalent width (EW) of \ion{Na}{1}~D absorption and extinction. We use our sequence of HARPS-N spectra, which resolve the Na absorption completely. The EW is measured directly from the spectra after removing any microtelluric absorption, taking the mean of the first HARPS-N spectra (for which we measure no decrease in Na component EW. We determine $E(B-V)=0.034$\,mag, which we multiply by a factor of 0.86 to correct for the conversion in ref\cite{Schlafly2011}; the final result is $E(B-V)=0.029\pm0.005$\,mag. Systematics dominate the uncertainty.

\paragraph{Photometric flux calibration:}
We calibrate the full spectral sequence to our photometric observations. First, we corrected the absolute-flux calibration by interpolating the $r$-band light curves using a fifth-order polynomial. We then corrected the relative-flux calibration by fitting a free linear function multiplying the flux of each spectrum. We find the best-fit parameters so that the interpolated $g,\,r,$ and $V$ photometry matches with the synthetic photometry calculated by integrating the spectra with the appropriate transmission filters acquired from the Spanish virtual observatory filter profile service\cite{svo2}. For the {\it HST} and the IR spectra, we do not apply a photometric relative-flux correction. 

\clearpage

\subsection{Reduction of XRT X-ray data}

The \swift\ satellite also obtained multiple epochs of SN\,2023ixf at X-ray energies with its X-ray telescope (XRT)\cite{Burrows2005} in photon-counting mode. In addition, \swift\ observed this field extensively between January 2005 and December 2019, long before the SN explosion, for 476\,ks in the windowed-timing (20\,ks) and the photon-counting (455\,ks) modes. We analysed all photon-counting data with the online tools of the UK \swift\ team that use the methods described by refs\cite{Evans2007, Evans2009, Evans2020} and the software package HEAsoft\cite{Heasoft}.

An image constructed from all pre-SN XRT data reveals no credible source within a radius of $19.8''$ from the SN position (for comparison, XRT's half-power radius\cite{Burrows2005} is $9''$ at 1.5\,keV). Some diffuse emission is present at the SN site. The inverse-variance-weighted count rate is $(3.8\pm0.5)\times10^{-4}~\rm ct\,s^{-1}$. Using the dynamic binning method of the \swift\ online tools, we measure the first $3\sigma$ detection of SN\,2023ixf with XRT on 2023-05-22 at 01:07:06. The uncertainty is +0.9\,day, $-1.8$\,day, reflecting the time interval of the XRT observations used to obtain this detection in the dynamic binning mode. The count rate at the time of discovery is $(10\pm3)\times10^{-4}\, \rm ct\,s^{-1}$, before removing the contribution from the diffuse emission. About 19\,days later, the light curve reached a maximum of $0.011 \pm 0.002\, \rm ct\,s^{-1}$. Since then, the light curve has been relatively stable, with a small oscillation in the X-ray flux, reaching a minimum on day 17, and peaking on day 23.5 again. The early X-ray rise is shown in Fig. \ref{fig:line_vs_xra}, while the full X-ray light curve is shown in Extended Data Figure \ref{fig:XRTlc}. The individual measurements are summarised in Supplementary Table \ref{tab:xrt}.

We convert the \swift XRT 0.3--10\,keV count rate to absorbed flux using the {\tt HEASARC} webPIMMS calculator\cite{webPIMMS}, assuming a thermal bremsstrahlung spectrum with a temperature of $T=34$\,keV and the hydrogen column $N(H)=5.6\times10^{22}\, \rm cm^{-2}$ as reported by {\it NuStar} in ref\cite{Grefenstette2023}. These values are broadly consistent with a conversion factor acquired by fitting the XRT data with an absorbed power-law spectrum.






\section{Supplementary Figures}

    

\clearpage

\begin{SIfigure}
 \centering
\begin{tabular}{cc}
\hspace*{-1.5cm}\includegraphics[width=15cm]{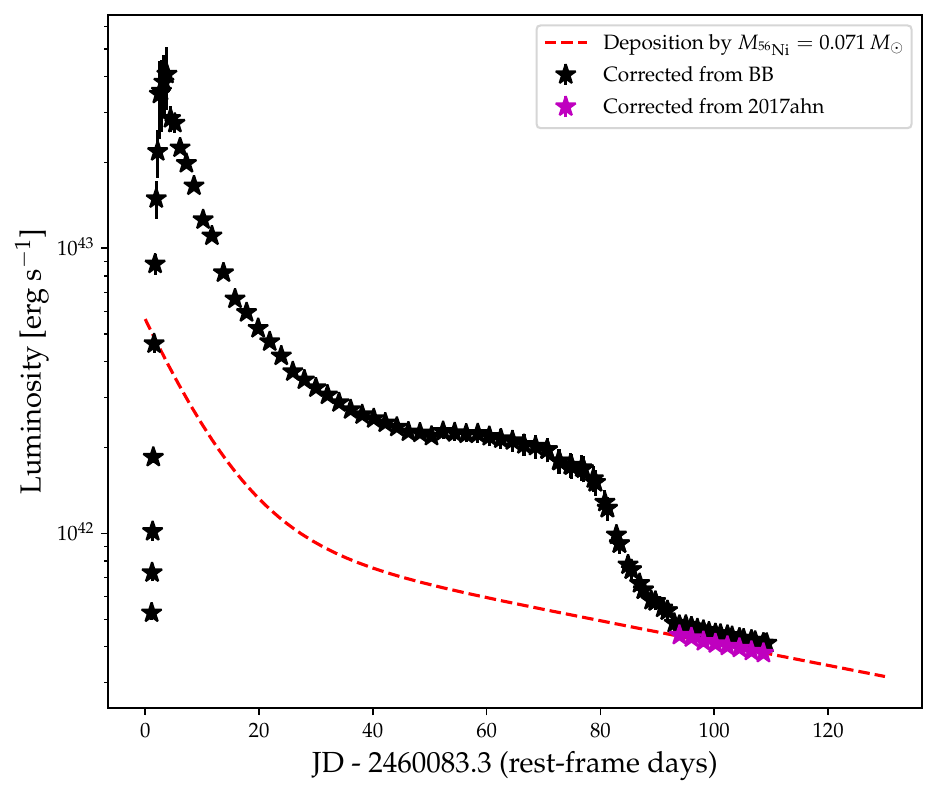}
\vspace*{-2cm}
\end{tabular}
\caption{Bolometric light curve of SN\,2023ixf. The black stars represent the reconstructed bolometric light curves using a blackbody extrapolation. The magenta points are the late-time bolometric luminosity reconstructed using the bolometric correction of SN\,2017ahn to the $uBgriz$ late-time pseudobolometric light curve. The dashed red line indicates the energy deposition from the best-fit $^{56}\rm Ni$ mass (labeled in the legend; see Methods $\S 2$) to the $t>90$\,day luminosity.\label{fig:bolometric_full}}
 \end{SIfigure}

\clearpage

\begin{SIfigure}
 \centering
\begin{tabular}{cc}
\hspace*{-1cm}\includegraphics[width=19cm]{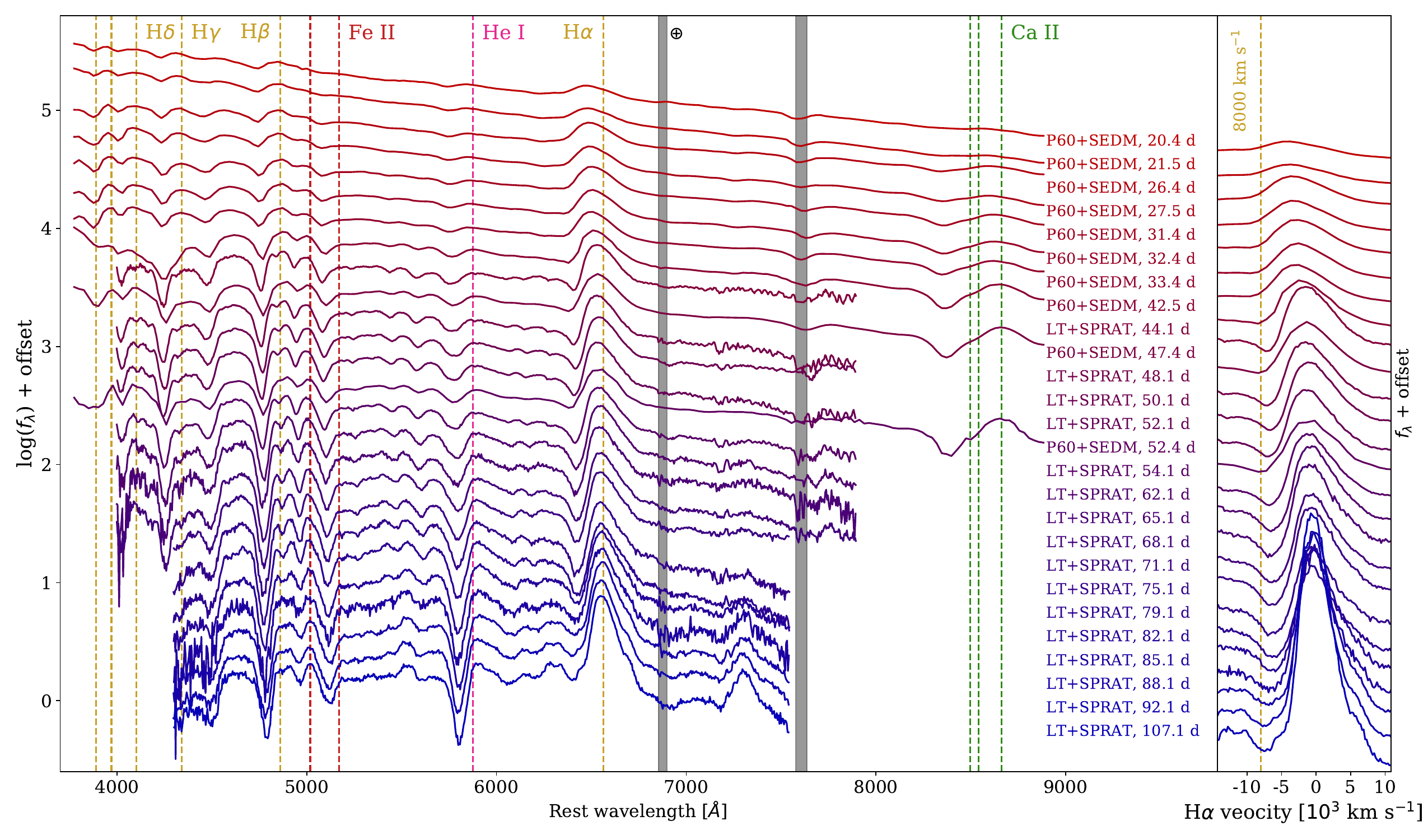}
\vspace*{-3cm}
\end{tabular}
\caption{Left panel: Photospheric phase spectra of SN\,2023ixf. The photospheric development of the SN is typical of Type II-P SNe. Right panel: A zoom-in of the H$\alpha$ P~Cygni profile with $8000 \kms$ H$\alpha$ marked, which we adopt as the early-time ejecta velocity.
  \label{fig:spec_phot}}
 \end{SIfigure}
 
\clearpage

\begin{SIfigure}
 \centering
\vspace*{-3cm}\hspace*{-3cm}\includegraphics[width=21cm]{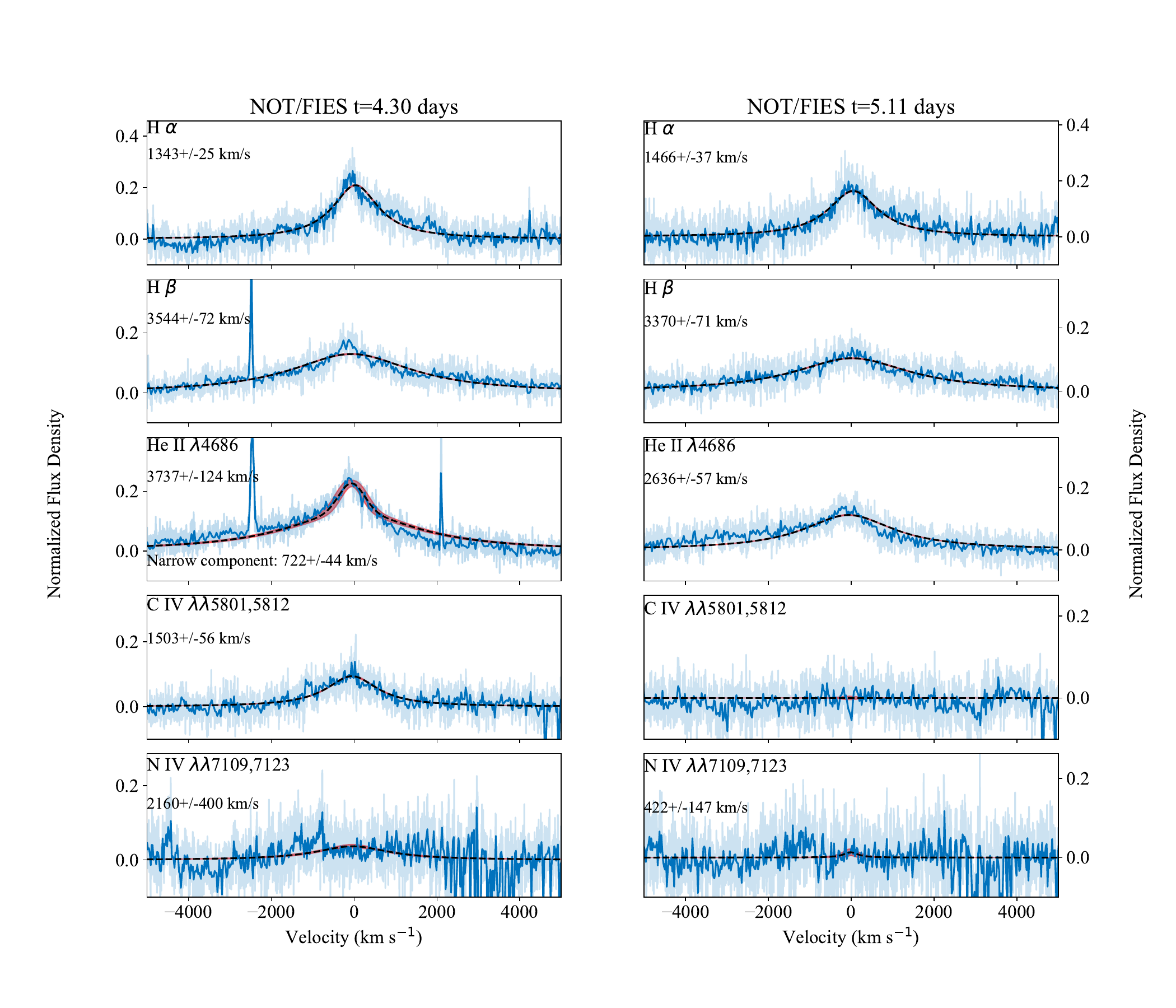}
\vspace*{-2cm}\caption{Narrow flash features from the $t=4.3$ and $5.11\,$ days NOT/FIES spectra. The narrow emission components on top of the broad wings no longer appear in most lines, with the exception of the day $4.3$ \HeII\ profile. Therefore, we only fit a narrow component to the \HeII\ at this epoch.\label{fig:broad_narrow_lines}}
\end{SIfigure}

\begin{SIfigure}
\hspace*{-1.5cm}\includegraphics[width=\textwidth]{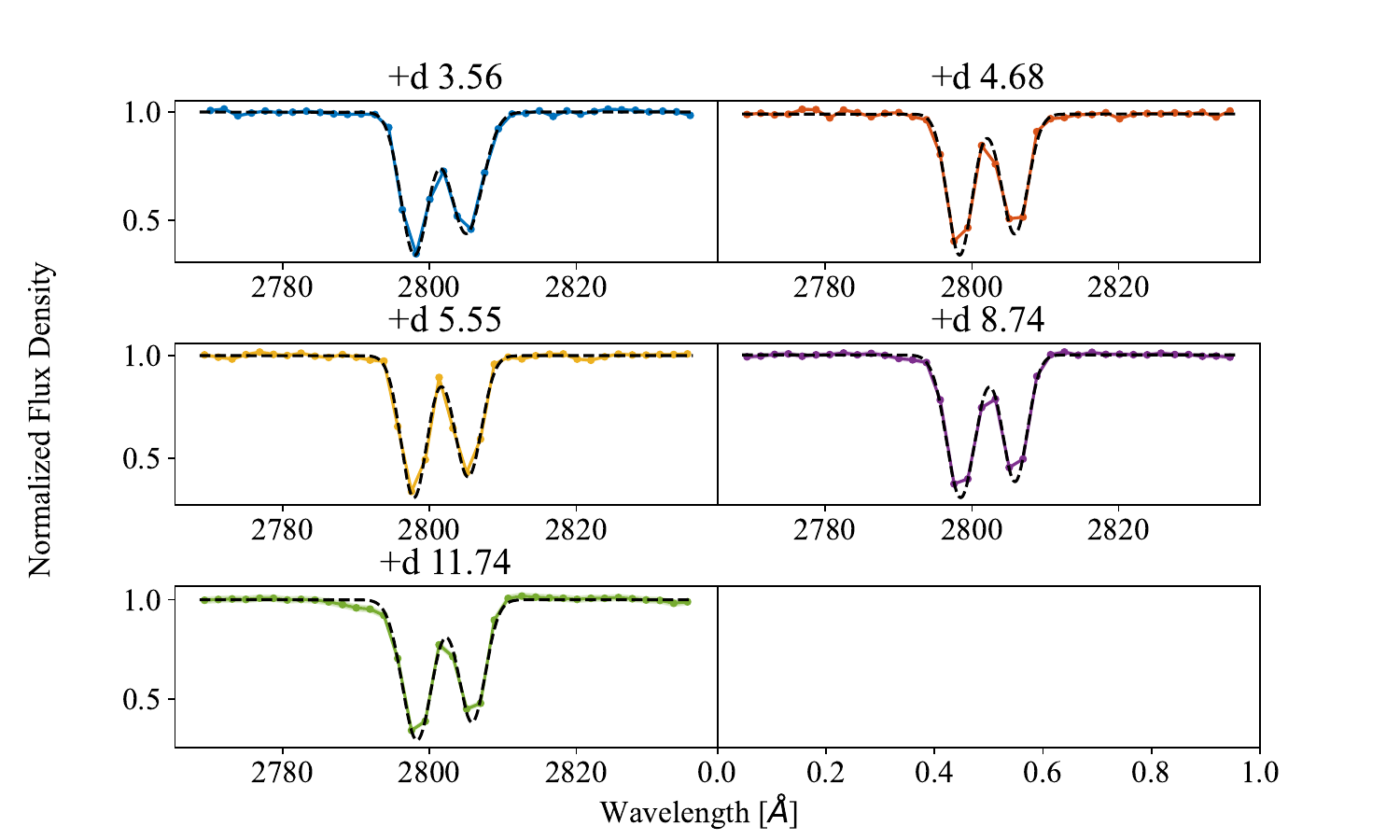}
\vspace*{-2cm}\caption{An example fit of a UV ISM absorption line. Here we fit a two-Gaussian model to the \MgII $\lambda\lambda 2800$ doublet. As the two lines are blended, the fit allows a measurement of the EW of each line. The colour curves are the raw {\it HST} data, while the black dashed line is the best-fit model. No change is observed in the line strength throughout all the UV observations. The flux uncertainties are smaller than the markers used.
\label{fig:absorption_lines}} 
  
\end{SIfigure}
 \clearpage

\begin{SIfigure}
\hspace{-2cm}\includegraphics[width=\textwidth]{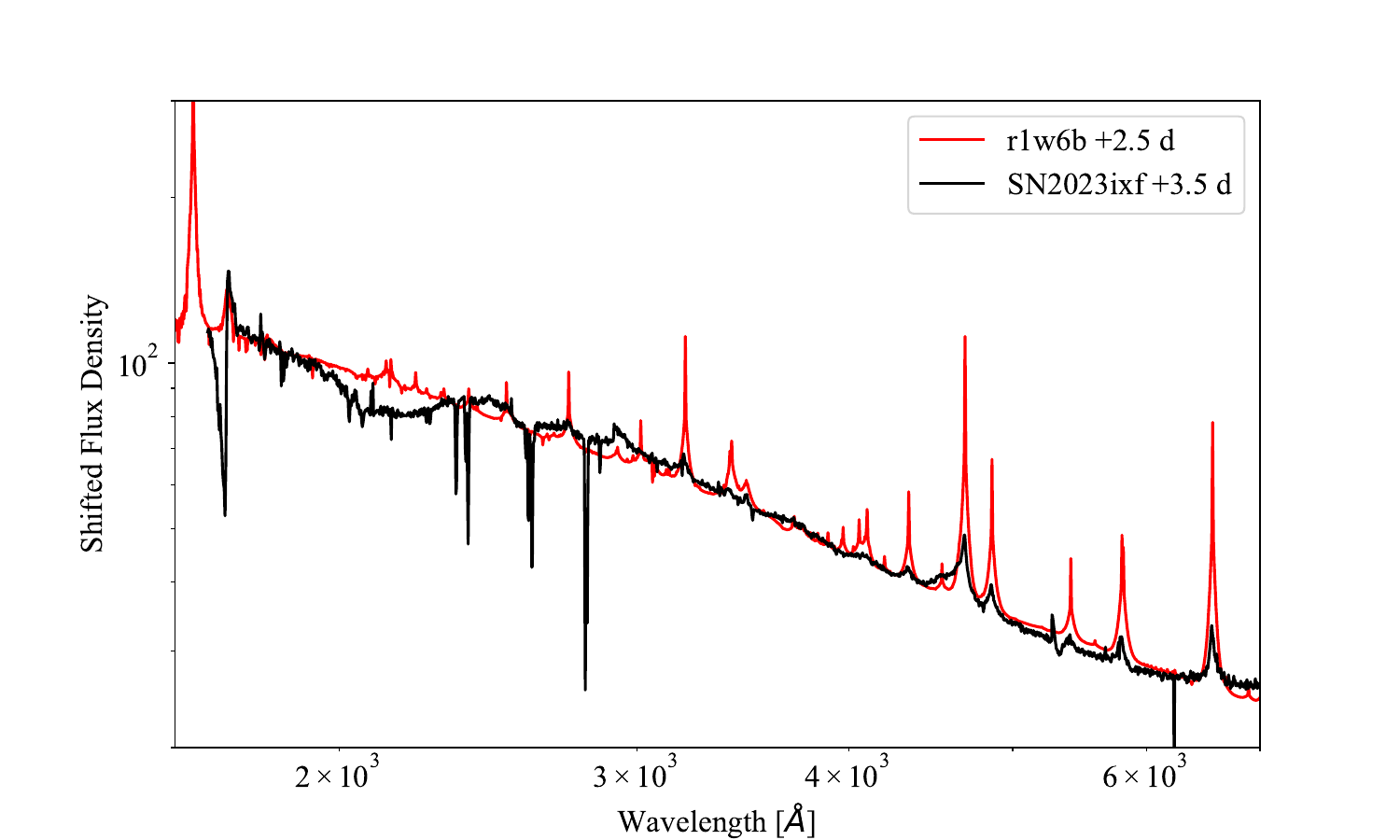}
\caption{A comparison between the r1w6b model (red) shown in ref\cite{Jacobson-Galan2023} to the first-epoch {\it HST} spectrum (black). While the model matches the optical narrow emission lines, only the NUV \NIV\ $\lambda1718$ feature is clearly observed in the {\it HST} spectrum. The slight deviation from a blackbody SED (in the $\sim 2100$\,\AA\ area) is likely due to the guiding issues at this epoch.
\label{fig:NUV_CMFGEN}}
    
\end{SIfigure}
\clearpage

\begin{SIfigure}
 \centering
\begin{tabular}{cc}
\includegraphics[width=\textwidth]{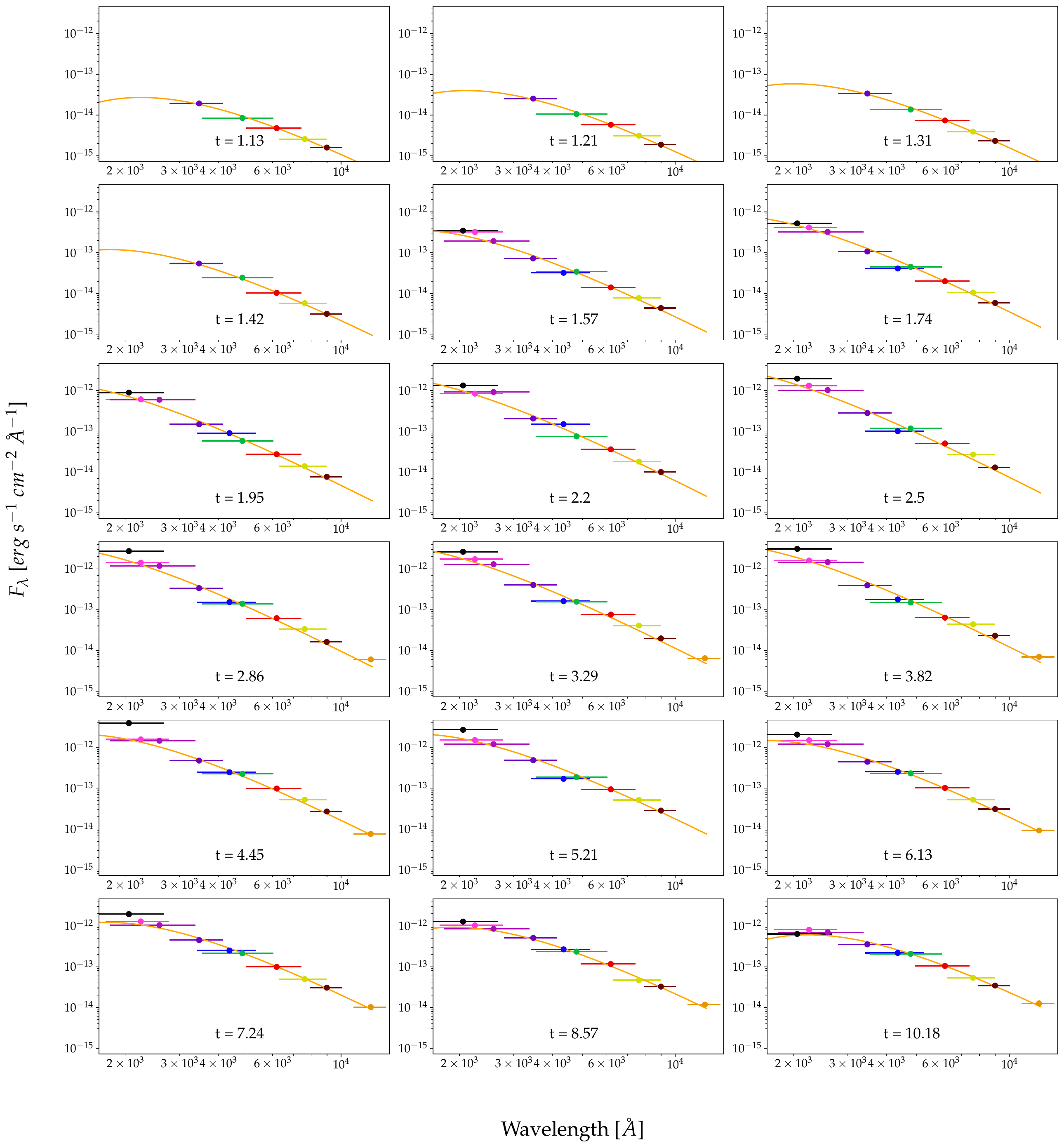}
\vspace*{-2cm}
\end{tabular}
\caption{Observed SEDs and best-fit blackbody. Individual panels show the interpolated SED for each corresponding epoch (text). Horizontal error bars represent the effective width of each filter. Units of time are days.
  \label{fig:BB_SEDs}}
 \end{SIfigure}

\clearpage

\begin{SIfigure}
 \centering
\begin{tabular}{cc}
\hspace*{-1.5cm}\includegraphics[width=15cm]{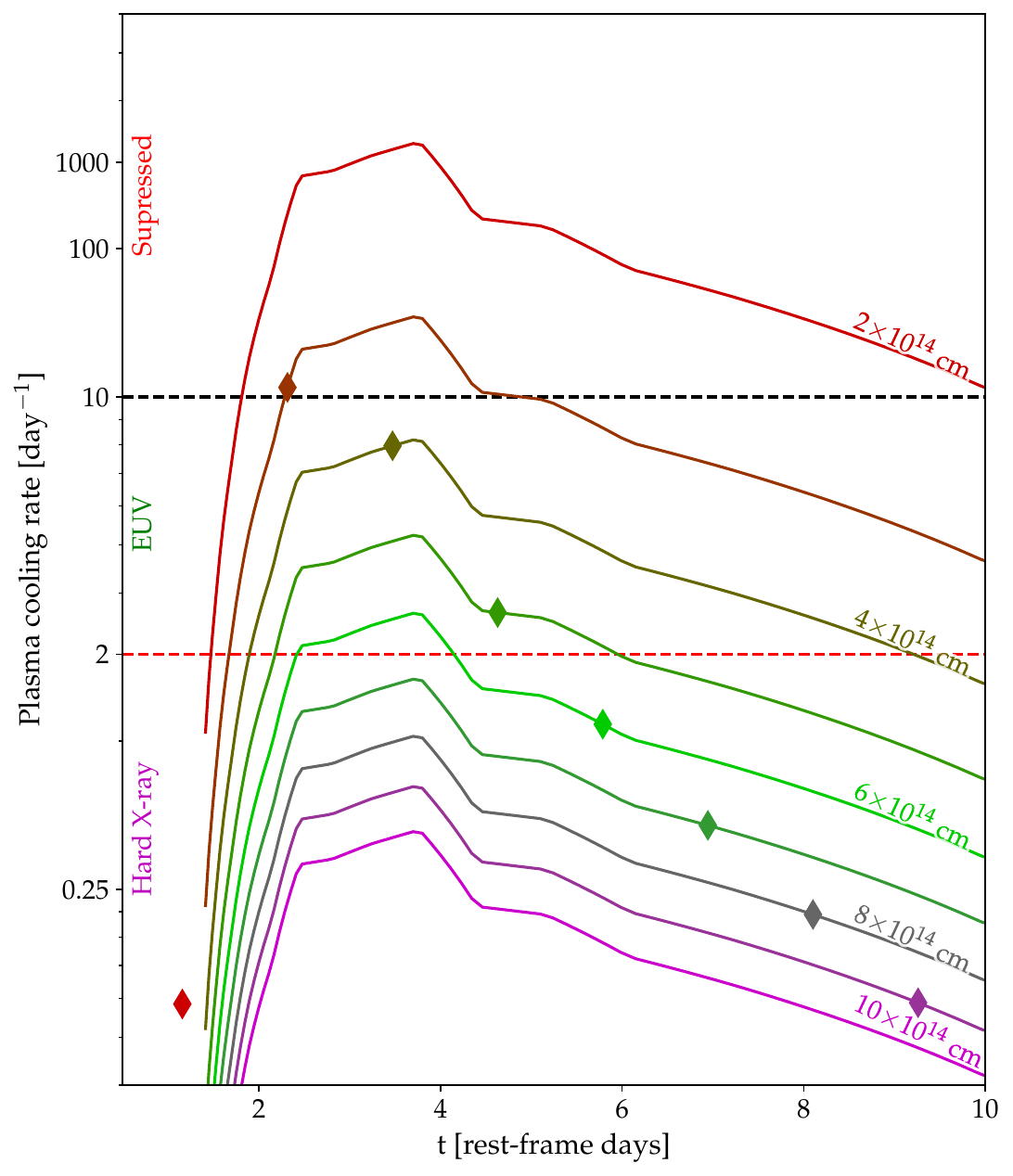}
\vspace*{-2cm}
\end{tabular}
\caption{The cooling rate $d\log_{10}E/dt$ of a shell of hot plasma above the photosphere via Compton scattering, calculated from the observed luminosity of the optical--UV emission component. The solid curves represent the cooling rate per day (i.e., by which factor the electrons will cool in one day) as a function of time. The solid times indicate the time at which the ejecta (moving at $R_{\rm bo}/2 + vt$ for $v=10^4 \kms$) would meet each location and the corresponding cooling rate. The diamonds indicate the time at which a shock moving at a velocity of $10^{9}$\,cm and starting from $R_{\rm bo}/2=10^{14}$\,cm would reach that radius, and its corresponding cooling rate. 
  \label{fig:cooling_rate}}
 \end{SIfigure}

\clearpage

\begin{SIfigure}
\hspace*{-2cm}\includegraphics[width=\textwidth]{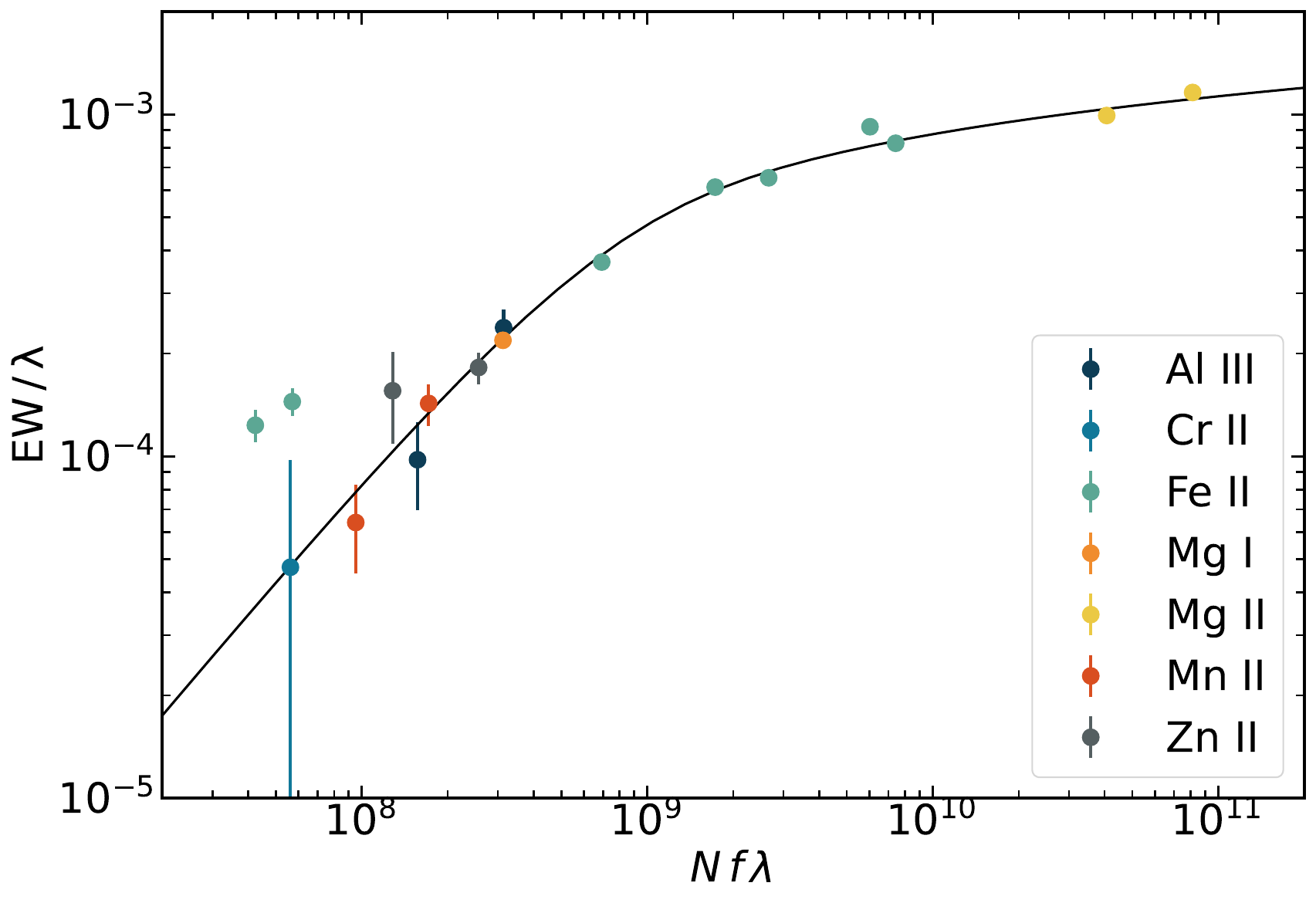}
\caption{Curve of growth of the ISM absorption lines in the {\it HST}/STIS spectrum. The fit to \ion{Fe}{2} lines provides an estimate of the instrumental and intrinsic line-broadening parameter. The abscissa is the product of the column density $N$, the oscillator strength $f$, and the wavelength $\lambda$ of the species. The ordinate is the ratio between the rest-frame EW and the wavelength of the species.
\label{fig:COG}}
\end{SIfigure}

\clearpage

\begin{SIfigure}
\hspace*{-2cm}\includegraphics[width=\textwidth]{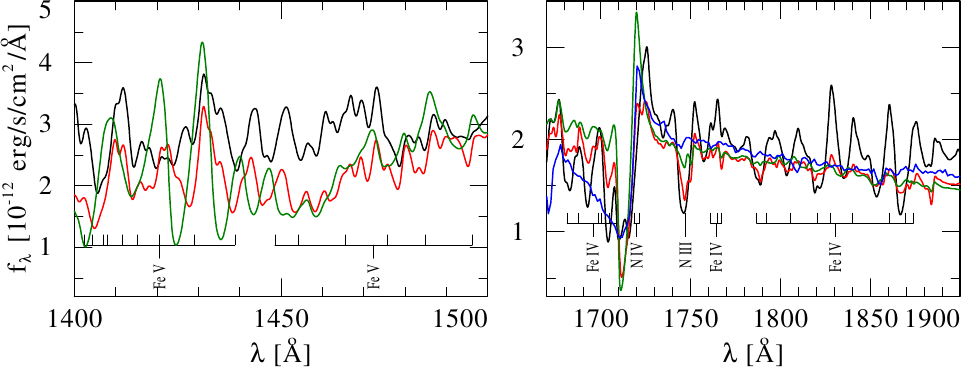}
\caption{{\bf Comparison with non-LTE PoWR models:} Plotted are three PoWR models and the UV {\it HST} observation (third epoch, noisy blue line). The PoWR models have $T_* = 35$, 40, and 45kK (black, red, and green curves, respectively). The PoWR models are scaled to match the observed flux-density levels. The left panel shows a region dominated by Fe\,{\sc v} lines, which are not accessible with the available data. The right panel shows a region dominated by the \NIVscat\ and \NIIIscat\ lines as well as a multitude of Fe\,{\sc iv} lines. The Fe\,{\sc iv} features vanish at $T_* \gtrsim 40\,$kK; the blueward region (shown in the left panel) is thus required to establish the iron abundance from such early-time UV spectra.\label{fig:PoWRTeff}}

\begin{SIfigure}
\hspace*{-2cm}\includegraphics[width=\columnwidth]{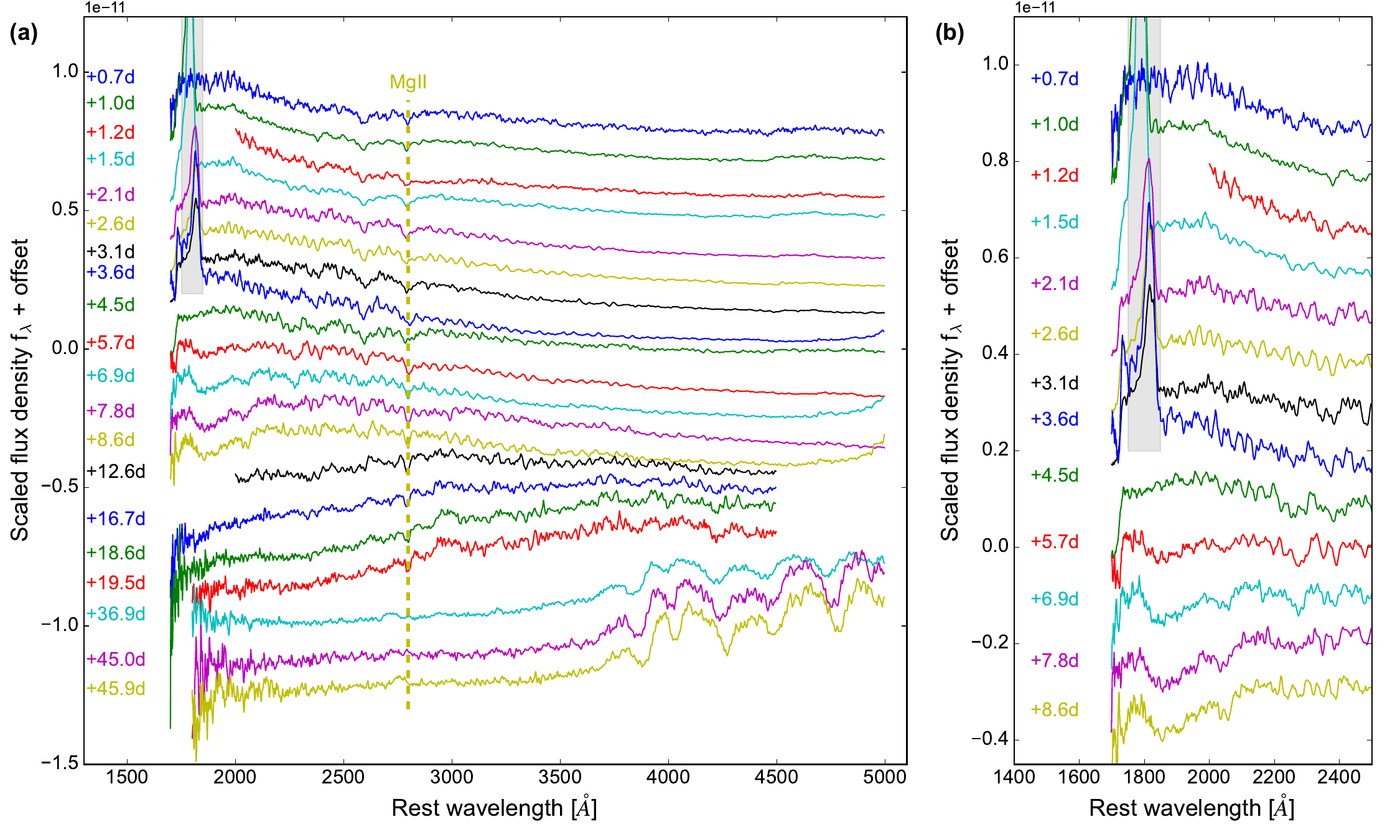}
\vspace*{-3cm}\caption{{\it Swift}/UVOT grism spectra of SN\,2023ixf. (a) The entire sequence of spectra. (b) Zoom-in view of early-phase ($<10$\,days after explosion) spectra at UV wavelengths ($\lambda<2500$\,\AA). The grey region indicates the wavelengths affected by zero-order contamination of a field star. \label{fig:uvot_grism}}
\end{SIfigure}

\clearpage
\end{SIfigure}

\clearpage
\section{Supplementary Tables}

\begin{deluxetable}{ccccc}
\tablecaption{ISM absorption line column density}
\tablewidth{0pt}
\tabletypesize{\footnotesize}
\setlength{\tabcolsep}{0pt}
\tablehead{
\colhead{Species} &
\colhead{Wavelength} &
\colhead{Oscillator} &
\colhead{Equivalent} &
\colhead{Column} \\
\colhead{} &
\colhead{\AA} &
\colhead{strength} &
\colhead{width} &
\colhead{density} \\
\colhead{} &
\colhead{} &
\colhead{}&
\colhead{\AA} &
\colhead{$\log\,N/\rm cm^{-2}$}
}
\setlength{\tabcolsep}{12pt}

\startdata

\ion{Al}{3}& 1854.7 & 0.5590 &$ 0.44\pm 0.06 $&  13.5\\
\ion{Al}{3}& 1862.8	& 0.2780 &$ 0.18\pm 0.05 $&  13.5\\
\ion{Cr}{2} & 2056.3	& 0.1050 &$ 0.10\pm 0.10 $&  13.4\\
\ion{Fe}{2} & 2249.9	& 0.0018 &$ 0.28\pm 0.03 $&  15.0\\
\ion{Fe}{2} & 2260.8	& 0.0024 &$ 0.33\pm 0.03 $&  15.0\\
\ion{Fe}{2} & 2344.2	& 0.1097 &$ 1.53\pm 0.03 $&  15.0\\
\ion{Fe}{2} & 2374.5	& 0.0282 &$ 0.88\pm 0.02 $&  15.0\\
\ion{Fe}{2} & 2382.8	& 0.3006 &$ 1.96\pm 0.02 $&  15.0\\
\ion{Fe}{2} & 2586.7	& 0.0646 &$ 1.59\pm 0.05 $&  15.0\\
\ion{Fe}{2} & 2600.2	& 0.2239 &$ 2.39\pm 0.05 $&  15.0\\
\ion{Mg}{1}	 & 2853.0	& 1.8300 &$ 0.62\pm 0.03 $&  12.8\\
\ion{Mg}{2} & 2796.4	& 0.6123 &$ 3.24\pm 0.04 $&  15.7\\
\ion{Mg}{2} & 2803.5	& 0.3054 &$ 2.78\pm 0.04 $&  15.7\\
\ion{Mn}{2} & 2576.9	& 0.3508 &$ 0.37\pm 0.05 $&  13.3\\
\ion{Mn}{2} & 2606.5	& 0.1927 &$ 0.17\pm 0.05 $&  13.3\\
\ion{Zn}{2} & 2026.1	& 0.5150 &$ 0.37\pm 0.04 $&  13.4\\
\ion{Zn}{2} & 2062.2	& 0.2529 &$ 0.32\pm 0.10 $&  13.4\\
\hline
\enddata
\label{tab:ism_lines}
\tablecomments{The equivalent widths are the weighted averages of all epochs.}
\end{deluxetable}

\clearpage

\begin{deluxetable}{cccccc}
\tablewidth{0pt}
\setlength{\tabcolsep}{2pt}
\tabletypesize{\footnotesize}
\tablecaption{Log of spectra taken for this study}
\tablehead{
\colhead{Telescope/Instrument} & 
\colhead{Observation time (UTC)} &
\colhead{Phase (days)} &
\colhead{Grating} &
\colhead{Slit (arcsec)} &
\colhead{Exposure time (s)}}

\startdata

LT/SPRAT & 2023--05--19 22:23:45 &1.13&Blue&1.8&1200 \\
LT/SPRAT & 2023-05-20 00:52:24&1.24 &Blue&1.8&1200\\
P60/SEDM & 2023-05-20 03:59:33 &   1.37 &IFU &          &1800 \\
Lick/Kast & 2023-05-20 04:23:31 &   1.38 &   300/7500 & 2.0 &      600 \\
Lick/Kast & 2023-05-20 04:23:31 &   1.38 &   600/4310 & 2.0 &      600 \\
Lick/Kast & 2023-05-20 04:42:14 &   1.40 &  600/5000 & 1.0 &     1200 \\
Lick/Kast & 2023-05-20 04:42:14 &   1.40 &  600/4310 & 1.0 &     1200 \\
P60/SEDM & 2023-05-20 07:34:02 &   1.52 &                           IFU &          &     1800 \\
Lick/Kast & 2023-05-20 08:48:29 &   1.57 & 1200/5000 & 1.0 &     3660 \\
Lick/Kast & 2023-05-20 08:48:29 &   1.57 & 600/4310 & 1.0 &     3660 \\
Lick/Kast & 2023-05-20 11:25:26 &   1.68 &  600/5000 & 1.0 &     1200 \\
Lick/Kast & 2023-05-20 11:25:26 &   1.68 &  600/4310 & 1.0 &     1200 \\
Lick/Kast & 2023-05-20 11:42:43 &   1.69 &   300/7500 & 2.0 &      600 \\
Lick/Kast & 2023-05-20 11:42:43 &   1.69 &   600/4310 & 2.0 &      600 \\
HCT/HFOSC & 2023-05-20 21:04:19 &   2.08 &                       Gr7,Gr8 &          &    2*900 \\
LT/SPRAT & 2023-05-20 23:20:53 &   2.17 &                          Blue &        1.8 &      600 \\
NOT/FIES & 2023-05-21 00:46:50 &   2.23 &                            F3 &          &     3600 \\
LT/SPRAT & 2023-05-21 02:55:58 &   2.32 &                          Blue &        1.8 &      600 \\
P60/SEDM & 2023-05-21 04:06:01 &   2.37 &                           IFU &          &      500 \\
P60/SEDM & 2023-05-21 07:33:12 &   2.51 &                           IFU &          &      500 \\
P60/SEDM & 2023-05-21 09:21:01 &   2.59 &                           IFU &          &      500 \\
Keck/LRIS & 2023-05-21 10:01:26 &   2.62 &            400/3400 &          1 &       40 \\
Keck/LRIS & 2023-05-21 10:01:26 &   2.62 &            400/8500 &          1 &       40 \\
LT/SPRAT & 2023-05-21 21:41:41 &   3.10 &                          Blue &        1.8 &      600 \\
 NOT/ALFOSC & 2023-05-21 21:48:38 &   3.11 &                       Grism\#4 &          1 &      600 \\
P60/SEDM & 2023-05-22 04:00:08 &   3.37 &                           IFU &          &      400 \\
P60/SEDM & 2023-05-22 04:14:46 &   3.38 &                           IFU &          &     1800 \\
P60/SEDM & 2023-05-22 06:55:48 &   3.49 &                           IFU &          &      400 \\
P60/SEDM & 2023-05-22 08:29:05 &   3.55 &                           IFU &          &      400 \\
HST/STIS & 2023-05-22 08:42:50 &   3.56 &                             &          &        \\
HCT/HFOSC & 2023-05-22 18:31:41 &   3.97 &                       Gr7,Gr8 &          &  600,700 \\
 NOT/ALFOSC & 2023-05-22 22:33:57 &   4.14 &                       Grism\#4 &          1 &      600 \\
NOT/FIES & 2023-05-23 02:18:54 &   4.30 &                            F3 &          &     3600 \\
Mayall/DESI & 2023-05-23 03:19:22 &   4.34 &                           VPH &          &  173.108 \\
P60/SEDM & 2023-05-23 04:01:34 &   4.37 &                           IFU &          &      400 \\
Keck/Deimos & 2023-05-23 05:33:00 &   4.43 &                         600ZD &          &       30 \\
Keck/Deimos & 2023-05-23 05:33:00 &   4.43 &                         1200G &          &       30 \\
P60/SEDM & 2023-05-23 06:52:47 &   4.49 &                           IFU &          &      400 \\
HST/STIS & 2023-05-23 11:30:56 &   4.68 &                             &          &        \\
 NOT/ALFOSC & 2023-05-23 21:33:06 &   5.10 &                       Grism\#4 &          1 &      600 \\
NOT/FIES & 2023-05-23 21:51:17 &   5.11 &                            F3 &          &     3600 \\
LT/SPRAT & 2023-05-23 22:48:33 &   5.15 &                          Blue &        1.8 &      600 \\
Mayall/DESI & 2023-05-24 03:36:43 &   5.35 &                           VPH &          &  150.479 \\
P60/SEDM & 2023-05-24 04:02:43 &   5.37 &                           IFU &          &      300 \\
HST/STIS & 2023-05-24 08:22:09 &   5.55 &                             &          &        \\
P60/SEDM & 2023-05-24 08:32:05 &   5.56 &                           IFU &          &      300 \\
 IRTF/SpecX & 2023-05-24 11:34:18 &   5.68 &                       ShortXD &     0.3x15 &  717.266 \\
 NOT/ALFOSC & 2023-05-25 02:18:50 &   6.30 &                       Grism\#4 &        1.3 &      540 \\
Mayall/DESI & 2023-05-25 03:18:20 &   6.34 &                           VPH &          &  318.394 \\
P60/SEDM & 2023-05-25 04:03:18 &   6.37 &                           IFU &          &      300 \\
P60/SEDM & 2023-05-25 06:42:58 &   6.48 &                           IFU &          &      300 \\
Mayall/DESI & 2023-05-26 03:21:59 &   7.34 &                           VPH &          &  384.802 \\
NOT/FIES & 2023-05-26 22:24:27 &   8.13 &                            F3 &          &     3600 \\
HST/STIS & 2023-05-27 12:59:03 &   8.74 &                             &          &        \\
Mayall/DESI & 2023-05-28 03:20:37 &   9.34 &                           VPH &          &  464.162 \\
Lick/Kast & 2023-05-28 04:30:53 &      &   300/7500 & 2.0 &      200 \\
Lick/Kast & 2023-05-28 04:30:53 &      &   600/4310 & 2.0 &      200 \\
Lick/Kast & 2023-05-28 04:37:38 &      &  600/5000 & 1.0 &      500 \\
Lick/Kast & 2023-05-28 04:37:38 &      &  600/4310 & 1.0 &      500 \\
Lick/Kast & 2023-05-28 10:23:30 &      &   300/7500 & 2.0 &      300 \\
Lick/Kast & 2023-05-28 10:23:30 &      &   600/4310 & 2.0 &      300 \\
Lick/Kast & 2023-05-28 10:30:37 &      &  600/5000 & 1.0 &     1200 \\
Lick/Kast & 2023-05-28 10:30:37 &      &  600/4310 & 1.0 &     1200 \\
NOT/ALFOSC & 2023-05-28 22:53:38 &  10.15 &                       Grism\#4 &          1 &      540 \\
NOT/FIES & 2023-05-28 23:14:24 &  10.17 &                            F3 &          &      810 \\
Mayall/DESI & 2023-05-29 03:17:48 &  10.34 &                           VPH &          &  602.708 \\
P60/SEDM & 2023-05-29 04:06:33 &  10.37 &                           IFU &          &     1800 \\
TNG/HARPS-N & 2023-05-29 21:51:09 &  11.11 &                       Echelle &          &     3000 \\
 NOT/ALFOSC & 2023-05-29 22:03:32 &  11.12 &                       Grism\#4 &          1 &      540 \\
NOT/FIES & 2023-05-29 22:34:25 &  11.14 &                            F3 &          &     3600 \\
LT/SPRAT & 2023-05-30 01:11:14 &  11.25 &                          Blue &        1.8 &      600 \\
LT/SPRAT & 2023-05-30 01:11:14 &  11.25 &                          Blue &        1.8 &      600 \\
Mayall/DESI & 2023-05-30 03:18:47 &  11.34 &                           VPH &          &  604.607 \\
HST/STIS & 2023-05-30 12:59:03 &  11.74 &                             &          &        \\
LT/SPRAT & 2023-05-30 22:38:52 &  12.14 &                          Blue &        1.8 &      600 \\
LT/SPRAT & 2023-05-30 22:38:52 &  12.14 &                          Blue &        1.8 &      600 \\
Mayall/DESI & 2023-05-31 03:20:31 &  12.34 &                           VPH &          &  603.903 \\
Mayall/DESI & 2023-06-01 03:20:41 &  13.34 &                           VPH &          &  602.202 \\
LT/SPRAT & 2023-06-01 22:19:25 &  14.13 &                          Blue &        1.8 &      600 \\
Mayall/DESI & 2023-06-02 03:20:51 &  14.34 &                           VPH &          &  609.524 \\
P60/SEDM & 2023-06-02 04:34:05 &  14.39 &                           IFU &          &      300 \\
LT/SPRAT & 2023-06-02 21:26:57 &  15.09 &                          Blue &        1.8 &      600 \\
NOT/FIES & 2023-06-03 01:09:18 &  15.25 &                            F3 &          &     3600 \\
Mayall/DESI & 2023-06-03 03:21:36 &  15.34 &                           VPH &          &  523.362 \\
P60/SEDM & 2023-06-03 04:09:23 &  15.37 &                           IFU &          &      300 \\
Mayall/DESI & 2023-06-04 03:21:29 &  16.34 &                           VPH &          &  603.402 \\
P60/SEDM & 2023-06-04 04:10:03 &  16.37 &                           IFU &          &      300 \\
P60/SEDM & 2023-06-05 04:10:48 &  17.37 &                           IFU &          &      300 \\
P60/SEDM & 2023-06-06 04:11:20 &  18.37 &                           IFU &          &      300 \\
LT/SPRAT & 2023-06-06 21:20:07 &  19.09 &                          Blue &        1.8 &      600 \\
P60/SEDM & 2023-06-07 07:05:25 &  19.50 &                           IFU &          &      300 \\
P60/SEDM & 2023-06-08 04:12:43 &  20.38 &                           IFU &          &      300 \\
TNG/HARPS-N & 2023-06-08 22:53:28 &  21.15 &                       Echelle &          &     3000 \\
 INT/IDS & 2023-06-08 23:59:28 &  21.20 &                         R300V &          1 &      120 \\
P60/SEDM & 2023-06-09 06:21:00 &  21.46 &                           IFU &          &      300 \\
P60/SEDM & 2023-06-14 04:16:14 &  26.38 &                           IFU &          &      400 \\
P60/SEDM & 2023-06-15 06:52:28 &  27.49 &                           IFU &          &      400 \\
P60/SEDM & 2023-06-19 04:18:30 &  31.38 &                           IFU &          &      400 \\
P60/SEDM & 2023-06-20 04:18:46 &  32.38 &                           IFU &          &      400 \\
P60/SEDM & 2023-06-21 04:18:50 &  33.38 &                           IFU &          &      400 \\
P60/SEDM & 2023-06-30 07:12:32 &  42.50 &                           IFU &          &     1800 \\
TNG/HARPS-N & 2023-06-30 21:16:27 &  43.09 &                       Echelle &          &     3600 \\
LT/SPRAT & 2023-07-01 22:28:19 &  44.14 &                          Blue &        1.8 &      800 \\
P60/SEDM & 2023-07-05 04:51:09 &  47.40 &                           IFU &          &      450 \\
LT/SPRAT & 2023-07-05 21:28:48 &  48.09 &                          Blue &        1.8 &      800 \\
LT/SPRAT & 2023-07-07 21:23:38 &  50.09 &                          Blue &        1.8 &      800 \\
LT/SPRAT & 2023-07-09 21:31:21 &  52.10 &                          Blue &        1.8 &      800 \\
P60/SEDM & 2023-07-10 04:18:10 &  52.38 &                           IFU &          &     1800 \\
LT/SPRAT & 2023-07-11 21:22:43 &  54.09 &                          Blue &        1.8 &      800 \\
TNG/HARPS-N & 2023-07-13 20:58:44 &  56.07 &                       Echelle &          &     5400 \\
LT/SPRAT & 2023-07-19 22:28:10 &  62.14 &                          Blue &        1.8 &      100 \\
LT/SPRAT & 2023-07-22 22:16:30 &  65.13 &                          Blue &        1.8 &      100 \\
LT/SPRAT & 2023-07-25 21:59:13 &  68.12 &                          Blue &        1.8 &      100 \\
LT/SPRAT & 2023-07-28 21:41:32 &  71.10 &                          Blue &        1.8 &      100 \\
LT/SPRAT & 2023-08-01 22:36:12 &  75.14 &                          Blue &        1.8 &      100 \\
LT/SPRAT & 2023-08-05 22:34:24 &  79.14 &                          Blue &        1.8 &      100 \\
LT/SPRAT & 2023-08-08 22:43:28 &  82.15 &                          Blue &        1.8 &      100 \\
TNG/HARPS-N & 2023-08-11 21:41:08 &  85.10 &                       Echelle &          &     5400 \\
LT/SPRAT & 2023-08-11 22:30:02 &  85.14 &                          Blue &        1.8 &      100 \\
LT/SPRAT & 2023-08-14 21:08:54 &  88.08 &                          Blue &        1.8 &      360 \\
LT/SPRAT & 2023-08-18 20:58:55 &  92.07 &                          Blue &        1.8 &      360 \\
LT/SPRAT & 2023-09-02 20:42:54 & 107.06 &                          Blue &        1.8 &      360 \\

\hline
\enddata

\label{tab:spectra}
\end{deluxetable}

\clearpage
\vspace{-0.5cm}\begin{deluxetable}{ccccccc}
\tablecaption{HST observation log}
\tabletypesize{\footnotesize}
\tablewidth{0pt}
\tablehead{visit & Exposure & Observation start time & Exposure Time (s) & Grating & Issues \\}
\startdata
    1 &        1 & 2023-05-22T05:02:43.083 &              &        G230LB &  Guide-star acquisition failed\\
    1 &        2 & 2023-05-22T05:14:41.097 &              &        G230LB &  Guide-star acquisition failed\\
    1 &        3 & 2023-05-22T06:30:02.090 &           648 &        G230LB &  \\
    1 &        4 & 2023-05-22T06:42:00.073 &           769 &        G230LB &  \\
    1 &        5 & 2023-05-22T06:56:51.083 &           769 &        G230LB &  \\
    1 &        6 & 2023-05-22T08:05:05.097 &           769 &        G230LB &  \\
    1 &        7 & 2023-05-22T08:19:04.073 &           769 &        G230LB &  \\
    1 &        8 & 2023-05-22T08:33:55.083 &           769 &        G230LB &  \\
    1 &        9 & 2023-05-22T09:40:08.073 &           769 &        G230LB &  \\
    1 &       10 & 2023-05-22T09:54:07.083 &           768 &        G230LB &  \\
    1 &       11 & 2023-05-22T10:08:57.097 &           768 &        G230LB &  \\
    1 &       12 & 2023-05-22T11:15:11.083 &           768 &        G230LB &  \\
    1 &       14 & 2023-05-22T11:36:22.090 &           492 &         G750L &  \\
    1 &       15 & 2023-05-22T11:45:44.073 &           492 &         G750L &  \\
    1 &       16 & 2023-05-22T12:50:14.090 &           492 &         G430L &  \\
    1 &       17 & 2023-05-22T12:59:36.073 &           574 &         G430L &  \\
    2 &        1 & 2023-05-23T04:48:30.090 &           648 &        G230LB &  \\
    2 &        2 & 2023-05-23T05:00:28.073 &           648 &        G230LB &  \\
    2 &        3 & 2023-05-23T06:15:49.097 &           648 &        G230LB &  \\
    2 &        4 & 2023-05-23T06:27:47.083 &           769 &        G230LB &  \\
    2 &        5 & 2023-05-23T06:42:38.090 &           769 &        G230LB &  \\
    2 &        6 & 2023-05-23T07:50:52.073 &           769 &        G230LB &  \\
    2 &        7 & 2023-05-23T08:04:51.083 &           769 &        G230LB &  \\
    2 &        8 & 2023-05-23T08:19:42.090 &           769 &        G230LB &  \\
    2 &        9 & 2023-05-23T09:25:55.083 &           769 &        G230LB &  \\
    2 &       10 & 2023-05-23T09:39:54.090 &           768 &        G230LB &  \\
    2 &       11 & 2023-05-23T09:54:44.073 &           768 &        G230LB &  \\
    2 &       12 & 2023-05-23T11:00:58.090 &           768 &        G230LB &  \\
    2 &       14 & 2023-05-23T11:22:09.097 &           492 &         G750L &  \\
    2 &       15 & 2023-05-23T11:31:31.083 &           492 &         G750L &  \\
    2 &       16 & 2023-05-23T12:36:01.097 &           492 &         G430L &  \\
    2 &       17 & 2023-05-23T12:45:23.083 &           574 &         G430L &  \\
    3 &        1 & 2023-05-24T04:34:15.083 &           648 &        G230LB &  \\
    3 &        2 & 2023-05-24T04:46:13.097 &           648 &        G230LB &  \\
    3 &        3 & 2023-05-24T06:01:34.090 &           648 &        G230LB &  \\
    3 &        4 & 2023-05-24T06:13:32.073 &           769 &        G230LB &  \\
    3 &        5 & 2023-05-24T07:36:37.097 &           769 &        G230LB &  \\
    3 &        6 & 2023-05-24T07:50:36.073 &           769 &        G230LB &  \\
    3 &        7 & 2023-05-24T08:04:35.083 &           769 &        G230LB &  \\
    3 &        8 & 2023-05-24T09:11:40.073 &            &        G230LB &  Failed due to guiding issues\\
    3 &        9 & 2023-05-24T09:25:39.083 &            &        G230LB &  Failed due to guiding issues\\
    3 &       10 & 2023-05-24T09:39:38.090 &            &        G230LB &  Failed due to guiding issues\\
    3 &       11 & 2023-05-24T10:46:43.083 &            &        G230LB &  Failed due to guiding issues\\
    3 &       12 & 2023-05-24T11:05:31.083 &            &        G230LB &  Failed due to guiding issues\\
    3 &       14 & 2023-05-24T11:26:42.090 &            &         G750L &  Failed due to guiding issues\\
    3 &       15 & 2023-05-24T12:21:45.097 &           492&         G750L &  \\
    3 &       16 & 2023-05-24T12:34:06.090 &           492&         G430L &  \\
    3 &       17 & 2023-05-24T12:43:28.073 &           574&         G430L &  \\
    4 &        1 & 2023-05-27T10:11:33.097 &           648&        G230LB &  \\
    4 &        2 & 2023-05-27T10:23:31.083 &           648&        G230LB &  \\
    4 &        3 & 2023-05-27T10:35:29.097 &           648&        G230LB &  \\
    4 &        4 & 2023-05-27T11:38:51.083 &           769&        G230LB &  \\
    4 &        5 & 2023-05-27T11:53:42.090 &           769&        G230LB &  \\
    4 &        6 & 2023-05-27T12:07:41.097 &           769&        G230LB &  \\
    4 &        7 & 2023-05-27T13:13:53.097 &           768&        G230LB &  \\
    4 &        8 & 2023-05-27T13:28:43.083 &           768&        G230LB &  \\
    4 &        9 & 2023-05-27T13:42:41.097 &           768&        G230LB &  \\
    4 &       11 & 2023-05-27T14:48:56.073 &           512&         G750L &  \\
    4 &       12 & 2023-05-27T14:58:38.090 &           512&         G750L &  \\
    4 &       13 & 2023-05-27T15:11:19.083 &           512&         G430L &  \\
    4 &       14 & 2023-05-27T15:21:01.097 &           512&         G430L &  \\
    5 &        1 & 2023-05-30T07:53:12.073 &           648&        G230LB &  \\
    5 &        2 & 2023-05-30T08:05:10.090 &           648&        G230LB &  \\
    5 &        3 & 2023-05-30T08:17:08.073 &           648&        G230LB &  \\
    5 &        4 & 2023-05-30T09:20:31.083 &           768&        G230LB &  \\
    5 &        5 & 2023-05-30T09:35:21.097 &           768&        G230LB &  \\
    5 &        6 & 2023-05-30T09:49:19.083 &           768&        G230LB &  \\
    5 &        8 & 2023-05-30T10:55:33.097 &           492&         G750L &  \\
    5 &        9 & 2023-05-30T11:04:55.083 &           492&         G750L &  \\
    5 &       10 & 2023-05-30T11:17:16.073 &           492&         G430L &  \\
    5 &       11 & 2023-05-30T11:26:38.090 &           574&         G430L &  \\
\enddata
\label{tab:HST_visits}
\end{deluxetable}

\clearpage

\begin{deluxetable}{ccccccc}
\tablecaption{Narrow-Line Velocities}
\tablewidth{0pt}
\tabletypesize{\footnotesize}
\tablehead{phase & instrument & line & velocity & broad wings velocity & line centre & reduced $\chi^{2}$ \\
}
\startdata

2.23 & NOT/FIES & H$\alpha$ & 91$\pm$2 & $2031\pm27$ & $-3.04$ & $0.96$ \\
2.23 & NOT/FIES & H$\beta$ & 121$\pm$5 & 4034$\pm$66 & -21.35 & 0.67 \\
2.23 & NOT/FIES & \HeII\ $\lambda 4686$ & 107$\pm$3 & 1420$\pm$44 & -34.79 & 1.88 \\
2.23 & NOT/FIES & \CIV\ $\lambda\lambda5801$ & 156$\pm$15 & 1751$\pm$111 & -43.33 & 0.59 \\
2.23 & NOT/FIES & \NIV\ $\lambda\lambda7109$ & 123$\pm$18 & 1326$\pm$105 & -42.16 & 0.75 \\
4.3 & NOT/FIES & H$\alpha$ & 0 & 1343$\pm$25 & 33.68 & 1.04 \\
4.3 & NOT/FIES & H$\beta$ & 0 & 3544$\pm$72 & -61.66 & 0.94 \\
4.3 & NOT/FIES & \HeII\ $\lambda 4686$ & $722\pm44$ & 3737$\pm$124 & -63.97 & 2.09 \\
4.3 & NOT/FIES & \CIV\ $\lambda\lambda5801$ & 0 & 1503$\pm$56 & -51.67 & 1.05 \\
4.3 & NOT/FIES & \NIV\ $\lambda\lambda7109$ & 0 & 2160$\pm$400 & -42.16 & 0.79 \\
5.11 & NOT/FIES & H$\alpha$ & 0 & 1466$\pm$37 & 45.67 & 1.02 \\
5.11 & NOT/FIES & H$\beta$ & 0 & 3370$\pm$71 & 17.04 & 0.91 \\
5.11 & NOT/FIES & \HeII\ $\lambda 4686$ & 0 & 2636$\pm$57 & -63.97 & 1.13 \\
8.13 & NOT/FIES & H$\alpha$ & 2095$\pm$341 & 0 & 13.31 & 1.38 \\
10.17 & NOT/FIES & H$\alpha$ & 3000$\pm$593 & 0 & 49.87 & 1.04 \\
11.14 & NOT/FIES & H$\alpha$ & 2032$\pm$200 & 0 & 64.69 & 1.35 \\
15.25 & NOT/FIES & H$\alpha$ & 2257$\pm$335 & 0 & -48.84 & 1.6
\enddata
\label{tab:velocity}
\end{deluxetable}

\begin{deluxetable}{ccccc}
\tablecaption{Narrow Line Fluxes}
\tabletypesize{\footnotesize}
\tablewidth{0pt}
\tablehead{Line & Phase & Instrument & Flux & Flux Uncertainty \\
 & [days] &  & [$\rm erg\,s^{-1}\,cm^{-2}\,s^{-1}$] & [$\rm erg\,s^{-1}\,cm^{-2}\,s^{-1}$] \\}
 \setlength{\tabcolsep}{12pt}
\startdata
\HeII~$\lambda$ 4686 &  1.13 &    LT/SPRAT & $3.57 \times 10^{-13}$ & $1.01 \times 10^{-14}$ \\
\HeII~$\lambda$ 4686 &  1.24 &    LT/SPRAT & $5.00 \times 10^{-13}$ & $1.38 \times 10^{-14}$ \\
\HeII~$\lambda$ 4686 &  1.38 &   Lick/Kast & $7.20 \times 10^{-13}$ & $1.36 \times 10^{-14}$ \\
\HeII~$\lambda$ 4686 &  1.40 &   Lick/Kast & $6.86 \times 10^{-13}$ & $1.63 \times 10^{-14}$ \\
\HeII~$\lambda$ 4686 &  1.57 &   Lick/Kast & $1.02 \times 10^{-12}$ & $1.53 \times 10^{-14}$ \\
\HeII~$\lambda$ 4686 &  1.68 &   Lick/Kast & $1.30 \times 10^{-12}$ & $2.33 \times 10^{-14}$ \\
\HeII~$\lambda$ 4686 &  1.69 &   Lick/Kast & $1.32 \times 10^{-12}$ & $1.93 \times 10^{-14}$ \\
\HeII~$\lambda$ 4686 &  2.17 &    LT/SPRAT & $2.29 \times 10^{-12}$ & $3.70 \times 10^{-14}$ \\
\HeII~$\lambda$ 4686 &  2.32 &    LT/SPRAT & $2.66 \times 10^{-12}$ & $4.88 \times 10^{-14}$ \\
\HeII~$\lambda$ 4686 &  2.62 &   Keck/LRIS & $3.17 \times 10^{-12}$ & $5.15 \times 10^{-14}$ \\
\HeII~$\lambda$ 4686 &  3.10 &    LT/SPRAT & $3.63 \times 10^{-12}$ & $8.14 \times 10^{-14}$ \\
\HeII~$\lambda$ 4686 &  4.34 & Mayall/DESI & $4.66 \times 10^{-12}$ & $2.15 \times 10^{-13}$ \\
\HeII~$\lambda$ 4686 &  5.15 &    LT/SPRAT & $3.99 \times 10^{-12}$ & $1.26 \times 10^{-13}$ \\
\HeII~$\lambda$ 4686 &  5.35 & Mayall/DESI & $3.45 \times 10^{-12}$ & $4.57 \times 10^{-13}$ \\
\HeII~$\lambda$ 4686 &  6.34 & Mayall/DESI & $1.53 \times 10^{-12}$ & $2.53 \times 10^{-13}$ \\
\HeII~$\lambda$ 4686 &  7.34 & Mayall/DESI & $7.72 \times 10^{-13}$ & $2.27 \times 10^{-13}$ \\
H$\alpha$ &  1.13 &    LT/SPRAT & $2.50 \times 10^{-13}$ & $4.66 \times 10^{-15}$ \\
H$\alpha$ &  1.24 &    LT/SPRAT & $2.98 \times 10^{-13}$ & $4.44 \times 10^{-15}$ \\
H$\alpha$ &  1.38 &   Lick/Kast & $3.97 \times 10^{-13}$ & $1.98 \times 10^{-15}$ \\
H$\alpha$ &  1.40 &   Lick/Kast & $3.90 \times 10^{-13}$ & $1.75 \times 10^{-15}$ \\
H$\alpha$ &  1.57 &   Lick/Kast & $4.86 \times 10^{-13}$ & $7.32 \times 10^{-15}$ \\
H$\alpha$ &  1.68 &   Lick/Kast & $4.89 \times 10^{-13}$ & $4.76 \times 10^{-15}$ \\
H$\alpha$ &  1.69 &   Lick/Kast & $5.45 \times 10^{-13}$ & $4.57 \times 10^{-15}$ \\
H$\alpha$ &  2.17 &    LT/SPRAT & $7.95 \times 10^{-13}$ & $2.45 \times 10^{-14}$ \\
H$\alpha$ &  2.32 &    LT/SPRAT & $8.19 \times 10^{-13}$ & $3.37 \times 10^{-14}$ \\
H$\alpha$ &  2.62 &   Keck/LRIS & $9.53 \times 10^{-13}$ & $2.90 \times 10^{-14}$ \\
H$\alpha$ &  3.10 &    LT/SPRAT & $9.86 \times 10^{-13}$ & $4.52 \times 10^{-14}$ \\
H$\alpha$ &  4.34 & Mayall/DESI & $9.75 \times 10^{-13}$ & $3.85 \times 10^{-14}$ \\
H$\alpha$ &  5.15 &    LT/SPRAT & $5.27 \times 10^{-13}$ & $7.27 \times 10^{-14}$ \\
H$\alpha$ &  5.35 & Mayall/DESI & $4.85 \times 10^{-13}$ & $1.53 \times 10^{-13}$ \\
H$\alpha$ &  6.34 & Mayall/DESI & $2.92 \times 10^{-13}$ & $8.41 \times 10^{-14}$ \\
\NIV~$\lambda \lambda 7109,7123$ &  1.13 &    LT/SPRAT & $3.97 \times 10^{-14}$ & $8.50 \times 10^{-15}$ \\
\NIV~$\lambda \lambda 7109,7123$ &  1.24 &    LT/SPRAT & $5.67 \times 10^{-14}$ & $8.28 \times 10^{-15}$ \\
\NIV~$\lambda \lambda 7109,7123$ &  1.38 &   Lick/Kast & $1.03 \times 10^{-13}$ & $4.10 \times 10^{-15}$ \\
\NIV~$\lambda \lambda 7109,7123$ &  1.40 &   Lick/Kast & $1.09 \times 10^{-13}$ & $8.68 \times 10^{-15}$ \\
\NIV~$\lambda \lambda 7109,7123$ &  1.68 &   Lick/Kast & $1.25 \times 10^{-13}$ & $1.21 \times 10^{-14}$ \\
\NIV~$\lambda \lambda 7109,7123$ &  1.69 &   Lick/Kast & $1.26 \times 10^{-13}$ & $2.78 \times 10^{-15}$ \\
\NIV~$\lambda \lambda 7109,7123$ &  2.17 &    LT/SPRAT & $1.87 \times 10^{-13}$ & $3.85 \times 10^{-14}$ \\
\NIV~$\lambda \lambda 7109,7123$ &  2.32 &    LT/SPRAT & $2.16 \times 10^{-13}$ & $2.55 \times 10^{-14}$ \\
\NIV~$\lambda \lambda 7109,7123$ &  2.62 &   Keck/LRIS & $2.01 \times 10^{-13}$ & $3.20 \times 10^{-14}$ \\
\CIV~$\lambda \lambda 5801,5812$ &  1.13 &    LT/SPRAT & $6.45 \times 10^{-14}$ & $4.12 \times 10^{-15}$ \\
\CIV~$\lambda \lambda 5801,5812$ &  1.24 &    LT/SPRAT & $8.78 \times 10^{-14}$ & $3.77 \times 10^{-15}$ \\
\CIV~$\lambda \lambda 5801,5812$ &  1.38 &   Lick/Kast & $1.10 \times 10^{-13}$ & $5.19 \times 10^{-15}$ \\
\CIV~$\lambda \lambda 5801,5812$ &  1.40 &   Lick/Kast & $1.12 \times 10^{-13}$ & $1.00 \times 10^{-14}$ \\
\CIV~$\lambda \lambda 5801,5812$ &  1.57 &   Lick/Kast & $6.96 \times 10^{-14}$ & $1.11 \times 10^{-14}$ \\
\CIV~$\lambda \lambda 5801,5812$ &  2.32 &    LT/SPRAT & $1.33 \times 10^{-13}$ & $4.21 \times 10^{-14}$ \\
\CIV~$\lambda \lambda 5801,5812$ &  2.62 &   Keck/LRIS & $4.05 \times 10^{-13}$ & $1.69 \times 10^{-14}$ \\
\CIV~$\lambda \lambda 5801,5812$ &  3.10 &    LT/SPRAT & $6.24 \times 10^{-13}$ & $4.38 \times 10^{-14}$ \\
\enddata
\label{tab:fluxes}
\end{deluxetable}

\clearpage

\begin{deluxetable}{cccc}
\tablecaption{EW of \ion{Na}{1}~D components}
\tabletypesize{\footnotesize}
\tablewidth{0pt}
\tablehead{
\colhead{Na line} &
\colhead{Component} &
\colhead{Phase}& 
\colhead{EW} \\
\colhead{} &
\colhead{[$\kms$]} &
\colhead{[days]}& 
\colhead{[$\rm \AA$]} \\
}
\setlength{\tabcolsep}{12pt}

\startdata
D1 &         $30 \kms$ &  11.11 &    $0.0114\pm0.0004$ \\
D2 &         $30 \kms$ &  11.11 &    $0.0049\pm0.0004$ \\
D1 &         $30 \kms$ &  21.15 &    $0.0108\pm0.0005$ \\
D2 &         $30 \kms$ &  21.15 &    $0.0045\pm0.0005$ \\
D1 &         $30 \kms$ &  43.09 &    $0.0074\pm0.0007$ \\
D2 &         $30 \kms$ &  43.09 &    $0.0029\pm0.0007$ \\
D1 &         $30 \kms$ &  56.07 &    $0.0088\pm0.0011$ \\
D2 &         $30 \kms$ &  56.07 &    $0.0027\pm0.0011$ \\
D1 &         $8 \kms$ &  11.11 &  $0.04705\pm0.00023$ \\
D2 &         $8 \kms$ &  11.11 &  $0.03251\pm0.00028$ \\
D1 &         $8 \kms$ &  21.15 &  $0.04659\pm0.00029$ \\
D2 &         $8 \kms$ &  21.15 &  $0.03191\pm0.00035$ \\
D1 &         $8 \kms$ &  43.09 &    $0.0447\pm0.0004$ \\
D2 &         $8 \kms$ &  43.09 &    $0.0300\pm0.0005$ \\
D1 &         $8 \kms$ &  56.07 &    $0.0437\pm0.0006$ \\
D2 &         $8 \kms$ &  56.07 &    $0.0294\pm0.0007$ \\
D1 &    Entire complex &  11.11 &    $0.1921\pm0.0011$ \\
D2 &    Entire complex &  11.11 &    $0.1265\pm0.0011$ \\
D1 &    Entire complex &  21.15 &    $0.1943\pm0.0013$ \\
D2 &    Entire complex &  21.15 &    $0.1226\pm0.0013$ \\
D1 &    Entire complex &  43.09 &    $0.1851\pm0.0017$ \\
D2 &    Entire complex &  43.09 &    $0.1269\pm0.0017$ \\
D1 &    Entire complex &  56.07 &    $0.1845\pm0.0026$ \\
D2 &    Entire complex &  56.07 &    $0.1307\pm0.0027$ \\
\enddata
\label{tab:NaD}
\end{deluxetable}

\clearpage

\begin{deluxetable}{ccc}
\tablewidth{0pt}
\tablecaption{Log of XRT observations}
\tabletypesize{\footnotesize}
\setlength{\tabcolsep}{0pt}
\tablehead{
\colhead{MJD} &
\colhead{Count rate} &
\colhead{$\rm Flux~(0.3-10~\rm keV)$} \\
\colhead{} &
\colhead{($10^{-3}~{\rm s}^{-1}$)} &
\colhead{($10^{-12}~\rm erg\,s^{-1}\,cm^{-2}\,s^{-1}$)}
}
\setlength{\tabcolsep}{12pt}

\startdata

$60086.05^{+ 0.87}_{-1.78}$ &   $0.98\pm 0.3$ &  $0.1\pm 0.03$ \\
$60088.64^{+ 0.99}_{-1.34}$ &  $4.17\pm 1.02$ & $0.44\pm 0.11$ \\
$60090.63^{+ 0.06}_{-0.02}$ &  $9.19\pm 2.55$ & $0.97\pm 0.27$ \\
$60091.54^{+ 0.01}_{-0.01}$ &  $11.0\pm 2.96$ & $1.16\pm 0.31$ \\
$60092.33^{+ 0.01}_{-0.01}$ & $11.22\pm 3.16$ & $1.18\pm 0.33$ \\
$60096.53^{+ 0.71}_{-1.03}$ &   $8.37\pm 2.1$ & $0.88\pm 0.22$ \\
$60099.37^{+ 1.04}_{-1.08}$ &  $6.91\pm 2.01$ & $0.73\pm 0.21$ \\
$60102.55^{+ 0.7}_{-0.24}$ &  $9.04\pm 1.98$ & $0.95\pm 0.21$ \\
$60105.77^{+ 2.91}_{-1.47}$ &  $11.3\pm 2.23$ & $1.19\pm 0.23$ \\
$60120.71^{+ 2.91}_{-2.58}$ &  $8.98\pm 1.95$ &  $0.95\pm 0.2$ \\
$60128.95^{+ 0.75}_{-0.38}$ &  $7.15\pm 1.49$ & $0.75\pm 0.16$ \\
$60149.61^{+ 11.55}_{-12.79}$ &  $3.84\pm 0.82$ &  $0.4\pm 0.09$ \\
$60168.91^{+ 2.58}_{-2.25}$ & $12.93\pm 2.74$ & $1.36\pm 0.29$ \\
$60185.87^{+ 5.35}_{-9.52}$ &  $3.08\pm 0.91$ &  $0.32\pm 0.1$ \\
$60203.69^{+ 5.31}_{-2.18}$ &  $6.93\pm 1.79$ & $0.73\pm 0.19$ \\
$60213.43^{+ 0.2}_{-0.35}$ &   $6.74\pm 1.9$ &  $0.71\pm 0.2$ \\
$60213.82^{+ 0.06}_{-0.06}$ &  $8.52\pm 5.89$ &  $0.9\pm 0.62$ \\

\enddata
\label{tab:xrt}
\tablecomments{The diffuse emission at the SN site of $(3.8\pm0.5)\times10^{-4}~\rm ct\,s^{-1}$ was subtracted from all measurements. The flux reports the brightness without accounting for absorption.}
\end{deluxetable}

\clearpage

\begin{deluxetable}{ccccccccc}
\tablecaption{Log of Blackbody fits}
\tablewidth{0pt}
\tabletypesize{\footnotesize}
\setlength{\tabcolsep}{1pt}
\tablehead{
Source &
JD &
$t$ &
$T_{\rm bb}$ &
$R_{\rm bb}$ &
$L_{\rm bb}$ &
$L_{\rm pseudo}$ &
$\rm L_{pseudo,extrap}$ &
$\chi^{2}/{\rm dof}$ \\
&
&
[days]&
$[\rm K]$ &
$[10^{14}\, \rm cm]$ &
$[10^{42}\, \rm erg\, \rm s^{-1}]$ &
$[10^{42}\, \rm erg\, \rm s^{-1}]$ &
$[10^{42}\,\rm erg\ \rm s^{-1}]$ &

}

\startdata
P & 2460084.43 & 1.13 & $12900\pm700$ & $1.63\pm0.1$ & $0.52\pm0.05$ & $0.2\pm0.2$ & $0.53\pm0.03$ & 1.61 \\
P & 2460084.51 & 1.21 & $13800\pm800$ & $1.67\pm0.11$ & $0.72\pm0.07$ & $0.26\pm0.26$ & $0.73\pm0.05$ & 1.58 \\
P & 2460084.61 & 1.31 & $14500\pm800$ & $1.79\pm0.11$ & $1.0\pm0.11$ & $0.33\pm0.33$ & $1.02\pm0.07$ & 1.38 \\
P & 2460084.73 & 1.42 & $16200\pm500$ & $1.92\pm0.06$ & $1.83\pm0.11$ & $0.5\pm0.5$ & $1.85\pm0.08$ & 0.5 \\
P & 2460084.87 & 1.57 & $20900\pm1400$ & $1.79\pm0.11$ & $4.36\pm0.6$ & $2.23\pm2.23$ & $4.63\pm0.33$ & 3.04 \\
P & 2460085.04 & 1.74 & $24300\pm1500$ & $1.86\pm0.11$ & $8.58\pm1.16$ & $3.29\pm3.29$ & $8.81\pm0.74$ & 4.41 \\
P & 2460085.25 & 1.95 & $26400\pm2700$ & $2.02\pm0.19$ & $14.1\pm3.22$ & $5.2\pm5.2$ & $14.94\pm2.23$ & 8.27 \\
P & 2460085.5 & 2.2 & $27300\pm3500$ & $2.26\pm0.26$ & $20.23\pm5.89$ & $7.49\pm7.49$ & $21.86\pm4.18$ & 11.03 \\
P & 2460085.8 & 2.5 & $30500\pm5400$ & $2.32\pm0.35$ & $33.13\pm13.79$ & $9.41\pm9.41$ & $34.75\pm10.55$ & 13.39 \\
P & 2460086.16 & 2.86 & $27900\pm5000$ & $2.79\pm0.43$ & $33.67\pm14.04$ & $11.48\pm11.48$ & $35.63\pm10.07$ & 18.44 \\
P & 2460086.6 & 3.29 & $27000\pm4300$ & $3.11\pm0.43$ & $36.33\pm13.4$ & $12.93\pm12.93$ & $38.29\pm9.35$ & 15.6 \\
S & 2460086.86 & 3.51 & $34300\pm13000$ & $2.63\pm0.36$ & $68.59\pm78.18$ & $20.48\pm20.48$ & $68.59\pm29.23$ & 3.72 \\
P & 2460087.12 & 3.82 & $26900\pm4400$ & $3.22\pm0.47$ & $38.85\pm14.64$ & $13.77\pm13.77$ & $40.85\pm10.2$ & 17.3 \\
P & 2460087.75 & 4.45 & $20600\pm2100$ & $4.45\pm0.42$ & $25.46\pm5.51$ & $15.09\pm15.09$ & $28.55\pm2.91$ & 11.14 \\
S & 2460087.98 & 4.63 & $25900\pm6200$ & $3.24\pm0.56$ & $33.81\pm19.24$ & $16.25\pm16.25$ & $33.81\pm29.23$ & 5.13 \\
P & 2460088.52 & 5.21 & $20200\pm1400$ & $4.7\pm0.32$ & $26.38\pm3.87$ & $13.86\pm13.86$ & $27.47\pm2.0$ & 7.33 \\
S & 2460088.85 & 5.49 & $23000\pm4000$ & $3.62\pm0.56$ & $26.14\pm9.16$ & $14.68\pm14.68$ & $26.14\pm29.23$ & 4.55 \\
P & 2460089.44 & 6.13 & $17800\pm900$ & $5.41\pm0.29$ & $20.75\pm2.13$ & $13.73\pm13.73$ & $22.53\pm0.9$ & 4.32 \\
P & 2460090.54 & 7.24 & $16700\pm800$ & $5.78\pm0.29$ & $18.38\pm1.68$ & $12.89\pm12.89$ & $19.86\pm0.64$ & 4.24 \\
P & 2460091.88 & 8.57 & $15100\pm700$ & $6.43\pm0.35$ & $15.4\pm1.33$ & $11.73\pm11.73$ & $16.57\pm0.42$ & 5.83 \\
S & 2460092.04 & 8.68 & $14600\pm1000$ & $5.67\pm0.77$ & $10.38\pm1.04$ & $8.92\pm8.92$ & $10.38\pm29.23$ & 3.99 \\
P & 2460093.48 & 10.18 & $13000\pm400$ & $7.53\pm0.32$ & $11.64\pm0.62$ & $9.98\pm9.98$ & $12.6\pm0.14$ & 4.76 \\
P & 2460095.03 & 11.72 & $11800\pm400$ & $8.53\pm0.37$ & $10.11\pm0.48$ & $9.26\pm9.26$ & $11.06\pm0.09$ & 6.24 \\
S & 2460095.04 & 11.68 & $11800\pm600$ & $7.24\pm0.67$ & $7.16\pm0.48$ & $6.53\pm6.53$ & $7.16\pm29.23$ & 3.15 \\
P & 2460097.06 & 13.75 & $10200\pm200$ & $10.06\pm0.27$ & $7.74\pm0.19$ & $7.21\pm7.21$ & $8.21\pm0.03$ & 3.64 \\
P & 2460099.1 & 15.78 & $9100\pm100$ & $11.51\pm0.34$ & $6.4\pm0.16$ & $5.94\pm5.94$ & $6.64\pm0.02$ & 4.57 \\
P & 2460101.13 & 17.81 & $8600\pm100$ & $12.25\pm0.37$ & $5.74\pm0.14$ & $5.33\pm5.33$ & $5.95\pm0.02$ & 5.03 \\
P & 2460103.16 & 19.85 & $8000\pm100$ & $13.22\pm0.43$ & $5.06\pm0.14$ & $4.68\pm4.68$ & $5.24\pm0.02$ & 6.18 \\
P & 2460105.19 & 21.88 & $7500\pm100$ & $14.14\pm0.51$ & $4.54\pm0.14$ & $4.18\pm4.18$ & $4.7\pm0.02$ & 7.91 \\
P & 2460107.23 & 23.91 & $7100\pm100$ & $14.83\pm0.58$ & $4.06\pm0.14$ & $3.69\pm3.69$ & $4.19\pm0.02$ & 9.72 \\
P & 2460109.26 & 25.94 & $6700\pm100$ & $15.59\pm0.7$ & $3.58\pm0.15$ & $3.21\pm3.21$ & $3.69\pm0.02$ & 12.8 \\
P & 2460111.29 & 27.97 & $6600\pm100$ & $15.82\pm0.71$ & $3.35\pm0.14$ & $2.98\pm2.98$ & $3.46\pm0.02$ & 12.71 \\
P & 2460113.33 & 30.0 & $6400\pm100$ & $16.22\pm0.75$ & $3.15\pm0.14$ & $2.77\pm2.77$ & $3.24\pm0.02$ & 13.54 \\
P & 2460115.36 & 32.03 & $6200\pm100$ & $16.92\pm0.86$ & $2.95\pm0.14$ & $2.59\pm2.59$ & $3.06\pm0.02$ & 16.08 \\
P & 2460117.39 & 34.06 & $5900\pm100$ & $17.81\pm1.08$ & $2.73\pm0.17$ & $2.4\pm2.4$ & $2.88\pm0.03$ & 20.29 \\
P & 2460119.42 & 36.1 & $5600\pm100$ & $18.6\pm1.28$ & $2.49\pm0.18$ & $2.24\pm2.24$ & $2.72\pm0.03$ & 24.28 \\
P & 2460121.46 & 38.13 & $5500\pm100$ & $19.06\pm1.46$ & $2.38\pm0.19$ & $2.13\pm2.13$ & $2.61\pm0.04$ & 28.45 \\
P & 2460123.49 & 40.16 & $5400\pm100$ & $19.77\pm1.58$ & $2.3\pm0.19$ & $2.04\pm2.04$ & $2.53\pm0.04$ & 30.09 \\
P & 2460125.52 & 42.19 & $5300\pm100$ & $20.09\pm1.75$ & $2.21\pm0.2$ & $1.96\pm1.96$ & $2.44\pm0.04$ & 33.6 \\
P & 2460127.56 & 44.22 & $5200\pm200$ & $20.17\pm1.95$ & $2.12\pm0.21$ & $1.88\pm1.88$ & $2.36\pm0.05$ & 38.72 \\
P & 2460129.59 & 46.25 & $5200\pm300$ & $20.16\pm2.6$ & $2.04\pm0.23$ & $1.8\pm1.8$ & $2.27\pm0.05$ & 42.47 \\
P & 2460131.62 & 48.28 & $4900\pm200$ & $22.0\pm2.67$ & $1.98\pm0.2$ & $1.69\pm1.69$ & $2.25\pm0.06$ & 34.94 \\
P & 2460133.66 & 50.32 & $4800\pm200$ & $22.49\pm2.85$ & $1.92\pm0.21$ & $1.64\pm1.64$ & $2.19\pm0.06$ & 38.95 \\
P & 2460135.69 & 52.35 & $4600\pm300$ & $24.98\pm3.94$ & $2.02\pm0.28$ & $1.24\pm1.24$ & $2.28\pm0.14$ & 46.1 \\
P & 2460137.72 & 54.38 & $4600\pm300$ & $25.09\pm4.04$ & $2.01\pm0.27$ & $1.22\pm1.22$ & $2.26\pm0.14$ & 44.38 \\
P & 2460139.76 & 56.41 & $4500\pm300$ & $26.01\pm4.25$ & $1.99\pm0.28$ & $1.19\pm1.19$ & $2.23\pm0.15$ & 46.96 \\
P & 2460141.79 & 58.44 & $4400\pm200$ & $27.52\pm4.58$ & $1.97\pm0.3$ & $1.16\pm1.16$ & $2.24\pm0.16$ & 51.22 \\
P & 2460143.82 & 60.47 & $4300\pm200$ & $28.28\pm4.85$ & $1.93\pm0.31$ & $1.11\pm1.11$ & $2.19\pm0.17$ & 55.29 \\
P & 2460145.85 & 62.5 & $4200\pm200$ & $28.5\pm4.88$ & $1.88\pm0.3$ & $1.08\pm1.08$ & $2.15\pm0.17$ & 53.92 \\
P & 2460147.89 & 64.53 & $4100\pm200$ & $29.72\pm5.19$ & $1.85\pm0.31$ & $1.04\pm1.04$ & $2.12\pm0.18$ & 56.32 \\
P & 2460149.92 & 66.57 & $4100\pm200$ & $29.99\pm5.47$ & $1.8\pm0.31$ & $0.99\pm0.99$ & $2.05\pm0.19$ & 61.02 \\
P & 2460151.95 & 68.6 & $4000\pm200$ & $31.16\pm5.62$ & $1.78\pm0.32$ & $0.95\pm0.95$ & $2.02\pm0.19$ & 62.18 \\
P & 2460153.98 & 70.63 & $3900\pm200$ & $31.87\pm5.67$ & $1.75\pm0.31$ & $0.89\pm0.89$ & $1.97\pm0.19$ & 60.14 \\
P & 2460156.02 & 72.66 & $3900\pm200$ & $30.67\pm5.56$ & $1.61\pm0.29$ & $0.8\pm0.8$ & $1.8\pm0.18$ & 58.37 \\
P & 2460158.05 & 74.69 & $3900\pm200$ & $30.73\pm5.6$ & $1.56\pm0.28$ & $0.77\pm0.77$ & $1.74\pm0.17$ & 59.24 \\
P & 2460160.08 & 76.72 & $3800\pm200$ & $32.0\pm5.73$ & $1.53\pm0.28$ & $0.72\pm0.72$ & $1.7\pm0.18$ & 57.98 \\
P & 2460162.12 & 78.75 & $3800\pm200$ & $31.35\pm5.4$ & $1.42\pm0.25$ & $0.64\pm0.64$ & $1.56\pm0.16$ & 55.14 \\
P & 2460164.15 & 80.79 & $3800\pm200$ & $28.76\pm4.72$ & $1.18\pm0.2$ & $0.52\pm0.52$ & $1.29\pm0.13$ & 49.96 \\
P & 2460166.18 & 82.82 & $3900\pm200$ & $24.19\pm3.62$ & $0.93\pm0.13$ & $0.4\pm0.4$ & $0.99\pm0.08$ & 36.35 \\
P & 2460168.22 & 84.85 & $3900\pm200$ & $21.17\pm2.75$ & $0.74\pm0.09$ & $0.32\pm0.32$ & $0.78\pm0.06$ & 25.59 \\
P & 2460170.25 & 86.88 & $3900\pm100$ & $20.0\pm2.39$ & $0.64\pm0.07$ & $0.27\pm0.27$ & $0.67\pm0.04$ & 20.96 \\
P & 2460172.28 & 88.91 & $3900\pm100$ & $18.91\pm2.2$ & $0.56\pm0.06$ & $0.23\pm0.23$ & $0.58\pm0.04$ & 20.19 \\
P & 2460174.32 & 90.94 & $3900\pm100$ & $18.24\pm2.06$ & $0.53\pm0.06$ & $0.22\pm0.22$ & $0.55\pm0.04$ & 19.34 \\
P & 2460176.35 & 92.97 & $4100\pm200$ & $14.93\pm1.8$ & $0.47\pm0.04$ & $0.2\pm0.2$ & $0.48\pm0.03$ & 13.03 \\
P & 2460178.38 & 95.0 & $4100\pm200$ & $14.91\pm1.79$ & $0.46\pm0.04$ & $0.2\pm0.2$ & $0.48\pm0.03$ & 12.95 \\
P & 2460180.41 & 97.04 & $4100\pm200$ & $14.72\pm1.78$ & $0.45\pm0.04$ & $0.19\pm0.19$ & $0.46\pm0.03$ & 13.25 \\
P & 2460182.45 & 99.07 & $4100\pm200$ & $14.51\pm1.81$ & $0.44\pm0.04$ & $0.19\pm0.19$ & $0.45\pm0.03$ & 14.17 \\
P & 2460184.48 & 101.1 & $4100\pm200$ & $14.49\pm1.83$ & $0.43\pm0.04$ & $0.18\pm0.18$ & $0.44\pm0.03$ & 14.8 \\
P & 2460186.51 & 103.13 & $4100\pm200$ & $14.22\pm1.79$ & $0.42\pm0.04$ & $0.18\pm0.18$ & $0.44\pm0.02$ & 14.38 \\
P & 2460188.54 & 105.16 & $4100\pm200$ & $14.08\pm1.93$ & $0.41\pm0.04$ & $0.18\pm0.18$ & $0.43\pm0.03$ & 17.22 \\
P & 2460190.58 & 107.19 & $4100\pm200$ & $14.04\pm1.79$ & $0.41\pm0.04$ & $0.17\pm0.17$ & $0.42\pm0.02$ & 15.13 \\
P & 2460192.61 & 109.22 & $4100\pm200$ & $13.83\pm2.52$ & $0.4\pm0.06$ & $0.17\pm0.17$ & $0.41\pm0.04$ & 17.73 \\
\hline
\enddata
\label{tab:bb_tab}
\tablecomments{P stands for photometry, S for spectroscopy. $L_{\rm bb}$ is the best-fit result, $L_{\rm pseudo}$ is calculated by integrating over the interpolated SED. $L_{\rm pseudo,extrap}$ is calculated from the interpolated SED with a bolometric extrapolation as a correction to the missing UV and IR bands.}
\end{deluxetable}

\clearpage

\begin{deluxetable}{ccc}
\tablewidth{0pt}
\tablecaption{Emission lines of prominent star-forming regions close to SN\,2023ixf. All measurements are not corrected for reddening.}
\tabletypesize{\footnotesize}
\tablehead{        & West  & North\\
Species & Flux      & Flux\\
        & $\left(10^{-17\,}\rm erg\,cm^{-2}\,s^{-1}\right)$ & $\left(10^{-17}\,\rm erg\,cm^{-2}\,s^{-1}\right)$\\}
\startdata
\setlength{\tabcolsep}{12pt}

H$\beta$                        & $147.1 \pm  10.1$  & $3664.5  \pm 25.8$\\ 
$[\rm O~III]\,\lambda$\,4960 & $22.1 \pm 8.6 $ & $ 2169.5 \pm 30.0$\\
$[\rm O~III]\,\lambda$\,5007 & $72.1 \pm  10.2$ & $6533.1  \pm 35.8$\\
H$\alpha$                        & $609.4\pm  4.8$  & $13947.2 \pm 62.7$\\
$[\rm N~II]\,\lambda$\,6584 & $141.4\pm  3.9$   & $2534.4  \pm 49.2$\\
\enddata
\label{tab:hii_lines}
\end{deluxetable}

\clearpage

\begin{deluxetable}{ccccc}
\tablewidth{0pt}
\tablecaption{Log of \swift grism spectra}
\tabletypesize{\footnotesize}
\tablehead{OBSID & Exposure ID & Start time & Exposure time & Astrometry$^*$ \\
 &&& (second) & }
\startdata
00032481003 & 1 & 2023-05-20T11:21:43 & 720 & g \\
00032481003 & 2 & 2023-05-20T15:43:21 & 1718 & g \\
00032481003 & 3 & 2023-05-20T17:17:32 & 849 & g \\
00032481003 & 4 & 2023-05-20T18:51:50 & 911 & g \\
00032481003 & 5 & 2023-05-20T22:26:19 & 826 & f \\
00032481008 & 1 & 2023-05-21T04:23:34 & 1717 & g \\
00032481008 & 2 & 2023-05-21T05:57:59 & 1720 & g \\
00032481008 & 3 & 2023-05-21T07:35:19 & 1501 & g \\
00016038002 & 1 & 2023-05-21T18:40:42 & 1662 & g \\
00016038002 & 2 & 2023-05-21T20:14:35 & 991 & g \\
00016038004 & 1 & 2023-05-22T07:22:45 & 1719 & g \\
00016038004 & 2 & 2023-05-22T08:57:53 & 989 & g \\
00016038010 & 1 & 2023-05-22T18:28:20 & 1717 & g \\
00016038010 & 2 & 2023-05-22T20:05:21 & 988 & g \\
00016038008 & 1 & 2023-05-23T07:17:18 & 1320 & f \\
00016038012 & 1 & 2023-05-24T05:23:39 & 901 & g \\
00016038012 & 2 & 2023-05-24T06:59:09 & 892 & g \\
00016043002 & 1 & 2023-05-25T05:13:19 & 901 & f \\
00016043002 & 2 & 2023-05-25T14:45:19 & 1008 & f \\
00016043004 & 1 & 2023-05-26T14:34:20 & 901 & f \\
00016043004 & 2 & 2023-05-26T16:19:19 & 901 & f \\
00016043006 & 1 & 2023-05-27T12:47:21 & 1599 & f \\
00016043008 & 1 & 2023-05-28T07:48:19 & 1331 & f \\
00016043010 & 1 & 2023-06-01T07:06:20 & 1322 & f \\
00016043013 & 1 & 2023-06-05T09:26:19 & 1331 & f \\
00016043015 & 1 & 2023-06-07T07:27:19 & 1322 & f \\
00016043017 & 1 & 2023-06-08T05:41:18 & 624 & f \\
00032481034 & 1 & 2023-06-25T10:38:18 & 1053 & f \\
00032481034 & 2 & 2023-06-25T15:11:19 & 1322 & f \\
00032481034 & 3 & 2023-06-25T21:43:19 & 1107 & f \\
00032481038 & 1 & 2023-07-03T13:39:19 & 1215 & f \\
00032481038 & 2 & 2023-07-03T23:16:19 & 892 & f \\
00032481040 & 1 & 2023-07-04T14:53:20 & 1267 & f \\
00032481040 & 2 & 2023-07-04T16:27:21 & 1160 & f \\
\enddata
\tablecomments{$^*$This flag indicates the astrometry calibration method used for spectrum extraction; ``f'' corresponds to cases in which the grism exposures are paired with a lenticular filter, and the lenticular filter image is used to refine the astrometry when extracting the spectrum; ``g'' corresponds to grism exposure without a paired lenticular filter image. }
\label{tab:grism_obsinfo}
\end{deluxetable}

 
\end{document}